\documentclass[11pt,a4paper]{article} 
\pdfoutput=1
\usepackage{jcappub}
\usepackage{enumerate}
\usepackage{epsfig}
\usepackage{wasysym} 
\usepackage{mathrsfs} 
\usepackage{graphicx} 
\usepackage{amsfonts} 
\usepackage{amsbsy} 
\usepackage{amscd}
\usepackage{slashed} 
\usepackage{multirow} 
\usepackage{gensymb}
\usepackage{subcaption}
\usepackage{natbib}
\allowdisplaybreaks
\usepackage[section]{placeins}
\usepackage{rotating}
\usepackage{booktabs}
\usepackage{color}
\definecolor{myred}{rgb}{0.6,0,0} 
\definecolor{myblue}{rgb}{0,0.2,0.4}
\definecolor{mygreen}{rgb}{0,0.9,0.1}
\definecolor{hc}{rgb}{.9,0.1,0.7}
\definecolor{hcout}{rgb}{.9,0.7,0.9}
\definecolor{Orange}{rgb}{0.25,0.75,0.03}
\definecolor{Brown}{rgb}{0.95,0.01,0.35}
\makeatletter

\newcommand{\fmslash}[2][0mu]{%
	\mathchoice
	{\fmsl@sh\displaystyle{#1}{#2}}%
	{\fmsl@sh\textstyle{#1}{#2}}%
	{\fmsl@sh\scriptstyle{#1}{#2}}%
	{\fmsl@sh\scriptscriptstyle{#1}{#2}}}
\newcommand{\fmsl@sh}[3]{%
	\m@th\ooalign{$\hfil#1\mkern#2/\hfil$\crcr$#1#3$}}
\makeatother

\newcommand{\lsim}{{\;\raise0.3ex\hbox{$<$\kern-0.75em\raise-1.1ex\hbox{$\sim$}}\;}}
\newcommand{\gsim}{{\;\raise0.3ex\hbox{$>$\kern-0.75em\raise-1.1ex\hbox{$\sim$}}\;}}

 \newcommand{\overbar}[1]{\mkern
	1.5mu\overline{\mkern-1.5mu#1\mkern-1.5mu}\mkern 1.5mu}

\usepackage{array}
\newcolumntype{C}[1]{>{\centering\arraybackslash$}p{#1}<{$}}

\usepackage{bm} 
\newcommand{\be}{\begin{equation}}
\newcommand{\ee}{\end{equation}}
\newcommand{\bes}{\begin{equation*}}
\newcommand{\ees}{\end{equation*}}
\newcommand{\bea}{\begin{eqnarray}}
\newcommand{\eea}{\end{eqnarray}}
\newcommand{\beas}{\begin{eqnarray*}}
	\newcommand{\eeas}{\end{eqnarray*}}
\title{A general study of decaying scalar dark matter: existing limits 
and projected radio signals at the SKA}

\author[a]{Koushik Dutta,}
\author[b]{Avirup Ghosh,} 
\author[c]{Arpan Kar,} 
\author[a]{Biswarup Mukhopadhyaya}

\affiliation[a]{Department of Physical Sciences, Indian Institute of Science Education and Research, Kolkata, Mohanpur - 741246, India.}  
\affiliation[b]{School of Physical Sciences, Indian Association for the Cultivation of Science, 2A and 2B Raja S.C. Mullick Road, Kolkata 700 032.}
\affiliation[c]{Center for Quantum Spacetime, Sogang University, Seoul 121-742, South Korea.}

\emailAdd{koushik@iiserkol.ac.in}
\emailAdd{avirup.ghosh1993@gmail.com}
\emailAdd{arpankarphys@gmail.com} 
\emailAdd{biswarup@iiserkol.ac.in}

\abstract{
We consider a decaying scalar dark matter (DM) with mass $m_\chi$ 
in the range 10 GeV - 10 TeV and vary the branching ratios of all 
possible two-body SM final states (excluding and including $\nu\bar{\nu}$) 
in the range $0\%-100\%$ to derive constraints on the total 
decay width $\Gamma$ using the data collected by several 
astrophysical and cosmological observations. We find that, 
$\Gamma \lesssim 10^{-26} - 10^{-27}\,{\rm s}^{-1}$ (excluding $\nu\bar{\nu}$) 
and $\Gamma \lesssim 10^{-24} - 10^{-26}\,{\rm s}^{-1}$ (including $\nu\bar{\nu}$) 
are allowed, depending on the values of $m_\chi$, which are most 
robust upper limits on $\Gamma$ for a generic decaying scalar DM. 
We then investigate the prospect of the upcoming Square Kilometre 
Array (SKA) radio telescope in detecting the DM decay induced radio 
signals originating inside the dwarf spheroidal (dSph) galaxies. 
We have classified the DM parameter space, allowed by the existing 
observations, independently of the branching ratio of each individual 
two-body SM final state, based on the detectability at the SKA. 
Excluding the $\nu\bar{\nu}$ decay mode, we find that, throughout 
the DM mass range considered, 
$\Gamma \gtrsim 10^{-30}\,{\rm s}^{-1} - 10^{-29}\,{\rm s}^{-1}$ 
is detectable for all possible branching ratio combinations 
at the SKA (assuming 100 hours of observation time), with conservative 
choices for the relevant astrophysical parameters. 
On the other hand, when arbitrary branching ratios are allowed 
also for the $\nu\bar{\nu}$ decay mode, DM decays can be probed 
independently of the branching ratio of each SM final state 
for $\Gamma \gtrsim 2 \times 10^{-29}\,{\rm s}^{-1}$, 
provided DM masses are greater than a few hundreds of GeV. 
}

\date{\today}


\begin{document}
\maketitle
\newpage
	
\section{Introduction}
\label{sec:intro}

Various astrophysical and cosmological observations over the last few 
decades point towards the fact that an unavoidably large fraction 
of the total energy density of our Universe (about 23.4$\%$) is 
constituted by a yet unknown material component dubbed as `dark 
matter (DM)'~\cite{Sofue:2000jx,Markevitch:2001ri,Markevitch:2003at,Planck:2015fie}. 
A popular idea is that the DM consists of one or more unidentified 
massive elementary particle(s). An often discussed candidate 
is some Weakly Interacting Massive Particle (WIMP)~\cite{Jungman:1995df,Arcadi:2017kky} 
whose mass lies in the GeV - TeV range. 
Mainly three experimental techniques, namely, direct search experiments, 
collider searches and indirect search observations, have been developed to 
detect such WIMP candidates. However, no confirmatory signals of 
the WIMP DM have been observed yet~\cite{Roszkowski:2017nbc}, and 
hence the need for the exploration of alternative scenarios of dark 
matter are largely felt.
Among various proposed DM scenarios outside the WIMP paradigm, 
Feebly Interacting Massive Particles (FIMPs)~\cite{Hall:2009bx}, 
Strongly Interacting Massive Particles (SIMPs)~\cite{Hochberg:2014dra},  
ELastically DEcoupling Relics (ELDERs)~\cite{Kuflik:2015isi} etc., are 
the most popular ones. The common feature shared by most of these 
scenarios is the negligibly small interactions between the dark matter 
and the Standard Model (SM) particles. Various signals of such DM candidates 
have been studied extensively in the 
literature~\cite{Belanger:2018sti,Choudhury:2008gb,Gupta:2007ui,Hessler:2016kwm,Daci:2015hca,Kuflik:2017iqs,Elor:2021swj}. 

Such DM candidates, including WIMPs, are usually assumed to be absolutely 
stable. This immediately raises another curiosity: can the frequently 
postulated symmetry disallowing DM decays be broken in such a 
way that the DM candidate may decay into SM particle 
pairs at extremely slow rates? The observed large scale  
structure of our Universe requires the DM particle to be long-lived 
only on a cosmological time scale, and the possibility that the DM decays  
with a lifetime much larger than the age of the Universe is thus 
not inconceivable~\cite{Takayama:2000uz,Buchmuller:2007ui,Ibarra:2007wg,Covi:2009pq,Endo:2013si,Ghosh:2019jzu,Carone:2011iw,Feng:2013vva,Kyae:2012vi,Park:2012xq,Mambrini:2015sia,Ando:2015qda,Choi:2013oaa,Cata:2016epa,Sun:2021jna}.
It is a likely proposition that the long-lived 
nature of the DM candidate is due to some continuous global 
symmetry~\cite{Yang:2013bda,Yang:2014jca,Bhattacharyya:2013rya}. 
Such a symmetry is liable to be broken at some level, albeit by a 
minuscule amount, thus causing extremely slow decays of the DM 
particle~\cite{Mambrini:2015sia,Banks:2010zn,Harlow:2018jwu}.

In such situations, if the only interactions of the DM candidate 
with the SM particles are those leading to DM decays, then the event 
rates in direct search or collider experiments are expected to be low. 
However, photons, antimatter particles and neutrinos, coming from 
the primary products of DM decays can still have substantial 
impacts on the data recorded in several cosmological and astrophysical 
observations. The mass of a decaying DM candidate is largely 
unconstrained~\cite{Kusenko:2012ch,Choi:2013oaa,Hambye:2016qkf,Heeba:2018wtf,Alcantara:2019sco}. 
Therefore, it is a common practice to phenomenologically restrict the 
space spanned by the DM mass 
and its decay width. 

For example, Ref.~\cite{Slatyer:2016qyl} has put upper limits on 
the decay rate of a decaying scalar DM with mass 
in the keV - TeV range using the CMB anisotropy data collected by the Planck 
collaboration~\cite{Planck:2015fie}. Specifically, the energy injections 
by the stable SM decay products of the DM candidate, into the photon-baryon  
plasma during the cosmic dark ages perturb the CMB anisotropy spectra, 
which is severely constrained by the Planck data~\cite{Planck:2015fie}.   
The isotropic gamma-ray background (IGRB), on the other hand, may receive 
contributions from the gamma-ray fluxes produced in the DM decays, 
in addition to the dominant contributions coming from the 
active galactic nuclei (AGN) and the star-forming galaxies. 
Thus the IGRB fluxes measured by SAS-2~\cite{osti_7051553}, 
EGRET~\cite{EGRET:1997qcq},  Fermi-LAT~\cite{Fermi-LAT:2010pat,Fermi-LAT:2014ryh} 
etc., can be compared against the gamma-ray fluxes induced by a 
decaying DM candidate 
to derive upper bounds on the DM decay width 
for the chosen value of the DM mass~\cite{Liu:2016ngs,Blanco:2018esa}. 

In the matter sector, antimatter particles are likely to form even more 
spectacular DM signals. Thus, the data from the antimatter 
searches performed by CAPRICE~\cite{Grimani:2002yz}, 
HEAT~\cite{Schubnell:2009gk}, AMS-01~\cite{AMS01:2007rrn},  
PAMELA~\cite{PAMELA:2008gwm}, AMS-02~\cite{AMS:2014xys,AMS:2015tnn,AMS:2019rhg} 
etc., can be used to set upper limits on the DM decay generated 
antimatter fluxes which in turn give stringent constraints on the 
parameter space of a decaying DM particle~\cite{Ibarra:2013zia,Lu:2015pta,John:2021ugy}. 
In addition, neutrino fluxes induced by the DM decay products are also detectable in 
several neutrino telescopes and hence, neutrino observations by  
Super-Kamiokande~\cite{Frankiewicz:2016nyr,Coy:2020wxp}, IceCube~\cite{Coy:2020wxp,Cohen:2016uyg,IceCube:2018tkk,IceCube:2021sog} 
etc., too, contribute to the existing limits on decaying DM. 
Multi-messenger analyses using the data taken by several gamma-ray, 
cosmic-ray and neutrino observations, are also available in the 
literature~\cite{Ishiwata:2019aet}.  

Yet another signature of decaying DM consists in the radio synchrotron 
emission from the $e^+e^-$ pairs originating in DM decays occurring 
inside DM dominated galaxies and clusters. These electrons (positrons) 
undergo energy loss via electromagnetic interactions in the interstellar 
medium and give rise to radio waves. Nearby dwarf spheroidal (dSph) 
galaxies are popular sources for studying such radio signals, because of 
their low-star formation rates and high mass-to-light 
ratios~\cite{Mateo:1998wg,Strigari:2006rd,Strigari:2007at,Strigari:2008ib} 
which ensure a comparatively lower astrophysical background 
and larger DM abundance inside them. 

Data taken by several existing radio observations have 
been used to draw constraints on the DM parameter space
\cite{Acero:2013xta,Natarajan:2013dsa,Natarajan:2015hma,Regis:2017oet,Cirelli:2016mrc,Regis:2014tga,Bhattacharjee:2020phk}. 
The prospect of the upcoming Square Kilometre Array (SKA) 
radio telescope is found to be quite encouraging in this 
regard~\cite{Colafrancesco:2015ola,Kar:2018rlm,Kar:2019cqo,Dutta:2020lqc,Chen:2021rea,Ghosh:2020ipv}.
In fact, Ref.~\cite{Ghosh:2020ipv} has shown that the SKA can probe much 
deeper into the parameter space of a decaying DM compared to the existing 
gamma-ray observations. The inter-continental baseline lengths of the SKA 
allow one to efficiently resolve the astrophysical foregrounds and its large 
effective area helps to achieve higher surface brightness sensitivity~\cite{SKA} 
compared to other existing radio telescopes. In addition, large frequency range 
of the SKA, i.e., 50 MHz - 50 GHz, has important implications for 
DM masses in the GeV - TeV range. 


A common practice in the existing studies of decaying DM is 
to assume that the DM decays into a specific two-body SM final state at a time 
with 100$\%$ branching ratio and then the DM parameter space is constrained 
by using the data of various cosmological and astrophysical 
observations~\cite{Slatyer:2016qyl,Cirelli:2012ut,Liu:2016ngs,Blanco:2018esa,Ibarra:2013zia,Lu:2015pta,Ishiwata:2019aet}. 
Any specific assumption about the branching fraction of each individual 
DM decay mode is equivalent to committing to a particular underlying DM model. 
However, in a generic model the DM candidate may decay to 
any arbitrary final state with \textit{a priori} undetermined branching ratios. 
Thus the limits obtained 
in such cases, are quite different from the limits derived assuming 100$\%$ 
branching ratio for any SM final state. 

We have taken an unbiased approach and \textit{allowed the branching 
fractions of all kinematically allowed two-body SM final states of 
the decaying scalar DM ($\chi$), to take arbitrary values in the range 
$0\% - 100\%$, while deriving the constraints on the total DM decay 
width ($\Gamma$), for the DM mass ($m_\chi$) in the range 10 GeV - 10 TeV}. 
We have additionally assumed that $\chi$ saturates the entire relic density of the 
Universe. A similar study for the annihilating WIMP DM can be found in~\cite{Leane:2018kjk}. 
It was found there that, \textit{the allowed region of the DM parameter space 
is substantially enlarged} when the above-mentioned approach is adopted.

In deriving the upper limit on total $\Gamma$, i.e., 
$\Gamma_{\rm max}$, we have used the data from the Planck CMB~\cite{Planck:2015fie}, 
Fermi-LAT isotropic gamma-ray background (IGRB)~\cite{Fermi-LAT:2014ryh} 
and AMS-02 positron flux~\cite{AMS:2014xys} observations. 
We have also utilized the observational data of the Super-Kamiokande 
neutrino flux measurement~\cite{Frankiewicz:2016nyr} while 
including 
DM decays to SM neutrinos (i.e., $\chi \rightarrow \nu\bar{\nu}$) in our analysis. 

We emphasize that, the $\nu\bar{\nu}$ decay mode can be constrained 
not only by the neutrino observations such as Super-Kamiokande~\cite{Frankiewicz:2016nyr}, 
but also by the data of Planck CMB~\cite{Planck:2015fie}, 
Fermi-LAT IGRB~\cite{Fermi-LAT:2014ryh} and AMS-02 positron 
flux~\cite{AMS:2014xys} observations. This is because, the final state 
$\nu\bar{\nu}$ pairs give rise to $e^+(e^-)$ and $\gamma$-ray photons 
via the radiation of electroweak gauge bosons. This process is suppressed 
by the $SU(2)_L$ gauge coupling and the gauge boson masses. As a result, 
the rates of $e^+(e^-)$ and $\gamma$ production are thus sizeable 
only when the (anti)neutrinos are energetic enough to produce on-shell 
$W$ and $Z$-bosons. 
Note that the $\nu\bar{\nu}$ final state was not considered in deriving 
the constraints on the total cross-section of annihilating WIMPs~\cite{Leane:2018kjk}. 
However, it should also be emphasized that the constraints on this 
final state coming from the data collected by the neutrino  
telescopes~\cite{Covi:2009xn,Frankiewicz:2016nyr,Coy:2020wxp,IceCube:2018tkk,IceCube:2021sog} 
may be stronger than those obtained from the data of other astrophysical  
observations~\cite{Slatyer:2016qyl}.

We have studied several scenarios differing from each other in the 
branching ratio attributed to the $\nu\bar{\nu}$ decay mode. 
In each case, the obtained value of $\Gamma_{\rm max}$ is independent 
of the specific branching ratios of individual SM final states 
and there exists no possible branching ratio combinations for which any 
value of $\Gamma$ greater than $\Gamma_{\rm max}$ is allowed by the existing data. 
It is found that, 
\textit{$\Gamma_{\rm max}$ obtained here can be considerably 
weaker compared to the limits obtained when 100$\%$ branching ratio 
is assumed for each individual DM decay mode.} 

In the next step, parts of the DM parameter space allowed by the existing 
astrophysical and cosmological observations are classified based on the 
detectability at the SKA assuming 100 hours of observation towards the 
Segue 1 (Seg 1) dSph. In each scenario, 
for every given $m_\chi$, we found the \textit{maximum} and 
the \textit{minimum} radio flux distributions 
by varying the branching ratios of all two-body SM final states in the range 
$0\% - 100\%$. The radio fluxes predicted for all branching ratio combinations 
lie between these \textit{maximum} and \textit{minimum} fluxes in every frequency 
bin. The SKA detectability is then determined depending on whether these 
\textit{maximum} and \textit{minimum} fluxes are above or below the SKA 
sensitivity level obtained assuming 100 hours of observation time.



The detectability at the SKA is presented by dividing the allowed portions 
of the $m_\chi-\Gamma$ plane into three separate regions marked as green, 
yellow and red. The green regions consist of those parameter 
points which are detectable for all possible branching ratio combinations 
of the kinematically allowed two-body SM final states of the DM 
while the yellow regions cover the parts of the DM parameter space that 
are only detectable for certain specific branching ratio combinations. 
Regions of the $m_\chi-\Gamma$ plane which are not detectable at the SKA 
(in the 100 hours of observation) for any possible branching ratio combinations, 
are marked red. 
Note that with more hours of observation time some parts of these non-detectable 
regions of the DM parameter space will also become accessible. We have also 
indicated how the variations of the astrophysical parameters affect our capability 
to probe the decaying DM parameter space. Our study shows that, 
\textit{the SKA will explore a much wider region of the decaying DM parameter 
space compared to the 
existing observations, not only when DM decays to visible SM 
final states are considered but also when $\chi \rightarrow \nu\bar{\nu}$ 
channel is included in the analysis.} 

Our study is `model-independent' in the sense that we have not adhered 
to any definite theoretical framework or specific terms in the DM Lagrangian, 
which determine the branching ratio of each individual decay mode. The following 
assumptions are in built in the study, largely for simplicity: 
\begin{itemize}
\item The scalar $\chi$ decays only into two-body SM final states.
\item $\chi$ decays into fermion pairs are flavour diagonal and also lepton-number conserving.
\item The hadronic showers of $\chi$ decays are faithfully simulated using  
$\mathtt{Pythia}$~\cite{Sjostrand:2006za}.	
\end{itemize}	
 
The paper is organized as follows: in Sec.~\ref{sec:DDMgen} we discuss 
a few salient features of our study followed by presenting the limits 
on individual DM decay modes coming from the Planck CMB, Fermi-LAT IGRB, 
AMS-02 positron flux and Super-Kamiokande neutrino flux observations. 
In Sec.~\ref{sec:modelindepexcons} we derive the 
upper limit on the total DM decay width using the data of the above-mentioned 
observations. 
A brief review of the generation of the radio synchrotron signals from 
the DM decays within the dwarf galaxies is given in 
Sec.~\ref{sec:radiosynch}, while in Sec.~\ref{sec:SKAmethodology} 
we outline the methodology we have adopted to classify the allowed part 
of the DM parameter space based on the detectability at the SKA. In 
Sec.~\ref{sec:SKAresults} we present the projected reach of the SKA. 
Finally, we summarize and conclude in Sec.~\ref{sec:summconc}.

\section{Decaying dark matter: a status review}
\label{sec:DDMgen}

As mentioned in the introduction, here we study the implications of 
the two-body decays of a scalar DM in the context of 
indirect search observations. In this section, we first give a brief 
review of some existing decaying DM scenarios with DM mass in the GeV - TeV 
range and thereby motivate the assumptions we have made in our study. 
Thereafter, we present the limits on 
$\Gamma$ as a function of $m_\chi$ obtained using the data coming from different 
existing astrophysical and cosmological observations assuming DM decays 
to a single channel at a time. This sets the stage for our subsequent 
analysis.
	
\subsection{Motivations and theoretical approach}
\label{sec:DMdecaytheory}

In most scenarios beyond the Standard Model (BSM), the DM 
candidate is stabilized by imposing an \textit{ad-hoc} symmetry 
which is somewhat inexplicable other than in a few well-motivated 
scenarios like the R-parity conserved Minimal Supersymmetric 
Standard Model (MSSM). Moreover, continuous global symmetries 
may be broken by interactions suppressed by the Planck  
scale~\cite{Banks:2010zn,Harlow:2018jwu,Mambrini:2015sia}. 
Therefore, scenarios where the GeV - TeV scale DM particles 
are stable in the cosmological time scales but decay afterwards, 
are not 
uncommon~\cite{Takayama:2000uz,Buchmuller:2007ui,Ibarra:2007wg,Covi:2009pq,Endo:2013si,Ghosh:2019jzu,Carone:2011iw,Feng:2013vva,Kyae:2012vi,Park:2012xq,Mambrini:2015sia,Ando:2015qda,Choi:2013oaa,Cata:2016epa,Sun:2021jna}. 

For example, in R-parity violating supersymmetric (SUSY) scenarios, 
gravitinos ($\tilde{G}$)~\cite{Takayama:2000uz,Buchmuller:2007ui,Ibarra:2007wg} 
and axinos ($\tilde{a}$)~\cite{Covi:2009pq,Endo:2013si} 
are two possible decaying DM candidates. Decaying gravitino lightest 
superparticles (LSPs) of $\mathcal{O}({\rm GeV})$ masses can serve as viable 
candidates for DM in several extensions of R-parity violating SUSY 
scenarios~\cite{Takayama:2000uz,Buchmuller:2007ui} with observable 
signatures in various gamma-ray observations~\cite{Ibarra:2007wg}. 
In addition, GeV scale axinos are also potential candidates for the 
decaying DM~\cite{Covi:2009pq} providing explanations for the 
Fermi-LAT 130 GeV gamma-ray excess~\cite{Endo:2013si}. This 
excess~\cite{Weniger:2012tx,Su:2012ft} can also be explained by 
the scalar partner $\tilde{\phi}$ of a chiral superfield 
$\Phi$ which serves as the DM and decays to $\gamma\gamma$ or $\gamma\,Z$ final 
states via global $U(1)_R$ breaking interactions~\cite{Kyae:2012vi}. 

Several scenarios of TeV-scale decaying scalar DM which 
decays to SM leptons~\cite{Arvanitaki:2008hq,Nardi:2008ix,Carone:2011iw,Feng:2013vva,Choi:2013oaa}  
have been proposed in the context of the positron excesses observed in 
PAMELA~\cite{PAMELA:2008gwm} and AMS-02~\cite{PhysRevLett.110.141102}. 
Apart from these, decays of GeV - TeV scale scalar DM particles via non-minimal 
couplings to the Ricci scalar (R)~\cite{Cata:2016epa} or via higgs-gravity 
portal interactions~\cite{Sun:2021jna} are also studied in the literature. 
In the SUSY context, right-handed sneutrinos ($\tilde{\nu}_R$) couple to 
left-handed sneutrinos ($\tilde{\nu}_L$) via Dirac neutrino Yukawa coupling 
and decays to SM leptons by means of the trilinear R-parity breaking interaction 
terms~\cite{Ando:2015qda}. When R-parity is conserved the heavier 
$\tilde{\nu}_R$ decays to the lighter $\tilde{\nu}_R$ and gives rise to a 
photon spectrum with sharp-spectral feature~\cite{Ghosh:2019jzu}.
As mentioned earlier, here we have varied the decaying DM mass in the range 
10 GeV - 10 TeV but refrain from assuming any specific mechanism for DM 
production. 
Most of the existing indirect search observations including the ones we have 
considered here (i.e., Planck CMB~\cite{Planck:2015fie}, Fermi-LAT 
IGRB~\cite{Fermi-LAT:2014ryh}, AMS-02 positron flux~\cite{AMS:2014xys} 
and Super-Kamiokande neutrino flux~\cite{Frankiewicz:2016nyr} measurements), 
are sensitive to the signals of WIMP annihilations. The spectra of photons, 
antimatter particles and neutrinos, originating from scalar DM particles 
decaying into two-body SM final states are similar to that coming from the 
WIMP annihilations and thus most stringent constraints from such observations 
can be derived for decaying DM particles in the GeV - TeV mass range~\cite{Ibarra:2013cra}.
Additionally, the next generation observations 
such as the upcoming SKA radio telescope 
is also sensitive to the signals of DM particles in the GeV - TeV mass  
range~\cite{Ghosh:2020ipv,Kar:2019cqo}. 
Therefore, our choice of the DM mass range is primarily motivated by 
phenomenological considerations and potential for detections rather 
than theoretical arguments.

For heavier DM particles there exist a number of gamma-ray, cosmic-ray 
and neutrino observations which provide quite strong 
constraints~\cite{Ishiwata:2019aet}.
The upcoming Cherenkov Telescope Array (CTA) will also be quite useful 
in constraining DM particles heavier than a few tens of 
TeV~\cite{Garny:2010eg,Pierre:2014tra,Viana:2019ucn}. 
On the other hand, in case of keV and MeV DM particles 
X-ray~\cite{Oldengott:2016yjc} 
and CMB~\cite{Slatyer:2016qyl} constraints are the most stringent ones, 
till date. In addition, gamma-ray searches by ACT~\cite{LargerACT:2006gvu},  
GRIPS~\cite{2012ExA....34..551G}, AdEPT~\cite{Hunter:2013wla}, 
COMPTEL~\cite{1999ApL&C..39..193W}, EGRET~\cite{Strong:2004de} etc. 
and positron observation of Voyager I~\cite{Krimigis1977,C.2013} 
have been used to constrain the parameter space of a MeV-scale decaying  
DM~\cite{Boddy:2015efa,Boudaud:2016mos}. Apart from these, future MeV 
$\gamma$-ray experiments like e-ASTROGRAM~\cite{e-ASTROGAM:2017pxr}, 
AMEGO~\cite{AMEGO:2019gny}, GRAMS~\cite{Aramaki:2019bpi} etc. will also have 
important implications for decaying DMs with masses in the MeV range. 
However, for such DM particles, detection at the SKA depends on other mechanism 
such as the inverse compton (IC) effect. 
The frequency distribution in such a case is different from what is 
predicted by simulating synchrotron emission in the dSph 
galaxies~\cite{Dutta:2020lqc}. This is kept beyond the scope of 
the present work.


Any specific theoretical scenario would clearly identify some preferred 
DM decay modes with known branching fractions, provided we know the Lagrangian 
for the concerned model (see, for example,~\cite{Ghosh:2020ipv}). However, 
without being governed by any such model, here, we consider a scalar DM in 
the mass range 10 GeV - 10 TeV, which accounts for the entire dark matter 
energy density of the Universe, and decays into all possible two-body SM final 
states, i.e., $\chi \rightarrow$ ${\rm SM}_{1}$ $\overbar{\rm SM}_2$, 
where ${\rm SM}_{1}$ $\overbar{\rm SM}_2$ belongs to the following set: 
$\{e^+e^-$, $\mu^+\mu^-$, $\tau^+\tau^-$, $b\bar{b}$, $t\bar{t}$, $q\bar{q}$, 
$W^+W^-$, $ZZ$, $\gamma\gamma$, $gg$, $hh$, $Z\gamma$, $Zh$, $\nu\bar{\nu}\}$. 
Note that, as mentioned in the introduction, unlike most of the studies 
available in the literature, we have also considered DM decays to 
$\nu\bar{\nu}$, in our analysis.

  
We would like to point out that in case of the $\nu\bar{\nu}$ final state 
$\nu_e, \nu_\mu$ and $\nu_\tau$ flavours are assumed to be produced with 
equal branching fractions, the sum of which determines the total branching 
ratio of the $\nu\bar{\nu}$ channel. Similarly, while considering the $q\bar{q}$ 
decay mode we have assumed that $u\bar{u},d\bar{d},c\bar{c}$ and $s\bar{s}$  
final states are produced with equal branching ratios so that their sum equals 
the total branching fraction attributed to the $q\bar{q}$ final state. 
For any given value of $m_\chi$ only the kinematically 
allowed two-body SM final states are taken into account. For example, if 
$m_\chi \lsim 160\,{\rm GeV}$, branching ratio for the $W^+W^-$ final 
state has been set to zero throughout the analysis. 

Furthermore, in our study we have implicitly assumed that 
there exists no BSM decay mode of the DM candidate such as the dark 
radiation. As we had mentioned earlier, we have also ignored any 
possible flavour violating decays (e.g., $\chi \rightarrow \mu\tau, 
d\bar{s}$) and lepton number violating decays 
(e.g., $\chi \rightarrow \nu\nu, \bar{\nu}\bar{\nu}$). 
These assumptions are usually valid in minimal models of decaying DM.   
However, inclusions of such decay modes would not change 
the results presented here substantially. 
In the line of several existing works on decaying DM (see, for  
example~\cite{Ibarra:2013cra,Blanco:2018esa,Ghosh:2020ipv} etc.), 
we have based our analysis on the postulate that all indirect signals in the 
form of $\gamma$-rays, electrons (positrons), $\nu\bar{\nu}$ and 
radio waves arise from \textit{DM decays only}, and the contributions of 
DM annihilations to such signals are negligible. Thus, using the data from 
different indirect search observations, the \textit{maximum amount} of flux 
arising from DM decay can be constrained for any $m_\chi$. 
These maximum values are realized in situations where DM annihilations 
into SM particle pairs have negligible rates. In these cases, the DM is then 
either produced non-thermally in the decay of superheavy states~\cite{Park:2012xq}, 
or freezes out via the participation of some hidden sector 
particles~\cite{Hambye:2016qkf,Teresi:2017yrp}. 
The observed relic density of our Universe is generated in this manner. 
On the other hand, DM decays to SM particle pairs can take place via several 
hitherto unknown effective interactions which are not necessarily correlated 
with the production mechanism of the DM candidate. Such effective 
interactions driving the DM decays are usually obtained on integrating out 
various heavy fields and thus the resulting decay processes are slow enough 
to evade the existing constraints from the indirect search observations.

\subsection{Indirect detection signals and existing constraints}
\label{sec:conseval}

Indirect detection of GeV - TeV scale decaying DM consists in detecting 
the gamma-ray photons, electrons-positrons (along with various 
electromagnetic signals originating from them) and neutrino-antineutrino pairs 
coming from the cascade decays of the SM particles which are produced as 
the primary decay products of the DM. Usually the data from several 
observations are used to constrain the DM parameter space assuming 100$\%$ 
branching ratio for each individual SM final state. However, our goal here 
is to obtain an 
upper limit on the total $\Gamma$ independent of the branching fraction of 
each individual DM decay modes. 
For this purpose, it is necessary to calibrate our analysis technique first. 





\begin{figure*}[htb!]
\centering
\centering
\includegraphics[width=7.6cm,height=7.4cm]{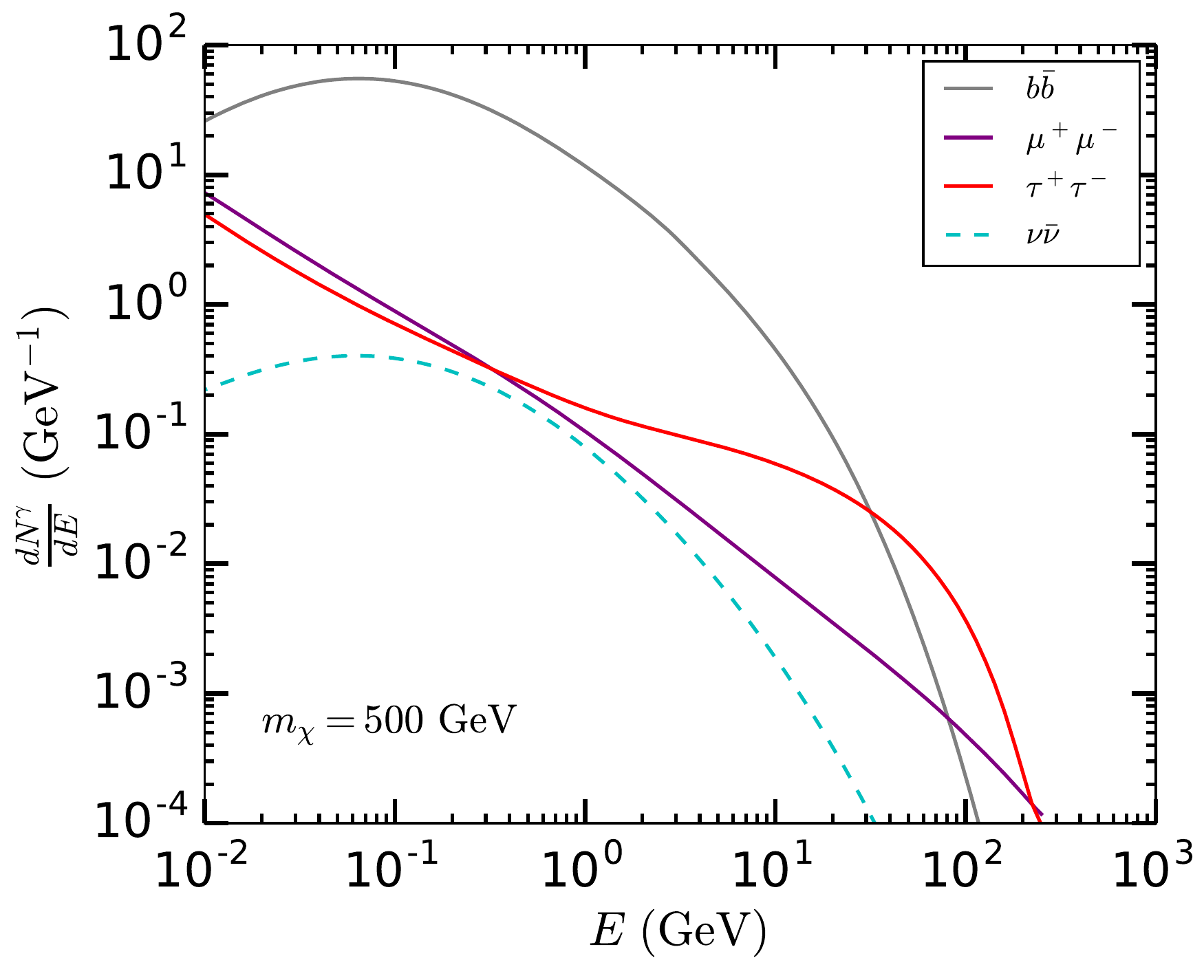}
\includegraphics[width=7.6cm,height=7.4cm]{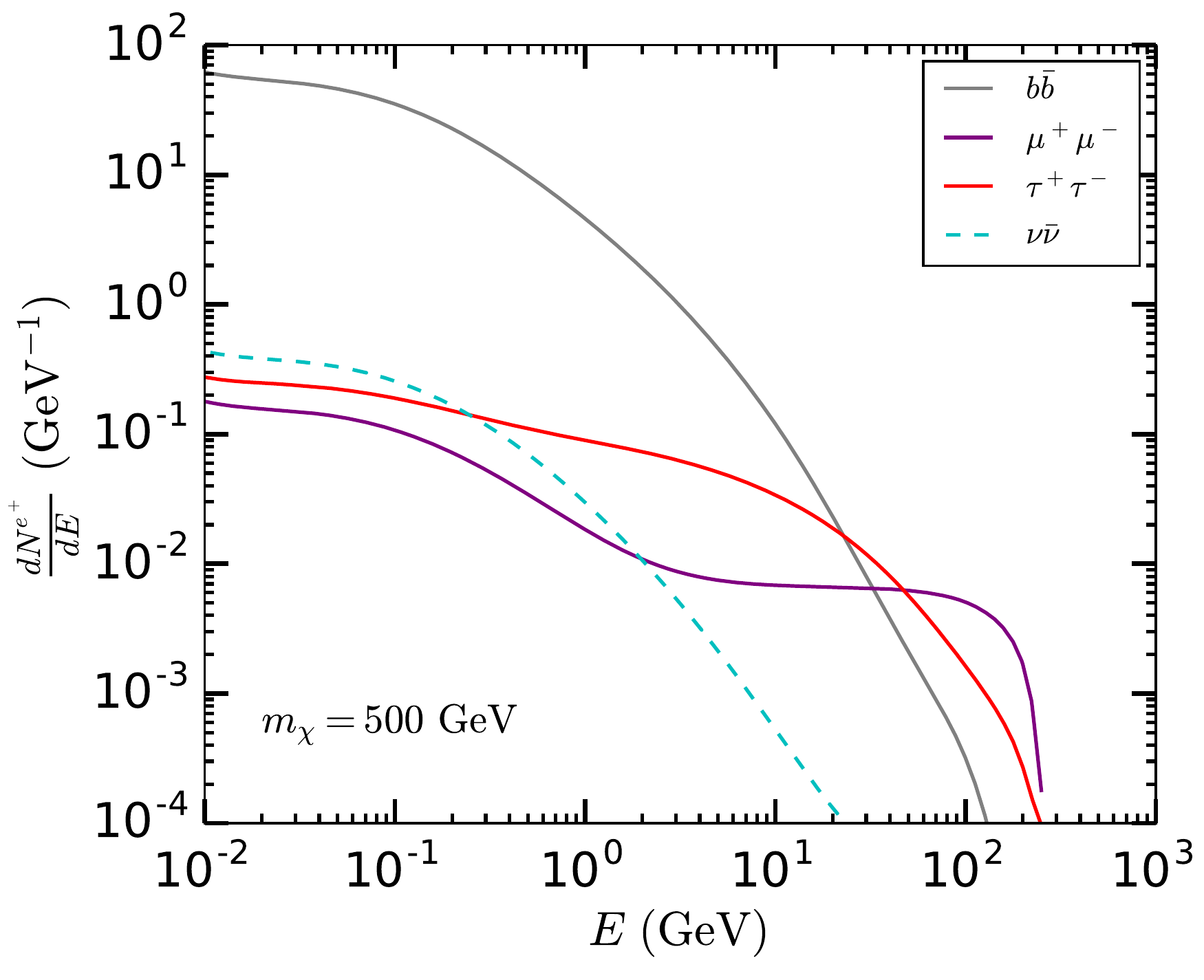}
\includegraphics[width=7.6cm,height=7.4cm]{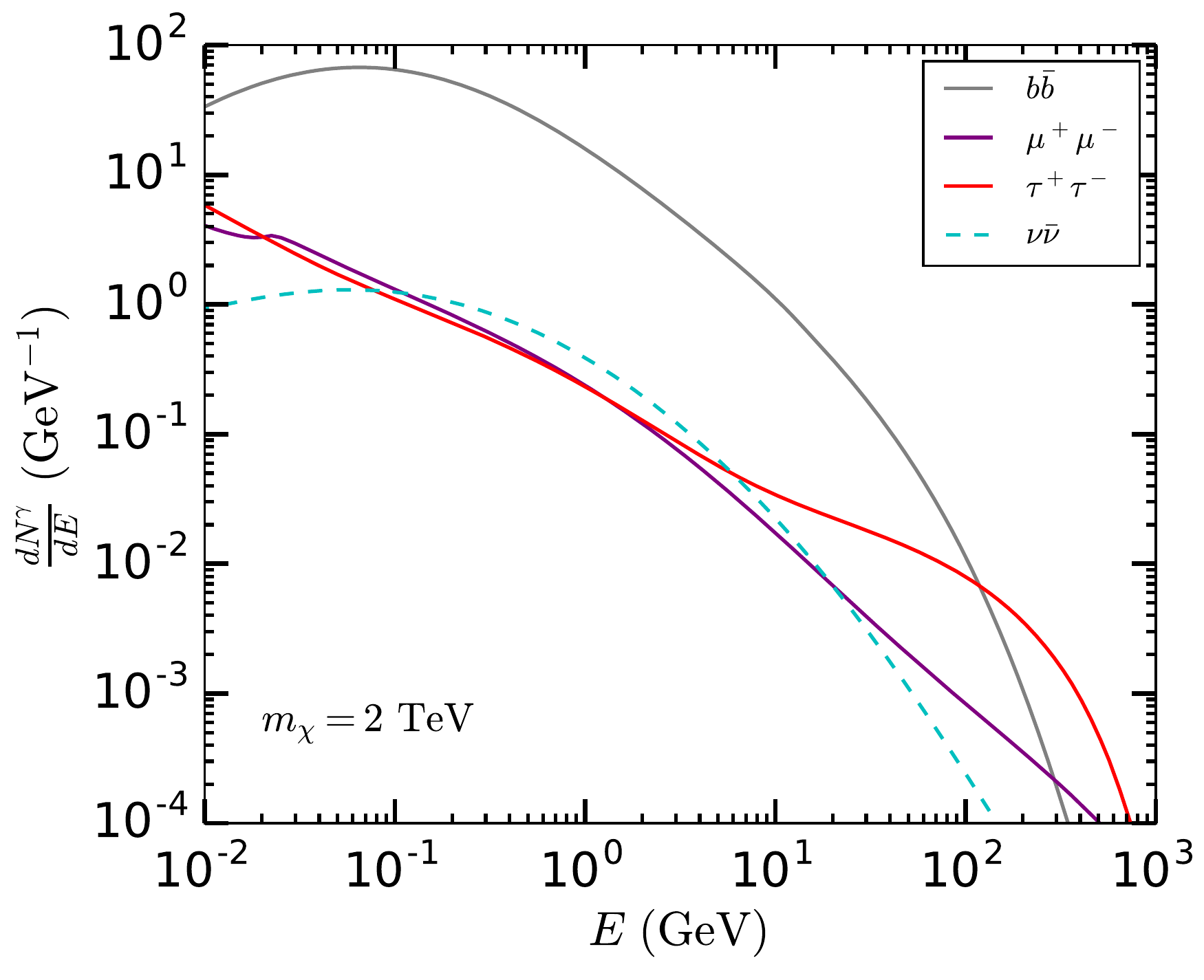}
\includegraphics[width=7.6cm,height=7.4cm]{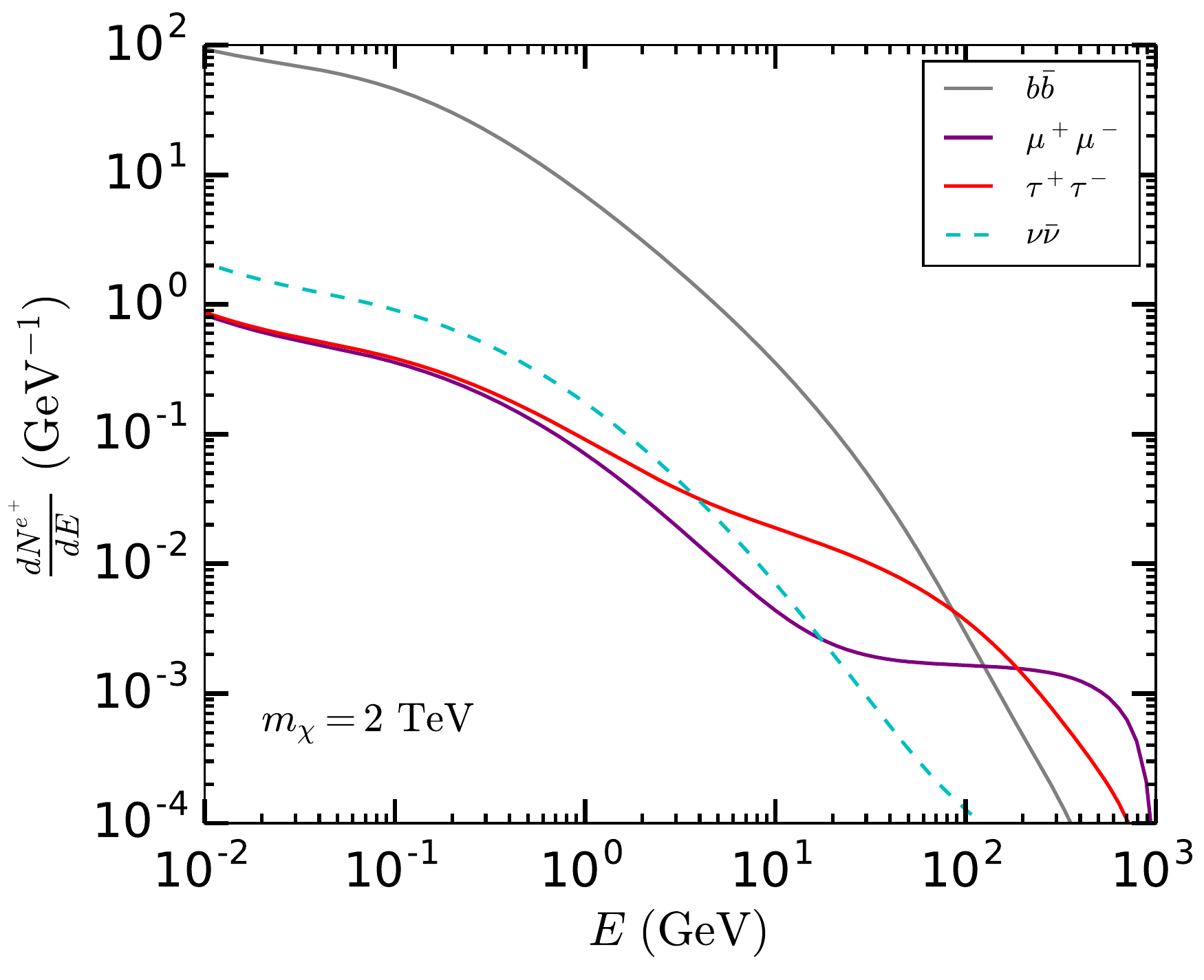}
\caption{{\it Top:} The photon spectra $dN^{\gamma}/dE$ (left panel) and positron 
spectra $dN^{e^+}/dE$ (right panel) are shown for 
$b\bar{b}$ (gray solid line), $\mu^+\mu^-$ (purple solid line), 
$\tau^+\tau^-$ (red solid line) and $\nu\bar{\nu}$ (cyan dashed line) 
final states assuming $100\%$ branching ratio 
for each individual channels for a benchmark value of $m_\chi = 500\,{\rm GeV}$.
{\it Bottom:} The corresponding distributions for $m_\chi = 2\,{\rm TeV}$ are shown. 
See the text for details.} 
\label{fig:dNdE_nunubar}
\end{figure*}

Note that, unlike most of the studies found in the literature, we have 
included the effect of the $\nu\bar{\nu}$ final state. As mentioned in the introduction, 
for this final state, the photons and electrons (positrons) are produced by the radiation of 
electroweak gauge bosons and hence, the resulting spectra are highly suppressed 
for $m_\chi$ lower than \textit{a few hundreds of GeV}. This is because, for such 
values of $m_\chi$ the final state $\nu\bar{\nu}$ pairs are of low energy. However, 
for heavier DM particles, the $\nu\bar{\nu}$ pairs are highly energetic, 
so that the on-shell $W$ and $Z$ bosons are abundantly radiated from them. Consequently, 
the produced $\gamma$ and $e^+(e^-)$ spectra are comparable to the corresponding spectra 
coming from other SM final states. 

In the top left panel of Fig.~\ref{fig:dNdE_nunubar}, 
considering $m_\chi = 500\,{\rm GeV}$, we have shown the $\gamma$-ray spectra $dN^{\gamma}/dE$ produced per DM decay, for the $b\bar{b}$ (gray solid line),  
$\mu^+\mu^-$ (purple solid line), $\tau^+\tau^-$ (red solid line) and 
$\nu\bar{\nu}$ (cyan dashed line) final states assuming $100\%$ branching ratio 
for each individual channel. In the top right panel, the 
$e^+$ spectra $dN^{e^+}/dE$ (same as the $e^-$ spectra $dN^{e^-}/dE$) 
for these DM decay modes are shown. It is evident that, the spectra coming from 
the $\nu\bar{\nu}$ final state are considerably suppressed compared to those 
from the $b\bar{b}$ final state throughout the energy range. In the higher energy 
bins, the fluxes originating from the $\nu\bar{\nu}$ channel are suppressed compared 
to the spectra of the $\mu^+\mu^-$ and $\tau^+\tau^-$ channels, too. 
As one decreases $m_\chi$, the spectra corresponding to the $\nu\bar{\nu}$ 
final state become even more subdominant. On the other hand, on increasing 
$m_\chi$, photon and positron (electron) fluxes coming from the 
$\nu\bar{\nu}$ decay mode are enhanced throughout the energy range, as shown in the 
bottom panels of Fig.~\ref{fig:dNdE_nunubar}, where 
$m_\chi = 2\,{\rm TeV}$ is assumed. The photon and positron (electron) 
spectra presented here are obtained from the 
$\mathtt{Pythia}$~\cite{Sjostrand:2006za} generated files provided by the  
$\mathtt{PPPC4DMID}$~\cite{Cirelli:2010xx,Ciafaloni:2010ti,Cirelli} and 
will be used in the upcoming analyses.


Keeping this in mind, in this section we go ahead to present the constraints on 
the DM parameter space for each individual observation, i.e., 
Planck CMB~\cite{Planck:2015fie}, Fermi-LAT IGRB~\cite{Fermi-LAT:2014ryh}, 
AMS-02 cosmic-ray positron flux~\cite{AMS:2014xys} and 
Super-Kamiokande neutrino flux~\cite{Frankiewicz:2016nyr} measurements, 
assuming the DM decays to any given two-body SM final state with 
100$\%$ branching ratio. This will enable us to include the effect of 
arbitrary branching fractions to all channels in our analysis.


\subsubsection{Planck CMB constraints}	

Energy injections to the photon-baryon fluid between recombination 
($z \sim 1100$) and reionization ($z \sim 6$) can alter the thermal 
history of the Universe and thereby causes perturbations to the CMB
anisotropy spectra. In scenarios with unstable DM candidates, the 
SM particles produced in the DM decays act as additional sources of 
energy injections during the cosmic dark ages to distort the CMB 
anisotropy spectra. On the other hand, the CMB is quite accurately 
measured by the Planck collaboration~\cite{Planck:2015fie} 
and any substantial distortion in the CMB spectrum is ruled out 
by the Planck data~\cite{Planck:2015fie}. Therefore, 
the data of the CMB observation by Planck~\cite{Planck:2015fie} 
can be employed to constrain the fluxes of $\gamma$, $e^-$ and 
$e^+$ coming from the DM decays. This in turn implies upper limits 
on 
$\Gamma$ for any $m_\chi$, since these 
fluxes of stable SM particles are proportional to the DM decay width 
for any given value of the DM mass. 

\begin{figure*}[htb!]
\centering
\centering
\includegraphics[width=7.6cm,height=7.4cm]{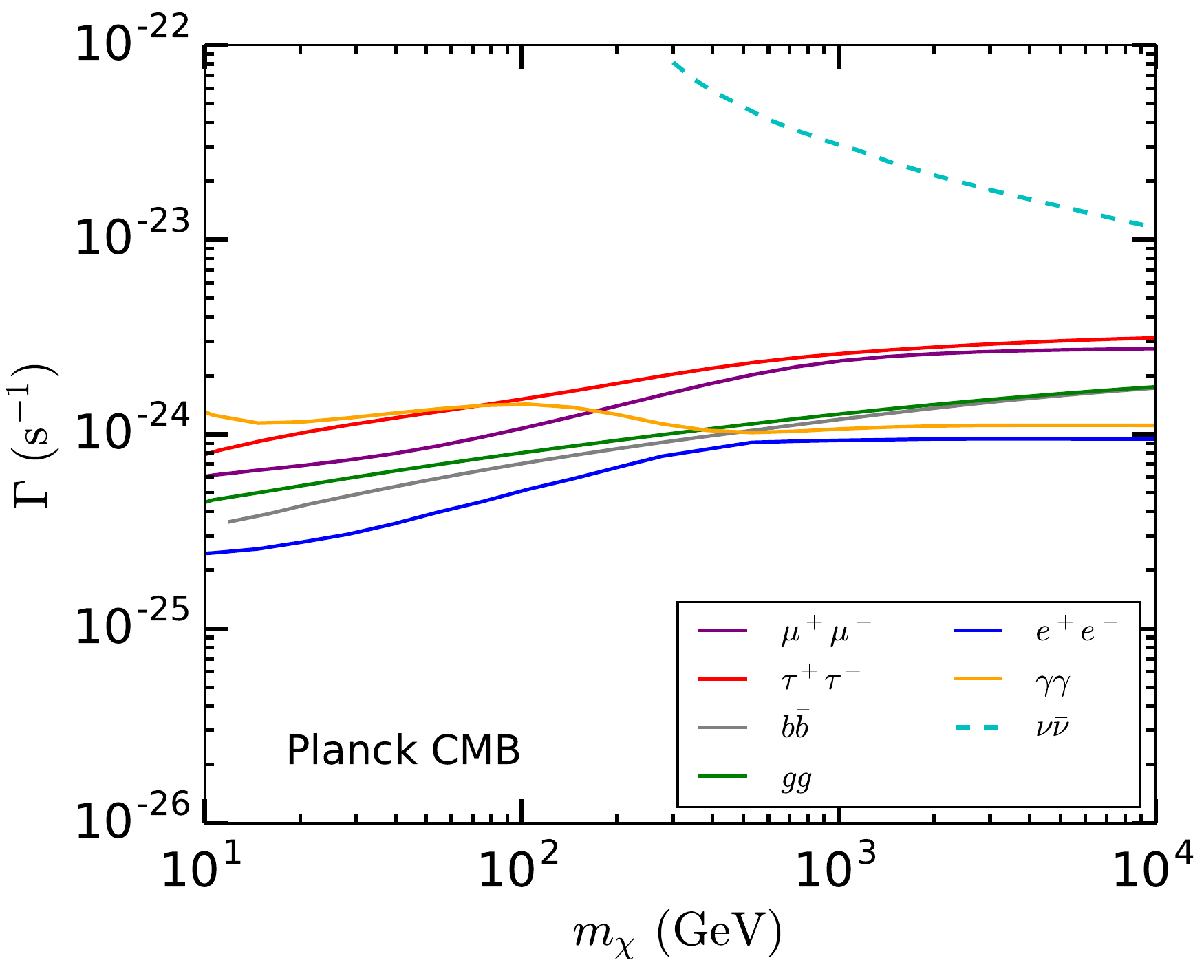}
\caption{95\% C.L. upper limits on $\Gamma$ assuming 
100\% branching ratio for each SM final state, are obtained from 
the Planck observation of the temperature and polarization 
anisotropies of CMB. 
Here, we show the constraints for seven SM final states for the 
purpose of illustration.}
\label{fig:Planck_CMB}
\end{figure*}

Here, we shall follow the methodology outlined in 
Ref.~\cite{Slatyer:2016qyl} to derive the Planck CMB~\cite{Planck:2015fie} 
constraints on 
$\Gamma$ for $m_\chi$ 
in the range 10 GeV - 10 TeV. Ref.~\cite{Slatyer:2016qyl} 
has used the technique of \textit{principal component analysis} (PCA) 
to derive constraints on a generic decaying scalar DM. The 95$\%$ 
confidence level (C.L.) upper limit on 
$\Gamma$ 
is given by~\cite{Slatyer:2016qyl}:
\begin{equation}
\Gamma \lesssim \left(\frac{\vec{N}.\vec{e}_1}{\vec{e}_1(30\,{\rm MeV}\,e^+e^-)}\right)^{-1}\,\left(2.6 \times 10^{25}\,{\rm s}\right)^{-1},
\label{eqn:CMBlimitfinal}
\end{equation}		
where $\vec{e}_1$ is the first principal component (PC) and $\vec{N}$ contains 
the information of the $e^+$ ($e^-$) or the $\gamma$ spectra for the considered 
DM decay mode:
\begin{equation}
\vec{N} = \left\{\sum_f\,B_f\frac{1}{m_\chi}E_i \frac{dN_f^{e^+}}{d\ln E_i}(E_i), \sum_f\,B_f\frac{1}{m_\chi}E_i \frac{1}{2}\frac{dN_f^{\gamma}}{d\ln E_i}(E_i) \right\},
\label{eqn:CMBlimitspec}
\end{equation}   
with $i= 1,2,...,N$ signifying the respective energy bins spanning 
the entire considered energy range 10 keV - 1 TeV~\cite{Slatyer:2016qyl}, 
$f$ being the two-body SM final state with branching ratio $B_f$, while, 
as mentioned earlier, $dN_f^{e^+}/d\ln E$ ($dN_f^{e^-}/d\ln E$) and 
$dN_f^{\gamma}/d\ln E$, respectively, represent the $e^+(e^-)$ and $\gamma$ 
spectra associated with per DM decay to the final state $f$. 

The first principal component $\vec{e}_1$ is the eigenvector corresponding 
to the largest eigenvalue of the marginalized Fisher matrix. The Fisher matrix 
is constructed out of the perturbations to the CMB anisotropy spectra 
caused by the basis energy injection models (see Ref.~\cite{Slatyer:2016qyl} 
for details). In Eqn.~\ref{eqn:CMBlimitfinal}, $\vec{e}_1(30\,{\rm MeV}\,e^+e^-)$ 
represents the value of the first PC for the $e^+e^-$ final state with an injection 
energy of 30 MeV which is chosen as a reference model in Ref.~\cite{Slatyer:2016qyl}. 
$\vec{N}.\vec{e}_1$ represents the projection of $\vec{N}$ on the first PC. 
For calculating $\vec{N}.\vec{e}_1$, in Eqn.~\ref{eqn:CMBlimitfinal}, 
the contributions of both $\gamma$ and $e^+(e^-)$ 
are taken into account. 
The first PC $\vec{e}_1$ for both $\gamma$ and $e^+(e^-)$ 
(for injection energies 10 keV - 1 TeV) are taken from~\cite{Slatyer:2016qyl} 
to which the readers are referred for the details of the analysis methodology. 


Using Eqn.~\ref{eqn:CMBlimitfinal}, we have derived the CMB constraints on 
the $m_\chi-\Gamma$ plane 
assuming DM decays to a specific SM final state with 100$\%$ branching ratio. 
We have shown our results for seven illustrative DM decay modes in 
Fig.~\ref{fig:Planck_CMB}. 
As mentioned earlier, the spectra of $e^+(e^-)$ and 
$\gamma$ coming from $\chi \rightarrow \nu\bar{\nu}$ decay mode 
are comparable to that originating from other SM final states only when 
$m_\chi$ is larger than \textit{a few hundreds of GeV} 
(see Fig.~\ref{fig:dNdE_nunubar}). As a result, the 
constraint on the $\nu\bar{\nu}$ channel also strengthens in this $m_\chi$ 
range (see cyan line in Fig.~\ref{fig:Planck_CMB}).

\subsubsection{Fermi-LAT IGRB constraints}
\label{sec:FLATlim}
		
Gamma-ray photons are produced as one of the end products of the 
decay cascades of the SM particles originated in the decays of 
GeV - TeV scale DM particles. In case of the hadronic decay modes, 
$\pi^0 \rightarrow \gamma\gamma$ acts as the source of such gamma-ray 
photons, while  for the leptonic decay channels, these photons are 
dominantly produced from the final state radiations. Direct decay of 
DM to $\gamma\gamma$ is also possible. In addition, $e^+e^-$ 
produced from the DM decays also upscatter photons of the interstellar 
radiation field (ISRF) to gamma-ray energies and contribute to the 
gamma-ray signals of decaying DM. 

For any given value of 
$m_\chi$, the gamma-ray flux is proportional to the DM decay width 
and larger the value of $\Gamma$ greater is the expected signal. 
Thus, the gamma-ray flux measured by any observation can be compared 
against the fluxes predicted from the decay of a DM particle of mass $m_\chi$ in 
any suitably chosen astrophysical source and the corresponding upper limit 
on $\Gamma$ can be derived. In case of a decaying DM scenario the gamma-ray 
flux from an extra-terrestrial source is proportional to a single power of 
the DM density in that source and broader the source is 
larger is the expected gamma-ray signal. Thus, for a decaying DM scenario 
most stringent upper limits from the gamma-ray observations are usually 
obtained when one considers the isotropic component of the signal, i.e., the 
extra-galactic component~\cite{Bertone:2007aw,Ibarra:2013cra}. As a result, 
in the case of a decaying dark matter Fermi-LAT IGRB constraints are more 
stringent than those coming from the Fermi-LAT observations of other gamma-ray 
sources such as the dSph galaxies~\cite{Baring:2015sza}.

The extra-galactic gamma-ray flux from the DM decays is given by~\cite{Liu:2016ngs,Cirelli:2012ut,Cirelli:2010xx},
\begin{eqnarray}
\frac{d\Phi^{\rm EG}}{dE_\gamma}&=&\frac{\Gamma}{4\pi m_\chi}\,c\,\Omega_{\rm DM}\,\rho_c \sum_f B_f \int_0^\infty \,dz\frac{1}{H(z)}e^{-\tau(E_\gamma,z)}\bigg[\frac{dN^\gamma_f}{dE_\gamma}(E_\gamma(1+z))\nonumber\\
&&+\frac{2}{E_\gamma(1+z)}\int_{m_e}^{m_\chi/2}dE_e\frac{P^{\rm CMB}_{\rm IC}(E_\gamma,E_e,z)}{b^{\rm CMB}_{\rm IC}(E_e)}\int_{E_e}^{m_\chi/2}d\tilde{E}_e\frac{dN^{e^+}_f}{d\tilde{E}_e}(\tilde{E}_e)\bigg],\nonumber\\
\label{eqn:FLATEGflux} 
\end{eqnarray}
where DM density $\Omega_{\rm DM}$, critical density $\rho_c$ and all other 
relevant parameters needed to evaluate the redshift dependent Hubble parameter
$H(z)$ 
are taken from Ref.~\cite{Planck:2015fie}. For the optical depth 
$e^{-\tau(E_\gamma,z)}$, the inverse compton scattering (ICS) 
power $P^{\rm CMB}_{\rm IC}$ and the energy loss term 
$b^{\rm CMB}_{\rm IC}$ we have used the parameterizations given 
in~\cite{Cirelli:2010xx,Buch:2015iya}.

\begin{figure*}[h!]
\centering
\includegraphics[width=7.6cm,height=7.4cm]{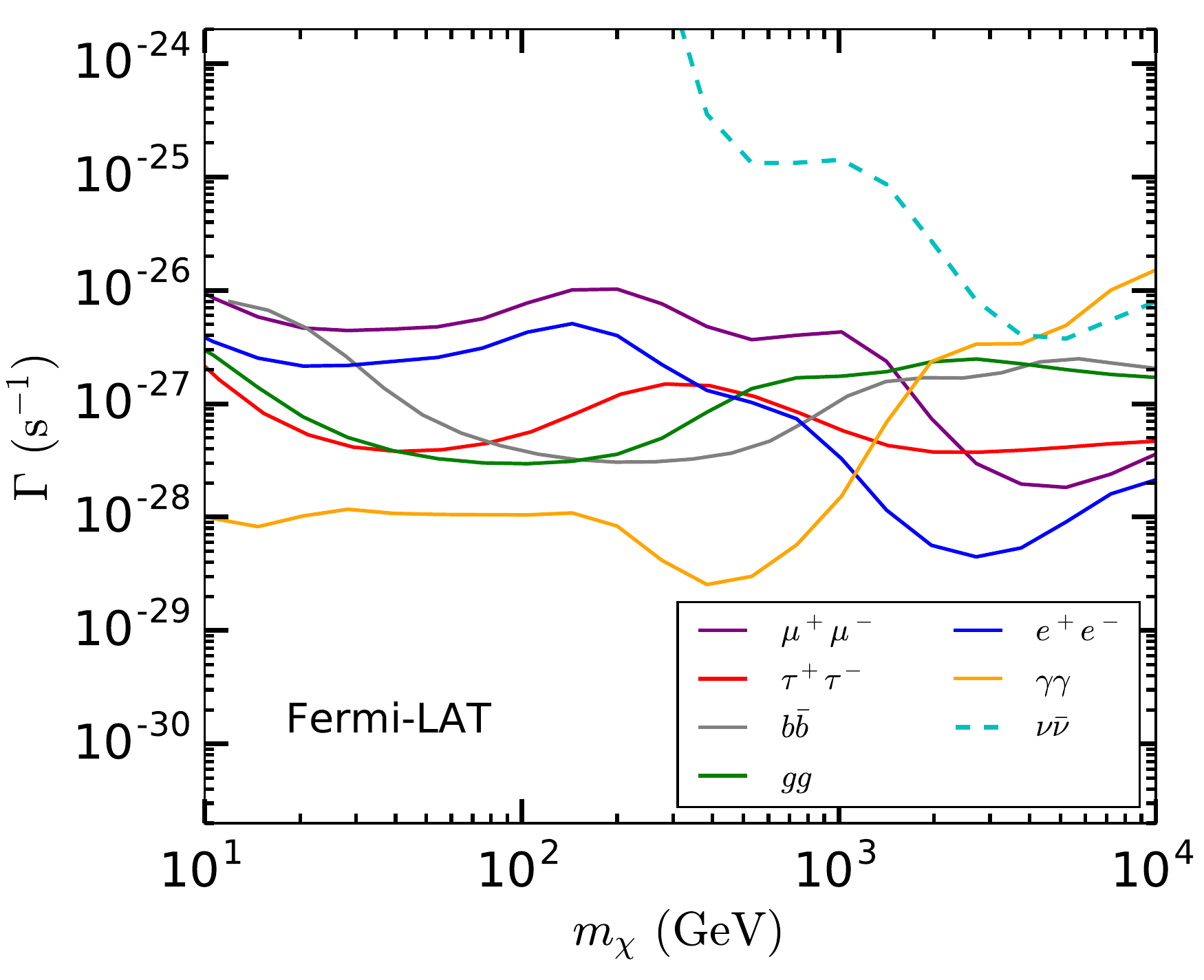}
\caption{95\% C.L. upper limits on $\Gamma$ (for seven illustrative decay modes)  
derived from the data of the Fermi-LAT 
IGRB 
observation	are shown. In deriving these limits 
100\% branching ratio is attributed to each individual DM decay channel. 
}
\label{fig:LAT_IGRB}
\end{figure*}

The diffuse isotropic gamma-ray background (IGRB) in the energy range 
100 MeV - 820 GeV~\cite{Fermi-LAT:2014ryh} is measured by the Fermi Large 
Area Telescope (Fermi-LAT) during its 50 months of observation time and 
provides the most stringent $\gamma$-ray constraints on the parameter 
space of a GeV - TeV scale decaying DM~\cite{Ando:2015qda,Liu:2016ngs,Blanco:2018esa}. 
In deriving these constraints we have followed the methodology 
of~\cite{Liu:2016ngs} and parameterized the background gamma-ray flux 
as~\cite{Fermi-LAT:2014ryh},
\begin{equation}
\frac{d\Phi^{\rm BG}}{dE_\gamma} = I_{100} \left(\frac{E_\gamma}{100\,{\rm MeV}}\right)^{-\beta}\,\exp\left(-\frac{E_\gamma}{E_c}\right).
\label{eqn:bgFLAT}
\end{equation} 
Then we perform a likelihood ratio test with the $\chi^2$ defined as: 
\begin{eqnarray}
\chi^2 = \sum_i \, \frac{\left(\Phi^{\rm BG}(E_i)+\Phi^{\rm EG}(E_i) - D_i \right)^2}{\sigma^2_i},
\label{eqn:chisqFLAT}
\end{eqnarray}
where $D_i$ is the Fermi-LAT IGRB data~\cite{Fermi-LAT:2014ryh}, 
$\Phi^{\rm BG}(E_i)$ is the expected number of background events 
and $\Phi^{\rm EG}(E_i)$ is the expected number of signal events 
in the $i$-th energy bin while $\sigma_i$ is the associated 
uncertainty~\cite{Fermi-LAT:2014ryh}. 
We then minimize the $\chi^2$ in Eqn.~\ref{eqn:chisqFLAT} with respect 
to the parameters $\{I_{100},\beta,E_c,\Gamma\}$ to obtain the 
best-fit values of these parameters which give us the 
best-fit value of $\chi^2$, i.e., $\chi^2_{\rm bf}$. 
Then all other parameters are kept fixed at their best-fit values 
while $\Gamma$ is increased until the $\chi^2$ increases by 2.71 
from its best-fit value, i.e.,
\begin{equation}
\chi^2 (\Gamma) = \chi^2_{\rm bf}+2.71,
\label{eqn:chisqFLATA}
\end{equation}  
which gives us the $95\%$ C.L. upper limit on $\Gamma$. For details 
of the analysis methodology adopted here we refer the reader to 
Ref.~\cite{Liu:2016ngs}. 


Assuming the DM decays to a specific SM final state with $100\%$ branching
ratio the $95\%$ C.L. upper limits on $\Gamma$ obtained using Eqn.~\ref{eqn:chisqFLATA} 
are shown 
in Fig.~\ref{fig:LAT_IGRB}, for seven different DM decay modes. 
As one can see from Fig.~\ref{fig:LAT_IGRB}, for $m_\chi \lsim 1\,{\rm TeV}$, 
the limit on the $\gamma\gamma$ final state is most stringent while the limit 
is weakest for $\mu^+\mu^-$. For $m_\chi$ greater than $\sim 1\,{\rm TeV}$ the 
constraint weakens considerably for $\gamma\gamma$ channel, since, the energies of 
the produced photons fall outside the 
sensitivity range of Fermi-LAT. One interesting point to note that 
the constraint on the $\nu\bar{\nu}$ channel 
becomes stronger than that for the $\gamma\gamma$ 
channel in the DM mass range $3\,{\rm TeV}$ - $10\,{\rm TeV}$ 
(cyan line in Fig.~\ref{fig:LAT_IGRB}). 
This is because, as mentioned previously, for the $\nu\bar{\nu}$ channel 
the $\gamma$-rays come from the electroweak gauge bosons which are abundantly 
radiated only when the DM mass is on the higher side (see Fig.~\ref{fig:dNdE_nunubar}). 
\subsubsection{AMS-02 positron constraints}		
\label{sec:AMSlim}

Among the stable SM particles produced as the final products of DM decay 
cascades positrons are another powerful probes for understanding the  
nature of DM interactions. The SM particles produced from the DM decays 
give rise to electrons-positrons via cascade decays and the produced 
positrons undergo diffusion and energy losses in the galactic medium before 
reaching the detectors devised to detect cosmic-ray positrons. 
The positron spectra coming from DM decays are governed by 
$\Gamma$ for any given $m_\chi$ and for specific choices of 
the astrophysical parameters governing the DM density distribution and 
positron propagation inside the Milky Way (MW) galaxy. Therefore, stringent upper 
limits on $\Gamma$ are derived by comparing the DM decay induced positron flux 
against the measured value of the cosmic-ray positron flux when all other parameters 
are kept fixed. The measurement of the cosmic-ray positron flux~\cite{AMS:2014xys} 
by the Alpha Magnetic Spectrometer (AMS) on the International Space Station (ISS) 
provides strong constraints on the parameter space of 
a decaying dark matter~\cite{Ibarra:2013zia,Lu:2015pta}. 
 
The propagation of the positrons produced from DM decays is governed by the 
diffusion-loss equation. After ignoring the convection and the diffusion 
terms in the momentum space which do not affect the positron spectra for the 
energy range of interest~\cite{Strong:2007nh}, this equation is given 
by~\cite{Leane:2018kjk}:
\begin{equation}
\frac{\partial N_i}{\partial t}= \vec{\nabla}.(D\vec{\nabla})N_i+\frac{\partial}{\partial p}(b(p,\vec{r}))N_i+Q_i(p,\vec{r})+\sum_{j>i} \beta n_{gas}(\vec{r})\sigma_{ji}N_j-\beta n_{gas}(\vec{r})\sigma_i^{\rm in}(E_k)N_i,
\label{eqn:difflossAMS}
\end{equation}
where $N_i(\vec{r},t)$ denotes the number density of the positrons 
and $p$ is its momentum. 
The DM decay contribution to the source term $Q_i(p,\vec{r})$ is given by, 
\begin{equation}
Q_\chi(p,\vec{r}) = \frac{\Gamma}{m_\chi}\sum_f\,B_f\,\frac{dN^{e^+}_f}{dE}(E)\rho_d(\vec{r}).
\label{eqn:sourceAMS}
\end{equation} 
In Eqn.~\ref{eqn:sourceAMS}, $\rho_d(\vec{r})$ represents the 
DM density distribution inside the MW galaxy which is assumed to 
follow the Navarro-Frenk-White (NFW) profile~\cite{Navarro:1995iw}: 
\begin{equation}
\rho_d(\vec{r}) = \rho_s\,\left(\frac{r}{r_s}\right)^{-1}\left(1+\frac{r}{r_s}\right)^{-2}.
\label{eqn:NFWprofile}
\end{equation}
This is parameterized by the scale radius $r_s = 20\,{\rm kpc}$ and the 
DM density at the location of the Sun (i.e., $r_\odot = 8.5\,{\rm kpc}$) 
which is $\rho_\odot = 0.25\,{\rm GeV}\,{\rm cm}^{-3}$~\cite{Leane:2018kjk}.
Following Ref.~\cite{Leane:2018kjk}, such a value of $\rho_\odot$, consistent 
with the allowed range of values reported in the 
literature~\cite{Iocco_2011,Bovy:2012tw}, is considered in 
order to derive a conservative upper limit on $\Gamma$.

The diffusion term $D(\rho,|\vec{r}|,z)$ in Eqn.~\ref{eqn:difflossAMS} 
is given by~\cite{Leane:2018kjk}:
\begin{equation}
D(\rho,|\vec{r}|,z) = D_0\,e^{|z|/z_t}\,\left(\frac{\rho}{\rho_0}\right)^\delta,
\label{eqn:diffusionAMS}
\end{equation}
where, $\rho = p/e$ represents the rigidity of the positrons with the reference 
rigidity $\rho_0 = 4\,{\rm GV}$~\cite{Leane:2018kjk}. Following~\cite{Leane:2018kjk}, 
we choose the diffusion coefficient $D_0 = 2.7 \times 10^{28}\,{\rm cm}^2{\rm s}^{-1}$, 
diffusion index $\delta = 0.6$ and $z_t = 4\,{\rm kpc}$, a conservative choice 
usually assumed for the propagation of charged particles inside the MW  
galaxy~\cite{Buch:2015iya,Lavalle:2014kca,Giesen:2015ufa,Evoli:2015vaa}. 
However, we have also compared below our results with those obtained 
for other commonly considered diffusion and NFW halo profile parameters. 
Here the diffusion zone is taken to be axisymmetric with 
thickness $2z_t$~\cite{Leane:2018kjk}. 
In Eqn.~\ref{eqn:difflossAMS}, $b(p,\vec{r})$ represents the energy loss term 
which depends on the magnetic field strength inside the MW galaxy. This is 
parameterized by the local magnetic field 
$B_\odot = 8.9\,\mu{\rm G}$~\cite{Leane:2018kjk}, 
which lies in the range, frequently used for the MW 
galaxy~\cite{John:2021ugy,Buch:2015iya,PhysRevLett.111.171101,Strong:2011wd}.
This value corresponds to local radiation field and magnetic field energy 
densities which are larger than the values used 
in~\cite{John:2021ugy,PhysRevLett.111.171101}. Therefore, such a choice leads 
to a comparatively higher energy loss rate for the positrons, thereby giving rise to a conservative upper limit on $\Gamma$. 
In the next step, Eqn.~\ref{eqn:difflossAMS} is solved using the cosmic-ray 
propagation code $\mathtt{DRAGON}$~\cite{Evoli:2008dv,Evoli:2016xgn} 
to obtain the final distribution of positrons which is expected to be 
observed in AMS-02. Following~\cite{Leane:2018kjk}, the effect of solar 
modulation is incorporated using the force-field 
approximation~\cite{Cholis:2015gna,Cavasonza:2016qem}  
with a modulation potential $\Phi = 0.6\,{\rm GV}$~\cite{Cholis:2015gna}. 

%
	
\begin{figure*}[h!]
\centering
\includegraphics[width=7.6cm,height=7.4cm]{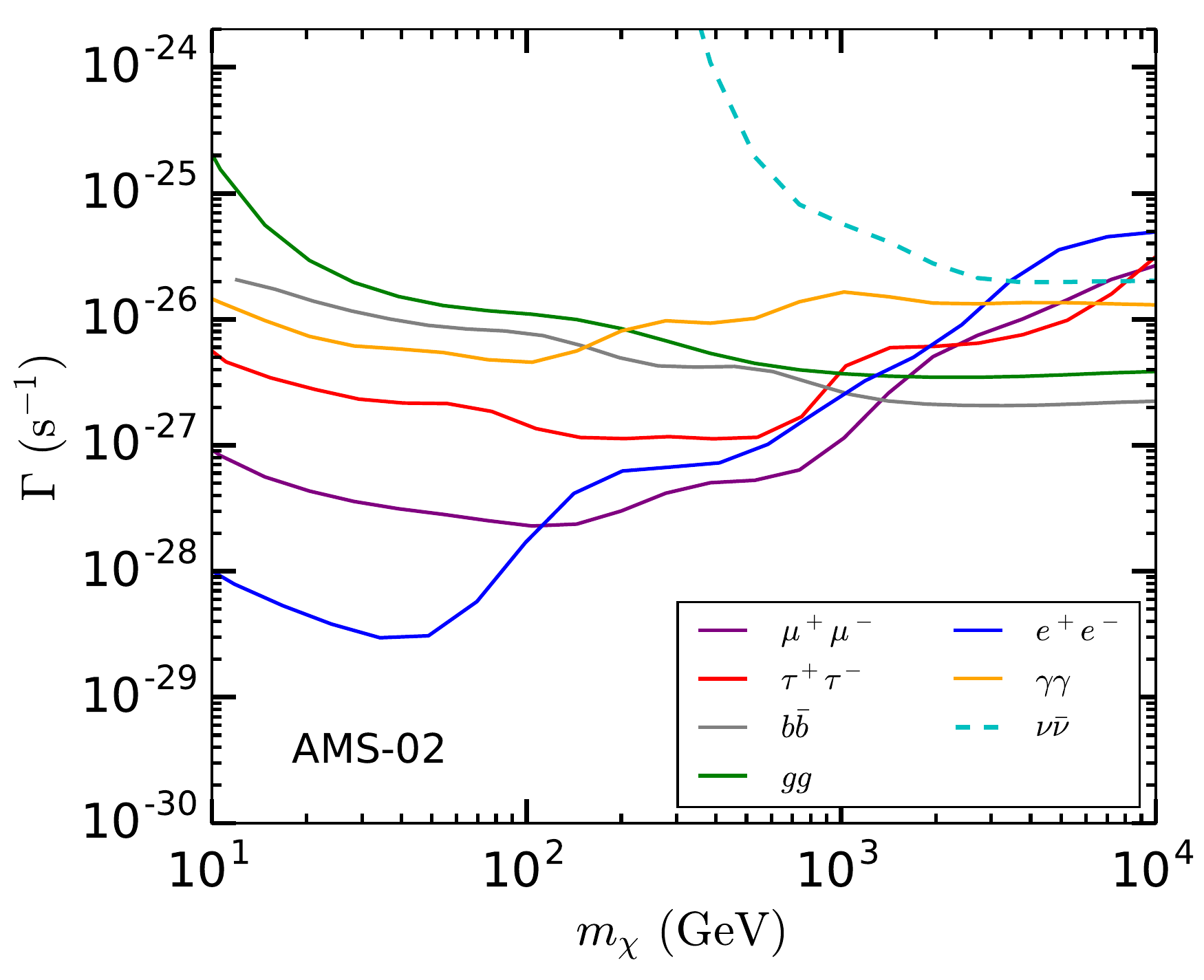}\\
\includegraphics[width=7.4cm,height=7.0cm]{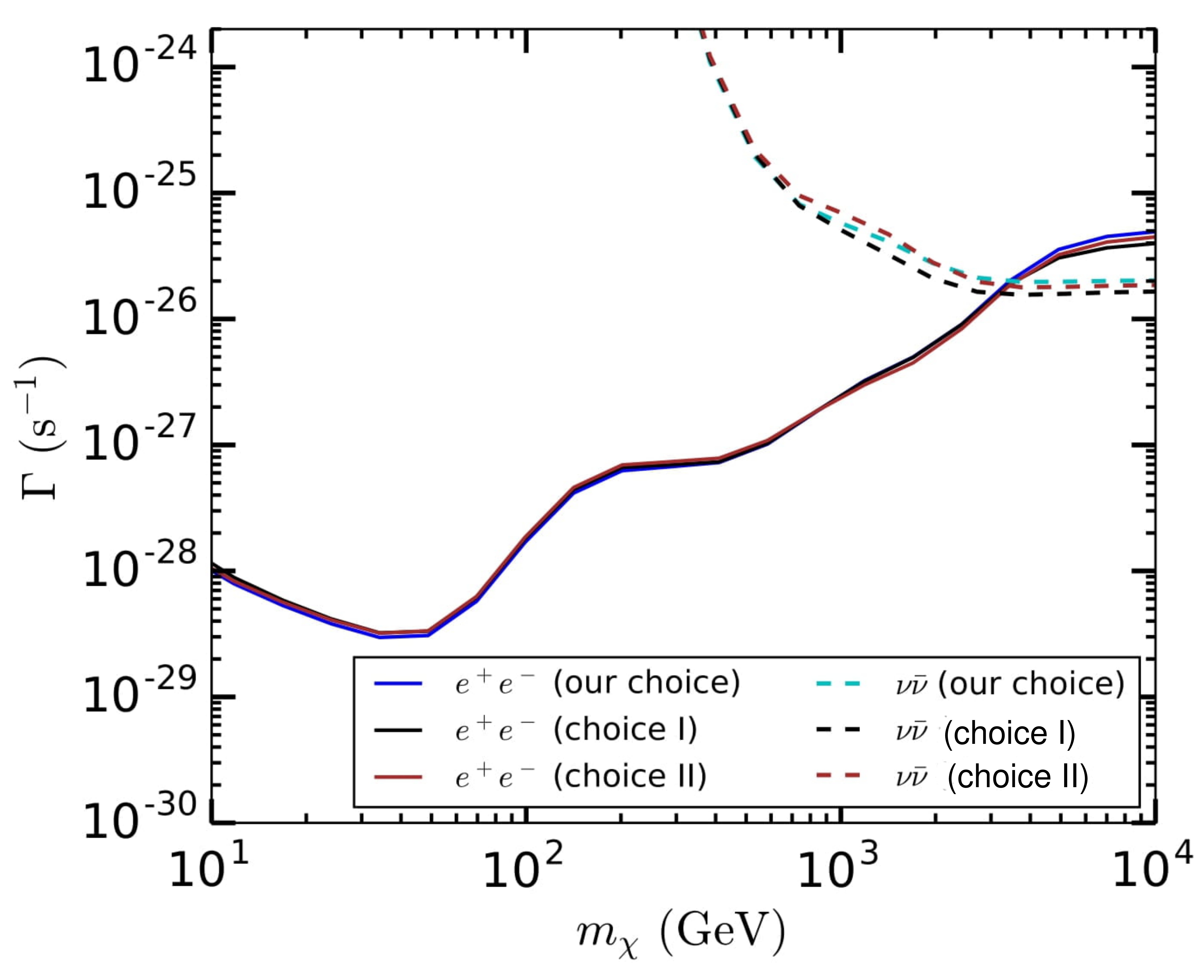}
\includegraphics[width=7.4cm,height=7.0cm]{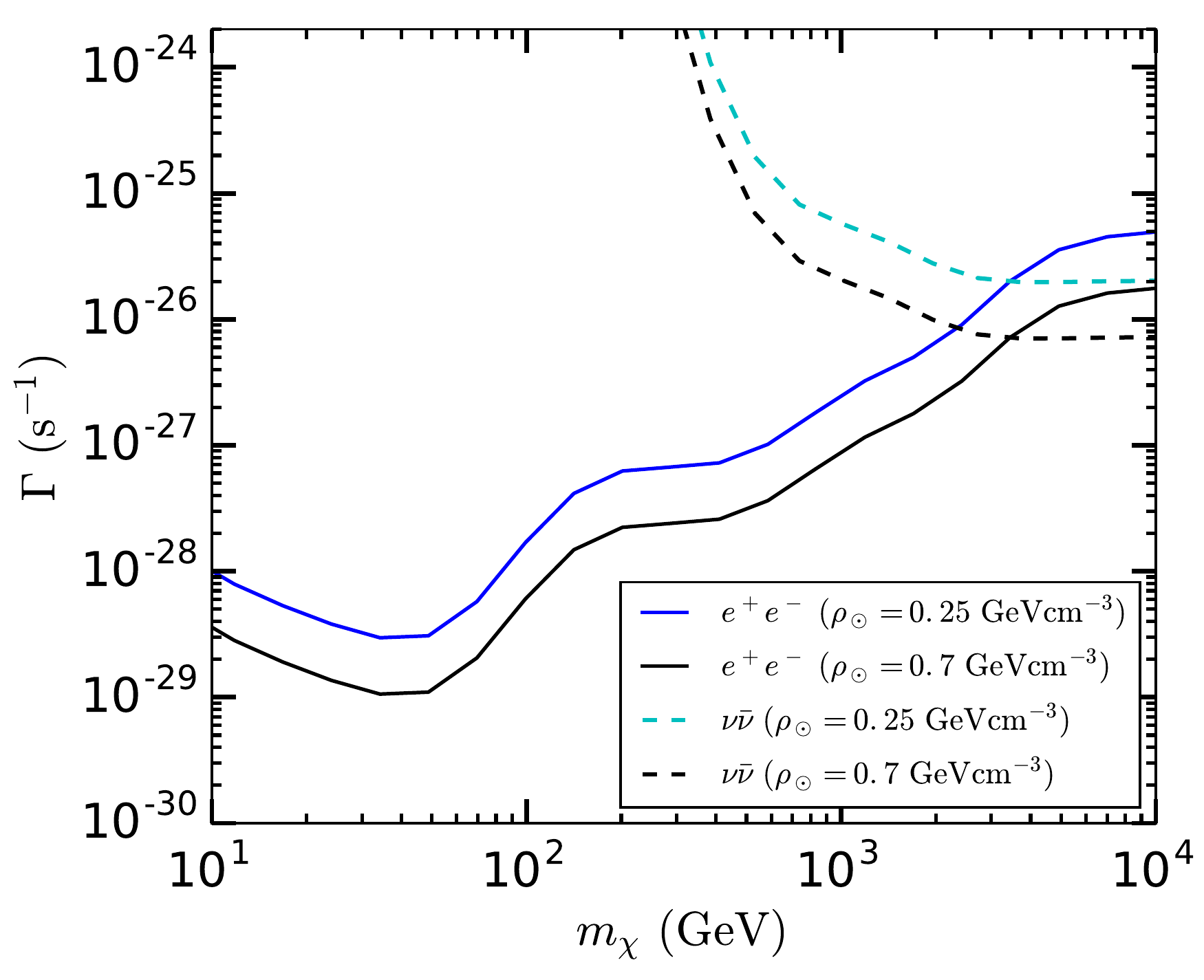}
\caption{{\it Top:} Cosmic-ray positron flux measurement of AMS-02 
is used to derive the 95\% C.L. upper limits on $\Gamma$ assuming 
100\% branching ratio for each individual SM final state. The results are 
obtained for $D_0 = 2.7 \times 10^{28}\,{\rm cm}^2{\rm s}^{-1}$, $\delta = 0.6$, 
$z_t = 4\,{\rm kpc}$ and $\rho_\odot = 0.25\,{\rm GeV}\,{\rm cm}^{-3}$. 
Here, the results for seven DM decay modes are presented for illustration. 
{\it Bottom:} Variation of these upper limits on $\Gamma$ (for $e^+e^-$ 
and $\nu\bar{\nu}$ final states), with the diffusion parameters 
(bottom left panel) and the local DM density $\rho_\odot$ (bottom right panel) 
are shown. See the text for details.}
\label{fig:AMS_positron}
\end{figure*}

In order to derive the $95\%$ C.L. upper limit on $\Gamma$ 
using the positron flux measured by AMS-02 we have adopted the 
methodology used in Ref.~\cite{Leane:2018kjk} to which the 
readers are referred for the details. 		
Similar to Ref.~\cite{Leane:2018kjk}, we assume the positron flux measured 
by AMS-02 arises solely from the astrophysical backgrounds and 
parameterize the $\log({\rm flux})$ as a degree 6 polynomial of 
$\log({\rm energy})$.
The $\chi^2$ is defined as follows:
\begin{equation}
\chi^2 = \sum_i \frac{\left(f^{model}_i(\alpha)-f^{data}_i \right)^2}{\sigma^2_i},
\end{equation}
where $i$ represents each individual energy bin of AMS-02 with 
$f^{data}_i$ signifies the central value of the measured positron flux~\cite{AMS:2014xys} 
and $f^{model}_i$ denotes the expected number of events in the $i$-th 
bin. $\sigma_i$ for each bin is obtained by adding the corresponding 
systematic and statistical uncertainties in quadrature~\cite{AMS:2014xys}. 
Next we determine the best-fit values of the associated 
parameters by minimizing this $\chi^2$ and obtained the best-fit $\chi^2_{\rm bf}$.  
Then we add the DM induced positron flux to this modelled background 
$f^{model}_i(\alpha)$ and vary the parameters of this function ($\{\alpha\}$) 
within $30\%$ of their best-fit values without DM. The DM decay width 
$\Gamma$ is increased until the $\chi^2$ increases by 2.71 from its best-fit 
value:
\begin{equation}
\chi^2 (\Gamma) = \chi^2_{\rm bf} + 2.71,
\label{eqn:chisqAMSA}
\end{equation}
so that the corresponding decay width gives the $95\%$ C.L. upper limit 
on $\Gamma$. Note that using the above-mentioned methodology we have 
recalculated the constraints for each individual SM final states for an 
annihilating WIMP and found that our limits match with the corresponding 
limits presented in Ref.~\cite{Leane:2018kjk}. 


From Eqn.~\ref{eqn:chisqAMSA}, the $95\%$ C.L. upper limits on 
$\Gamma$ are obtained for $100\%$ branching ratio attributed to each individual channel. 
These limits are shown in the top panel of Fig.~\ref{fig:AMS_positron} where we have 
reported the limits obtained for seven illustrative decay modes of the DM assuming 
$D_0 = 2.7 \times 10^{28}\,{\rm cm}^2{\rm s}^{-1}$, $\delta = 0.6$, $z_t = 4\,{\rm kpc}$ 
and $\rho_\odot = 0.25\,{\rm GeV}\,{\rm cm}^{-3}$. For $m_\chi$ in the range 
$10\,{\rm GeV}$ to $\sim 2\,{\rm TeV}$, the constraints are most stringent for 
DM decays to electron and muon final states. For $m_\chi \gtrsim 2\,{\rm TeV}$, 
the positron spectra coming from $e^+e^-$ and $\mu^+\mu^-$ final states are highly 
energetic and fall outside the sensitivity range of AMS-02 and hence the corresponding 
limits weaken. We have also shown the limit for $\chi \rightarrow \nu\bar{\nu}$ channel 
and found that this decay mode is reasonably constrained for $m_\chi \gsim 4\,{\rm TeV}$ 
(see cyan line in Fig.~\ref{fig:AMS_positron}; top panel). 
This is because, similar to the earlier cases, here, too, the resulting flux of 
positrons are comparable to the corresponding fluxes coming from other 
SM decay modes when $m_\chi$ is \textit{larger than a few TeV} 
(see Fig.~\ref{fig:dNdE_nunubar}). 
DM decays to gluons and $b\bar{b}$ pairs are most weakly constrained for 
$m_\chi \sim 10\,{\rm GeV}$ - $100\,{\rm GeV}$, while the $\gamma\gamma$ 
is the least constrained channel for $m_\chi \sim 100\,{\rm GeV}$ - $4\,{\rm TeV}$. 

In the bottom panel of Fig.~\ref{fig:AMS_positron} we have shown the variation 
of the upper limits on $\Gamma$ with the diffusion parameters (bottom left) and the local 
DM density (bottom right), respectively. Here, we have considered $e^+e^-$ and 
$\nu\bar{\nu}$ channels for the sake of illustration. In the bottom 
left panel, we have used two different sets of diffusion parameters, namely, 
$D_0 = 3.4 \times 10^{27}\,{\rm cm}^2{\rm s}^{-1}$, $\delta = 0.7$, 
$z_t = 4\,{\rm kpc}$ (choice I; shown in black) and 
$D_0 = 2.3 \times 10^{28}\,{\rm cm}^2{\rm s}^{-1}$, 
$\delta = 0.46$, $z_t = 15\,{\rm kpc}$ (choice II; shown in brown), in addition 
to our choice for these parameters which was mentioned earlier. 
Both of these choices are commonly used for the MW 
galaxy~\cite{Buch:2015iya,Lavalle:2014kca,Giesen:2015ufa,Evoli:2015vaa}. 
From this figure it is evident that the limits obtained by us are quite robust against the 
variation of the diffusion parameters. 
In the bottom right panel, apart from our choice of 
$\rho_\odot = 0.25\,{\rm GeV}\,{\rm cm}^{-3}$ 
we have also considered the maximum allowed value 
$\rho_\odot = 0.7 \,{\rm GeV}\,{\rm cm}^{-3}$ and shown the resulting limits for 
$e^+e^-$ and $\nu\bar{\nu}$ final states (black lines), since astronomical 
observations constrain $\rho_\odot$ in the range $0.25\,{\rm GeV}\,{\rm cm}^{-3} - 
0.7 \,{\rm GeV}\,{\rm cm}^{-3}$~\cite{Bovy:2012tw}. 
From this figure, it is clear that, as one increases $\rho_\odot$, the limits 
on $\Gamma$ strengthen and for our choice of $\rho_\odot = 0.25\,{\rm GeV}\,{\rm cm}^{-3}$ 
one obtains the most conservative upper limits on $\Gamma$.   



\subsubsection{Super-Kamiokande constraints}
\label{sec:SuperKlim}		

In addition to photons and electrons-positrons, SM neutrinos are also 
produced as the end products of the DM decay cascades. Similar to the 
gamma-ray fluxes the neutrino fluxes induced by the DM decays, too, are 
proportional to 
$\Gamma$ for any given value of $m_\chi$. 
Therefore, the decay width for any chosen value of the DM mass 
is constrained by comparing the expected DM decay generated neutrino flux 
from any suitably chosen astrophysical target with the measured flux of 
the SM neutrinos. Among several neutrino observations, 
Super-Kamiokande~\cite{Frankiewicz:2016nyr} provides the strongest 
constraints in case of DM decays for $m_\chi$ less than a 
TeV~\footnote{For higher DM masses constraints from IceCube observation 
become relevant. However, we have checked that, for 
$500\,{\rm GeV} \lesssim m_\chi \lesssim 10\,{\rm TeV}$, 
the IceCube constraints are weaker in comparison to those coming from AMS-02 and 
Fermi-LAT observations~\cite{Coy:2020wxp,Cohen:2016uyg,IceCube:2018tkk,IceCube:2021sog} which are already taken into account.}.
Assuming the DM decays to each individual SM final state with
$100\%$ branching ratio we have derived the $90\%$ C.L. upper limits on 
$\Gamma$ 
using the neutrino flux data from the MW galaxy measured by the 
Super-Kamiokande collaboration~\cite{Frankiewicz:2016nyr}. 


\begin{figure*}[htb!]
\centering
\includegraphics[width=7.6cm,height=7.4cm]{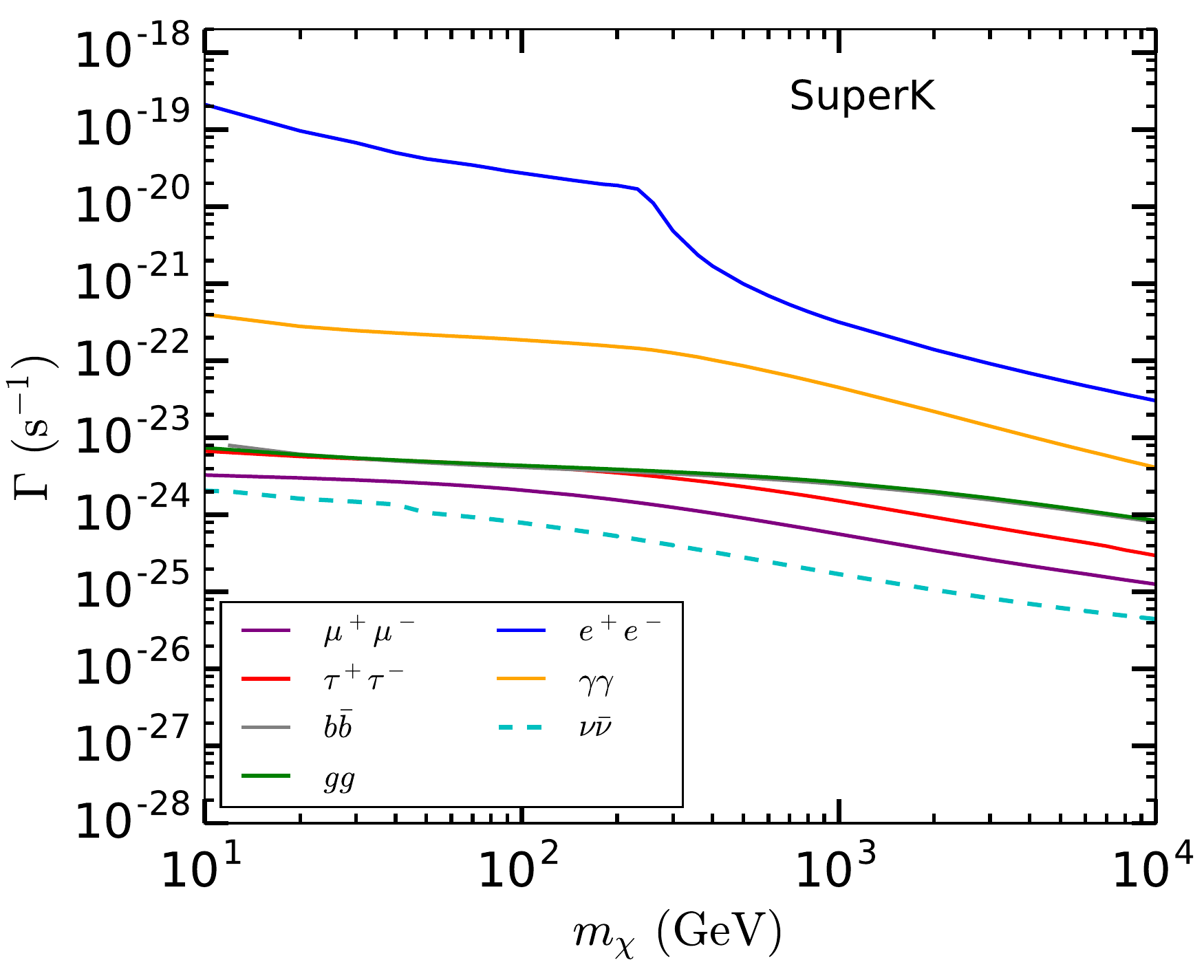}
\caption{90$\%$ C.L. upper limits on the DM decay width $\Gamma$ obtained 
using the measurement of the neutrino flux from the Milky Way 
(MW) galaxy by the Super-Kamiokande collaboration 
are shown for a 100$\%$ branching ratio attributed to each individual SM 
decay mode. Here the results for seven illustrative decay modes are shown.
}
\label{fig:SuperK}
\end{figure*}

In deriving these limits we have followed the ON-OFF procedure outlined 
in~\cite{Frankiewicz:2016nyr}. The ON and the OFF source regions are 
chosen exactly in accordance with Ref.~\cite{Frankiewicz:2016nyr}. 
The ON source region is a circular region centered around 
the galactic center (GC) with half-opening angle $80^{\circ}$ while the 
OFF source region is another circular region of the same size but offset 
by $180^{\circ}$ in Right Ascension with respect 
to the GC (see Ref.~\cite{Frankiewicz:2016nyr} for details). 
The neutrino+anti-neutrino flux distributions coming from the ON and the OFF source 
regions are obtained as follows: 
\begin{eqnarray}
\label{eqn:nudist}
\frac{d\Phi_{\rm ON,OFF}}{dE_\nu} &=& \,{\rm J}_{\rm ON,OFF}\frac{\Gamma}{4\pi\,m_{\chi}}\sum_f B_f \frac{2}{3}\,\frac{d N^{\nu}_f}{dE_\nu}(E_\nu), 
\label{eqn:SKflux}
\end{eqnarray}
where, $d N^{\nu}_f/dE_\nu$ represents the $\nu$ spectrum 
(same as the $\bar{\nu}$ spectrum) 
produced per DM decay to the SM final state $f$ and is obtained 
from~\cite{Sjostrand:2006za,Cirelli:2010xx,Ciafaloni:2010ti,Cirelli}. 
The factor of 1/3 in Eqn.~\ref{eqn:SKflux} represents the fact that 
we have assumed $e,\mu$ and $\tau$ flavours are equally populated. 
The effect of the neutrino oscillation during the propagation throughout 
the galaxy is incorporated following the procedure described in 
Ref.~\cite{Super-Kamiokande:2020sgt}. 
In Eqn.~\ref{eqn:SKflux}, ${\rm J}_{\rm ON,OFF}$ are the 
astrophysical J-factors for the ON and the OFF source 
regions~\cite{Frankiewicz:2016nyr}, respectively:
\begin{equation}
J_{\rm ON,OFF} = \underset{\Delta \Omega_{\rm ON,OFF}}{\int}d\Omega\underset{l.o.s}{\int}ds\,\rho_d(\vec{r}),
\label{eqn:JONOFF}
\end{equation}
where the NFW profile $\rho_d(\vec{r})$ (given in Eqn.~\ref{eqn:NFWprofile}) 
is parameterized by $r_s = 20\,{\rm kpc}$ and 
$\rho_\odot = 0.3\,{\rm GeV\,cm}^{-3}$~\cite{Frankiewicz:2016nyr,Frankiewicz:2018ixf}. 
In Eqn.~\ref{eqn:JONOFF}, $s$ is the line-of-sight ($l.o.s$) coordinate, 
$\Delta \Omega_{\rm ON,OFF}$ signify the solid 
angle subtended by the ON and the OFF source regions, respectively. 

Thereafter, the number of upward going muon (UP$\mu$) type signal events 
($N^{\rm UP\mu}_{\rm ON,OFF}$) and the 
number of contained muon events ($N^{\rm con}_{\rm ON,OFF}$) are 
calculated using the formulae provided in~\cite{Covi:2009xn}. 
Following Ref.~\cite{Frankiewicz:2016nyr} detector livetimes of
4527 days for the UP$\mu$ events and 4223.3 days for the contained muon 
events have been assumed. The total number of signal events are thus 
given by $N_{\rm ON,OFF} = N^{\rm UP\mu}_{\rm ON,OFF} + N^{\rm con}_{\rm ON,OFF}$.
The background flux of atmospheric neutrinos being isotropic the  
difference between the number of events expected from the ON and the OFF  
regions are essentially same as the difference between the number of  
expected signal events, i.e., $\Delta N = N_{\rm ON} - N_{\rm OFF}$. 
This number is then compared against the $90\%$ C.L. upper limit on 
the above-mentioned difference (provided in~\cite{Frankiewicz:2016nyr}) 
to derive the $90\%$ C.L. upper limit on $\Gamma$ as shown in 
Fig.~\ref{fig:SuperK}. 


In Fig.~\ref{fig:SuperK}, the $90\%$ C.L. upper limits on $\Gamma$ for 
seven decay channels are shown. Note that, unlike other 
observations, here the constraint on $\chi\rightarrow \nu\bar{\nu}$ channel 
is the strongest one over the entire DM mass range considered 
(see Figs.~\ref{fig:Planck_CMB}, \ref{fig:LAT_IGRB} and \ref{fig:AMS_positron}). 
This constraint is substantially stronger than those obtained from the 
previously considered observations for DM mass less than 
\textit{a few hundreds of GeV}. This is because, in this case 
the neutrinos (anti-neutrinos) themselves are detected without relying 
on the photons or the electrons-positrons originating from them. 
On the other hand, for the $e^+e^-$ and $\gamma\gamma$ final 
states the detectable $\nu\,(\bar{\nu})$ come from the 
radiation of electroweak gauge bosons, which being suppressed 
the corresponding constraints are the weaker ones 
(shown by blue and orange lines in Fig.~\ref{fig:SuperK}).     

\section{Analysis of existing data: a generic approach}
\label{sec:modelindepexcons}


We now proceed to derive new limits on the total DM decay width from 
existing astrophysical and cosmological data, allowing $\chi$ 
to decay into all possible SM particle pairs with arbitrary branching ratios. 
This is clearly different from most approaches taken so far, where 
constraints from different observations have been derived assuming 
$100\%$ branching ratio for each individual DM decay mode taken one 
at a time (see Sec.~\ref{sec:conseval}). 

In a generic model of decaying scalar dark matter, the DM particle may in 
principle decay into multiple two-body SM final states with different 
branching ratios. These branching fractions are decided by the 
parameters of the model. \textit{In such cases, the DM 
decay induced fluxes of $\gamma$-rays, electrons (positrons) and 
neutrinos (antineutrinos) are obtained by summing over the 
fluxes arising from each individual channels weighted by their respective 
branching fractions. Thus the shapes of these resulting flux distributions 
are quite different from that obtained in the case of DM decays to a 
single SM final state with $100\%$ branching ratio. The spectral 
shapes of these distributions govern the bin-by-bin fluxes and 
thereby affect the constraints arising from different observations.} 

\begin{figure*}[htb!]
\centering
\centering
\includegraphics[width=7.6cm,height=7.4cm]{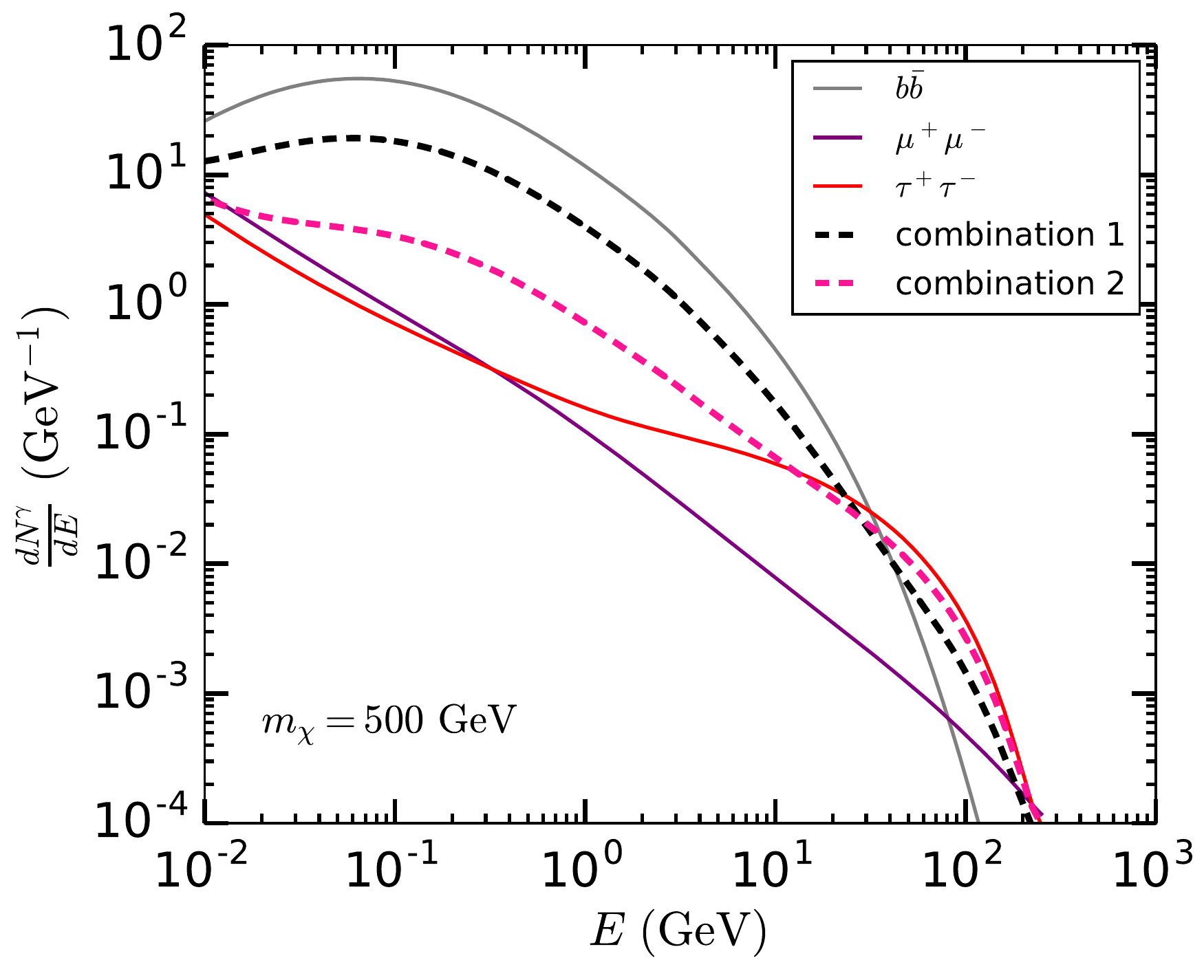}
\includegraphics[width=7.6cm,height=7.4cm]{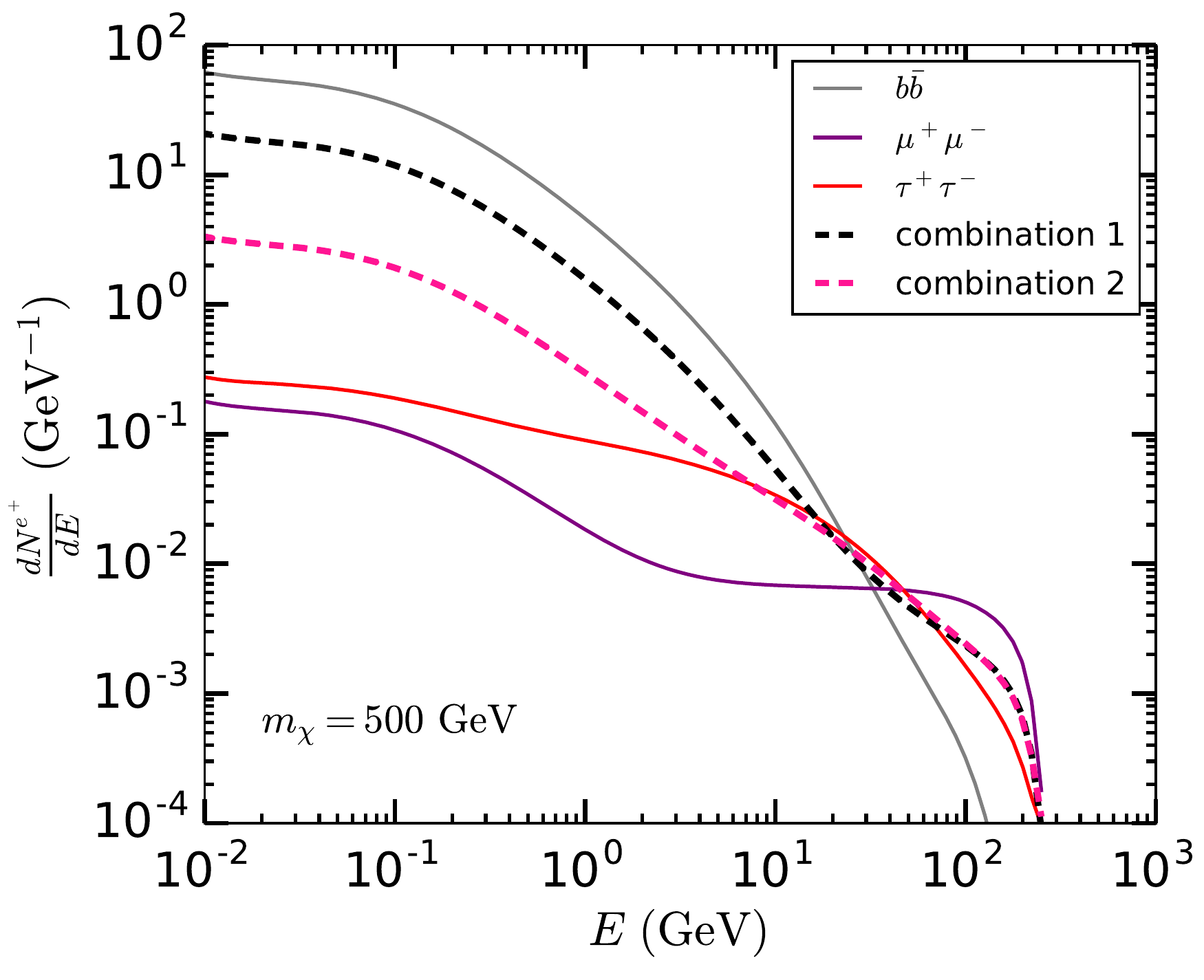}
\caption{For $m_\chi = 500\,{\rm GeV}$, the photon spectra 
$dN^{\gamma}/dE$ (left panel) and positron spectra 
$dN^{e^+}/dE$ (right panel) are shown for $b\bar{b}$ (gray solid line), 
$\mu^+\mu^-$ (purple solid line) and $\tau^+\tau^-$ (red solid line) final 
states assuming $100\%$ branching ratio for each individual channel. The corresponding 
energy distributions assuming equal branching fraction for each of the above-mentioned 
decay modes (black dashed lines) and the distributions obtained when 
$5\%$ branching ratio is attributed to $b\bar{b}$, $25\%$ branching ratio to 
$\mu^+\mu^-$ and $70\%$ branching ratio to $\tau^+\tau^-$ (magenta dashed lines) 
are also shown in each panel. See the text for details.}
\label{fig:dNdE_BRcomb_compare}
\end{figure*}

This is illustrated in Fig.~\ref{fig:dNdE_BRcomb_compare}, where, we 
have presented the photon spectra $dN^{\gamma}/dE$ (left panel) and the positron 
spectra $dN^{e^+}/dE$ (right panel) for two different branching ratio combinations 
of $b\bar{b}$, $\mu^+\mu^-$ and $\tau^+\tau^-$ final states, considering 
$m_\chi = 500\,{\rm GeV}$. For combination 1, shown by the black dashed lines 
in each panel, equal branching fraction, i.e., $33.33\%$, is attributed to each 
individual DM 
decay mode. While, the magenta dashed lines are obtained for combination 2, 
for which we have assumed $5\%$ branching ratio to $b\bar{b}$, $25\%$ branching 
ratio to $\mu^+\mu^-$ and $70\%$ branching ratio to $\tau^+\tau^-$ channels. For 
comparison, we have also shown the corresponding spectra obtained for $100\%$ 
branching ratio to $b\bar{b}$ (gray solid line), $\mu^+\mu^-$ (purple solid line) and 
$\tau^+\tau^-$ (red solid line) final states. Clearly, in every energy bin, the 
fluxes predicted by these combinations are different from what one obtains for any 
particular final state with $100\%$ branching ratio and thus for each observation 
the resulting limits are also different. Also, note that the spectra obtained for 
combination 1 are quite different from those obtained for combination 2.
This explains why the limit on the total decay width of the DM 
should be derived using these resulting flux distributions, the calculation 
of which requires a precise knowledge of the branching ratios of each allowed 
decay mode.


In fact, studies exist in the literature where DM decays are 
parameterized by higher-dimensional operators and constraints 
on the DM parameter space are derived assuming DM decays to all 
possible SM final states allowed by any given higher-dimensional 
operator~\cite{Ghosh:2020ipv}. 
In this case, branching ratio for any given decay mode is decided 
by the higher-dimensional operator which is responsible for the 
corresponding decay. However, it is possible to adopt a more general 
approach where one does not need to know the Lagrangian governing the DM 
interactions. 
For example, in the context of 2$\rightarrow$2 $s$-wave annihilations of a WIMP  
DM, Ref.~\cite{Leane:2018kjk} has derived a robust 
lower limit of $\sim 20\,{\rm GeV}$ on $m_\chi$ using the 
data of Planck CMB observation~\cite{Planck:2015fie}, AMS-02 
positron flux measurement~\cite{AMS:2014xys} and 
Fermi-LAT gamma-ray measurement from the dSph 
galaxies~\cite{Fermi-LAT:2015att,Fermi-LAT:2016uux}. In deriving 
this limit arbitrary branching fractions are assumed for each of the 
DM annihilation final states (excluding $\nu\bar{\nu}$). 
The most important conclusion of such study is substantial 
relaxation of the lower limit on $m_\chi$ which is otherwise of 
the order of $\sim 100\,{\rm GeV}$ when one considers DM 
annihilations to individual SM final states, e.g., $b\bar{b}$, 
$\tau^+\tau^-$, with 100$\%$ branching ratio. Motivated by this, 
here, we shall derive a robust 
constraint on the DM parameter space 
utilizing the data of Planck CMB~\cite{Planck:2015fie}, 
Fermi-LAT IGRB~\cite{Fermi-LAT:2014ryh}, AMS-02 positron~\cite{AMS:2014xys} 
and Super-Kamiokande neutrino flux measurements~\cite{Frankiewicz:2016nyr} 
by allowing arbitrary branching fraction to each individual DM decay mode.

\subsection{Analysis methodology}
\label{sec:modelindepexconsmethod}


In order to derive the upper limit on the total decay width 
of $\chi$ that decays to all possible kinematically 
allowed two-body SM final states, i.e., 
$\chi \rightarrow$ ${\rm SM}_{1}$ $\overbar{\rm SM}_2$ (as 
mentioned in Sec.~\ref{sec:DMdecaytheory}) with arbitrary branching 
fractions we take the following steps:

\begin{itemize}
\item We have varied $m_\chi$ in the range 10 GeV - 10 TeV. 
For any given value of $m_\chi$, we scan over all possible branching 
ratio combinations of the kinematically allowed two-body SM final states 
${\rm SM}_{1}$ $\overbar{\rm SM}_2$, to derive the maximum allowed value 
of the total DM decay width. 
While scanning over the possible branching 
ratio combinations the branching fraction corresponding to each two-body 
SM final state is varied in the range 0$\%$ to 100$\%$ with an incremental 
change of $2\%$ in each successive step. Additionally, we have also 
ensured that for each such combination the branching fractions of all 
channels add up to unity. 
		

\item For any value of $m_\chi$ and specific branching ratio 
combination, we derive the 95$\%$ C.L. upper limits on $\Gamma$ for 
each individual observations following the procedures described 
in Sec.~\ref{sec:conseval}. Thus, given the $m_\chi$ and the 
branching ratio combination, we have three different upper limits on 
$\Gamma$ corresponding to Planck CMB~\cite{Planck:2015fie}, 
Fermi-LAT IGRB~\cite{Fermi-LAT:2014ryh} and AMS-02 positron 
flux~\cite{AMS:2014xys} measurements. 
The \textit{strongest} among these upper limits represents the value 
of $\Gamma$ that is allowed by all the observations considered. 
In addition, when $\chi \rightarrow \nu\bar{\nu}$ is included in our 
analysis, the upper limit on $\Gamma$ coming from Super-Kamiokande neutrino 
flux measurement~\cite{Frankiewicz:2016nyr} is also 
taken into consideration. 
		


\item Given any $m_\chi$, for each branching ratio combination, 
we obtained the $\Gamma$ value that is consistent with all observations. The 
\textit{weakest} one among these allowed values of $\Gamma$ obtained 
for different branching ratio combinations represents the maximum 
allowed value of the total DM decay width, i.e., $\Gamma_{\rm max}$, 
for the considered value of $m_\chi$. 
Therefore, there exists no branching ratio combination for which 
a value of $\Gamma > \Gamma_{\rm max}$ is allowed by the existing 
observational data. The branching ratio combination for which the allowed 
value of $\Gamma$ is the weakest, i.e., the combination corresponding to 
$\Gamma_{\rm max}$, is defined as the threshold branching ratio 
combination ($B_f$). 
\end{itemize}
For any $m_\chi$ value $\Gamma_{\rm max}$ and $B_f$ together define 
the most weakly constrained decaying scalar DM scenario assuming DM 
decays to SM particle pairs only. The above-mentioned methodology is 
applied to derive $\Gamma_{\rm max}$ for four different situations:

\begin{itemize}
\item \underline{Case 1:} DM decays to visible SM final states with arbitrary 
branching fractions, while the branching ratio of the $\nu\bar{\nu}$ final 
state is identically set to zero.
\item \underline{Case 2:} DM decays to all possible SM final states 
including $\nu\bar{\nu}$. In this case the branching fraction for 
the $\nu\bar{\nu}$ decay mode is also allowed to vary in the range 
0$\% - 100\%$. 
\item \underline{Case 3:} DM decays to $\nu\bar{\nu}$ with exactly 
50$\%$ branching ratio while the branching ratios of other SM decay 
modes are allowed to vary freely such that their branching ratios add 
up to 0.5. In this case, the incremental changes in the branching ratios 
of all non-neutrino SM final states are $1\%$ in each step. 
\item \underline{Case 4:} DM decays to $\nu\bar{\nu}$ with 100$\%$ 
branching ratio and the branching ratios for all other decay modes are 
set to zero.
\end{itemize}


\noindent 
Actually, Case 3 and Case 4 are two sub-classes of Case 2. 
However, our results are very sensitive to the exact branching ratio of the 
$\nu\bar{\nu}$ final state and thus we have considered 
Case 3 and Case 4, separately. 

In the next subsection, we present $\Gamma_{\rm max}$ for all of the aforementioned 
cases and the threshold branching fractions $B_f$ for Case 1 and Case 2.

\subsection{General constraints on decaying DM}
\label{sec:modelindepexconscons}

\begin{figure*}[htb!]
\centering
\includegraphics[width=7.6cm,height=7.4cm]{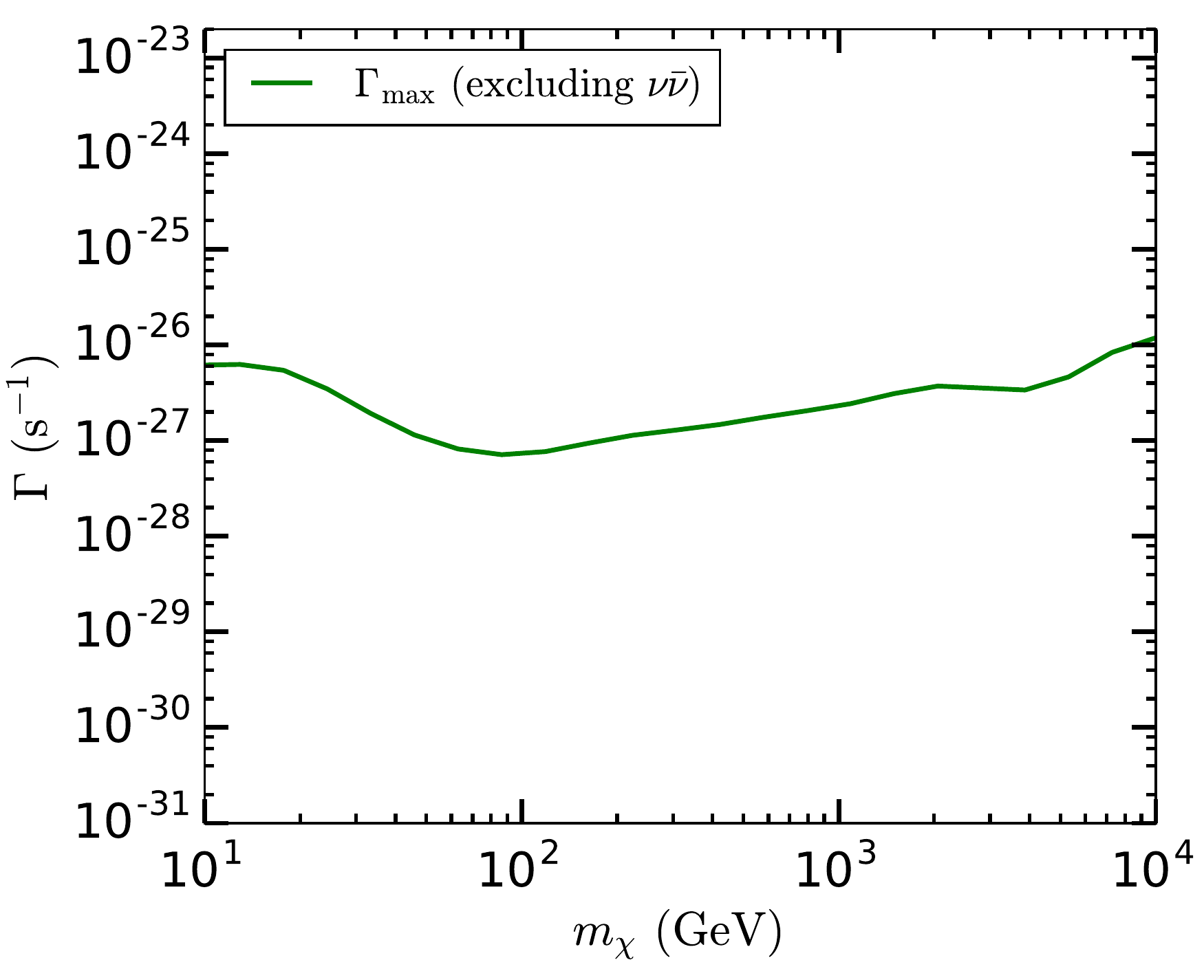}\hspace{0mm}
\includegraphics[width=7.6cm,height=7.4cm]{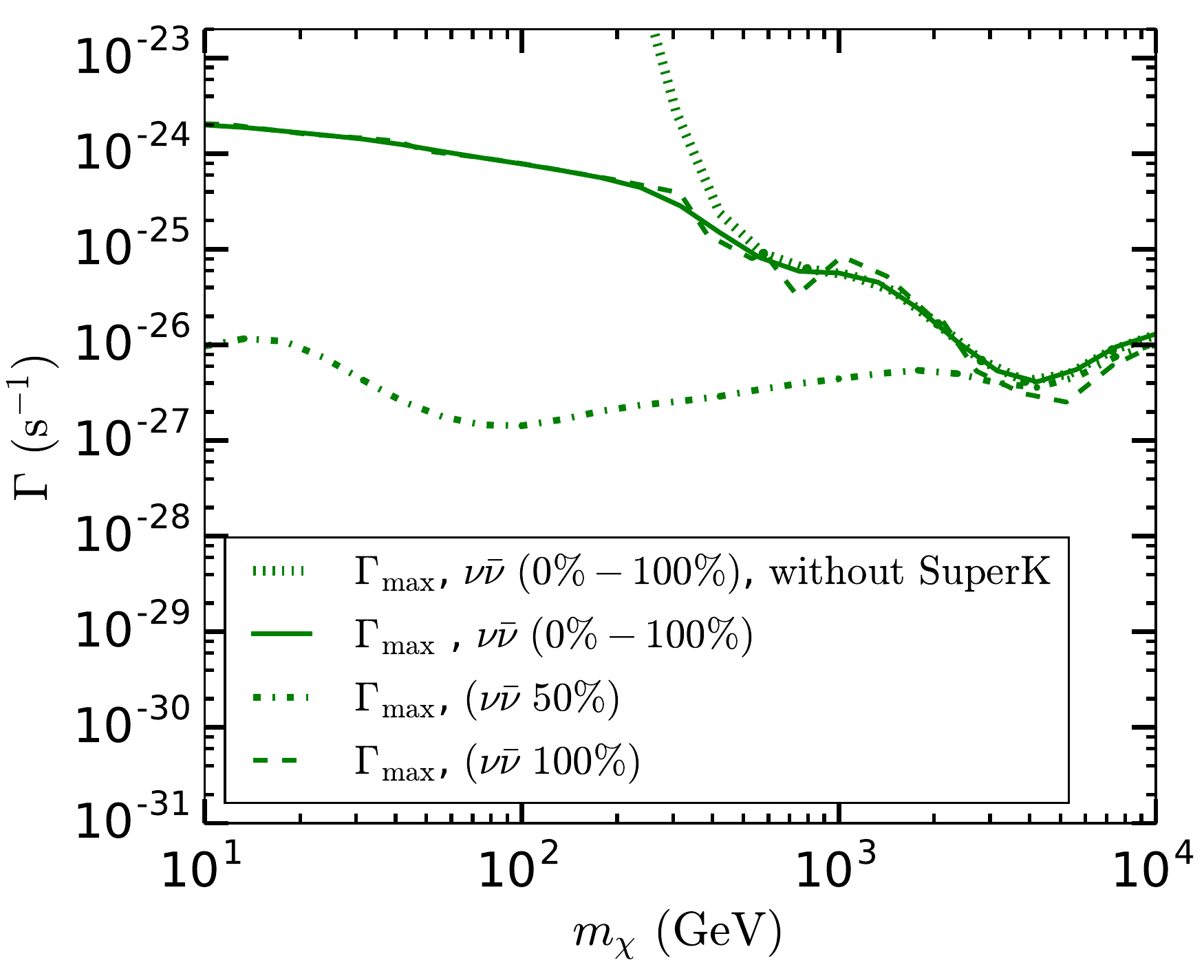}
\caption{
{\it Left:} Varying the branching ratios of all SM final states 
(excluding $\nu\bar{\nu}$) in the range $0\% - 100\%$ and 
combining the data of Planck CMB, 
Fermi-LAT IGRB 
and AMS-02 positron 
observations the maximum allowed  value of the total DM decay 
width, i.e., $\Gamma_{\rm max}$, is obtained for $m_\chi$ in 
the range 10 GeV - 10 TeV.
For any $\Gamma > \Gamma_{\rm max}$ there exists no branching ratio 
combination which is allowed by all the aforementioned observations.    
{\it Right:} The branching ratio of $\nu\bar{\nu}$ final state is also 
varied in the range $0\% - 100\%$ and the data of the above-mentioned 
observations are used along with the data of Super-Kamiokande 
neutrino flux 
measurement to obtain the $\Gamma_{\rm max}$ shown by 
the green solid line. The green dotted line is obtained without 
including the Super-Kamiokande data. 
%
Combining all four observations the green dashed dotted and the 
green dashed lines are derived which represent 
the $\Gamma_{\rm max}$ for DM decays to 
$\nu\bar{\nu}$ with precisely 50$\%$ and 100$\%$ branching ratios, 
respectively. See the text for details.}
\label{fig:Gamma_max}
\end{figure*}

\begin{figure*}[htb!]
\centering
\includegraphics[width=7.6cm,height=7.4cm]{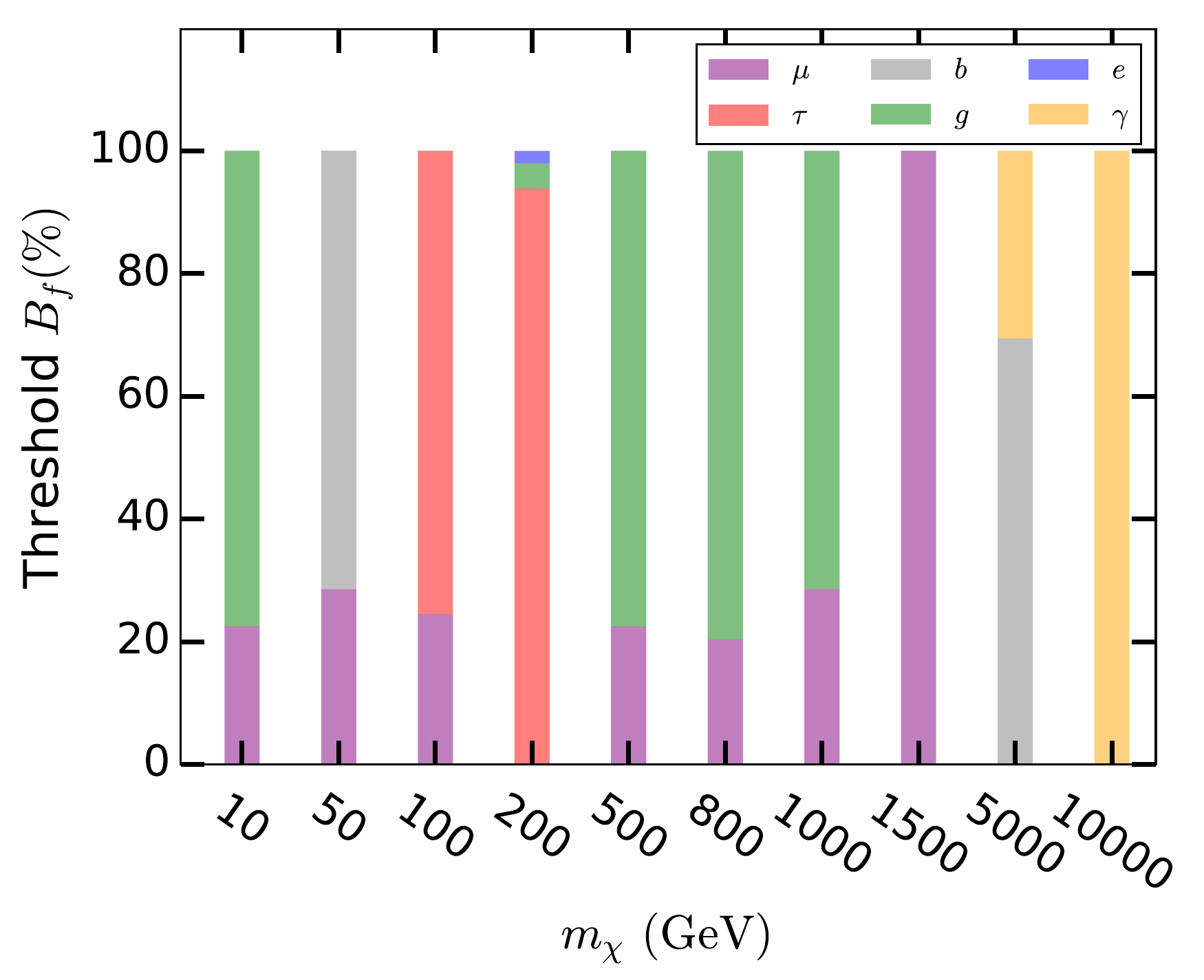}\hspace{0mm}
\includegraphics[width=7.6cm,height=7.4cm]{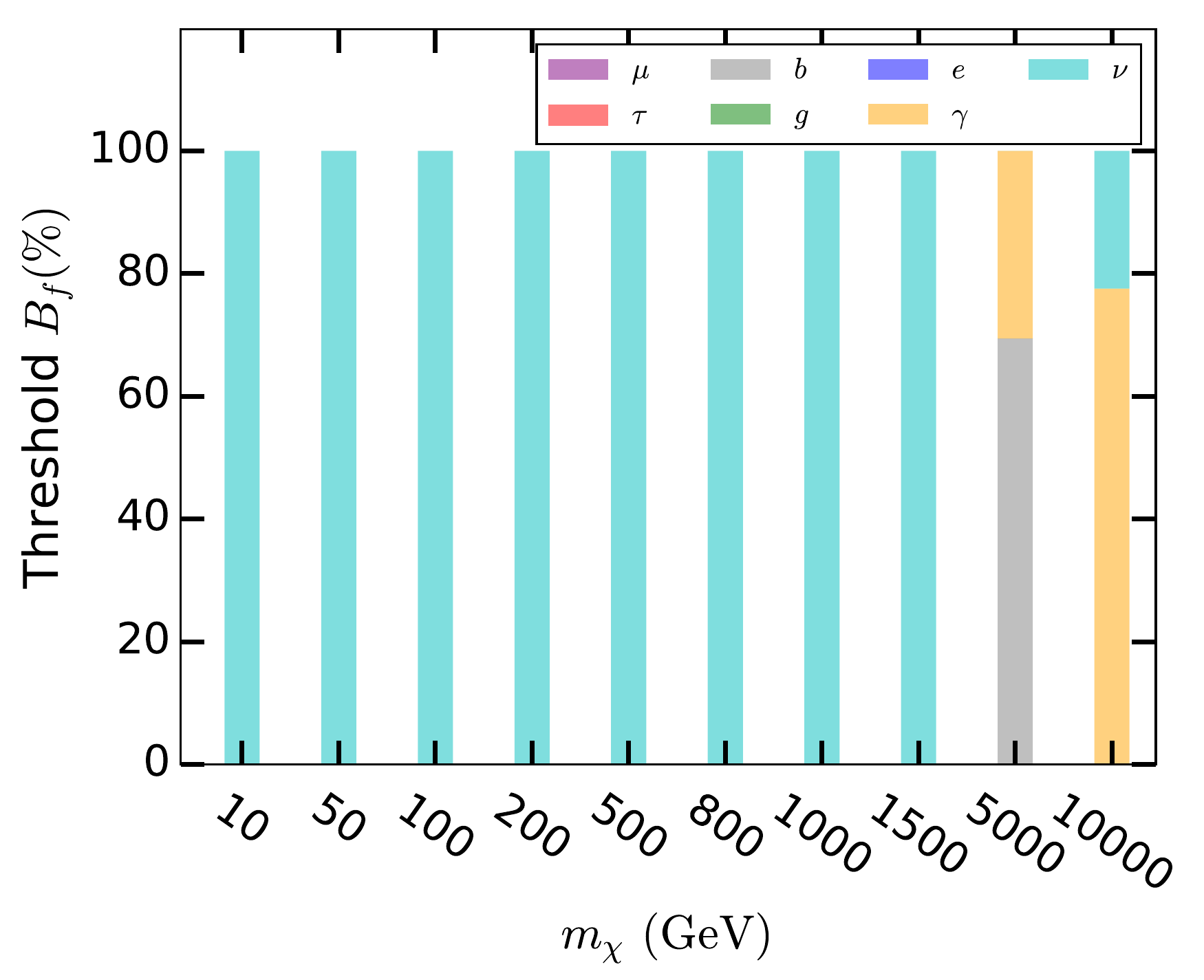}
\caption{{\it Left:} One representative set of threshold branching ratio 
combinations ($B_f$'s) that correspond to the $\Gamma_{\rm max}$ obtained for 
Case 1, i.e., excluding the $\nu\bar{\nu}$ decay mode 
(shown in Fig.~\ref{fig:Gamma_max}; left panel) are shown for various DM masses.
{\it Right:} The threshold $B_f$'s (a representative set) for the 
$\Gamma_{\rm max}$ obtained in Case 2, i.e., when the branching ratio for 
$\chi \rightarrow \nu\bar{\nu}$ is also allowed to vary in the range 
$0\% - 100\%$ and the data of all four observations are used, are 
presented (solid line in Fig.~\ref{fig:Gamma_max}; right panel).   
See the text for details.
}
\label{fig:Gamma_max_leftBr}
\end{figure*}

For the range of the DM masses we have studied, $\Gamma_{\rm max}$'s for 
the four above-mentioned cases are shown in Fig.~\ref{fig:Gamma_max} and 
the threshold branching fractions, $B_f$'s for Cases 1 and 2 are 
presented in Tab.~\ref{tab:BPthreshold} and in Fig.~\ref{fig:Gamma_max_leftBr}. 

In the left panel of Fig.~\ref{fig:Gamma_max}, we show $\Gamma_{\rm max}$ 
for Case 1 by the green solid line. 
In deriving this constraint we have used the data collected by 
Planck~\cite{Planck:2015fie}, Fermi-LAT~\cite{Fermi-LAT:2014ryh} and  
AMS-02~\cite{AMS:2014xys}. 
Note that when we considered the DM decays to a single channel with 
100$\%$ branching ratio the existing constraints are quite strong. 
For example, for decays to $\gamma\gamma$ final state one finds that 
$\Gamma \lsim 3 \times 10^{-29}\,{\rm s}^{-1}$ is allowed for 
$m_\chi \sim 400\,{\rm GeV}$ (see Fig.~\ref{fig:LAT_IGRB}), while for 
the $e^+e^-$ channel the upper limit on $\Gamma$ is 
$\sim 3 \times 10^{-29}\,{\rm s}^{-1}$ for $m_\chi \sim 40\,{\rm GeV}$ 
(see Fig.~\ref{fig:AMS_positron}). 
On the other hand, \textit{allowing arbitrary branching ratios to different 
channels relax the constraints substantially, e.g., $\Gamma \lsim 
10^{-27}\,{\rm s}^{-1}$ is consistent with all observations for all 
values of $m_\chi$ in the range 10 GeV - 10 TeV 
(Fig.~\ref{fig:Gamma_max}; left panel).}

The limits for the Cases 2, 3 and 4 are shown in the right panel of 
Fig.~\ref{fig:Gamma_max}. In obtaining the dotted line, the branching 
ratios of all decay channels including the $\chi \rightarrow \nu\bar{\nu}$ 
channel 
are varied in the range $0\% - 100\%$ and the data of 
Planck CMB~\cite{Planck:2015fie}, Fermi-LAT IGRB~\cite{Fermi-LAT:2014ryh} 
and AMS-02 positron flux~\cite{AMS:2014xys} observations have been used. 
It is clear that, no upper limit on $\Gamma$ is obtained in this case,  
for DM masses below \textit{a few hundreds of GeV}.
This is because, for such values of $m_\chi$, the low energy 
$\nu\bar{\nu}$ pairs produced from DM decays give rise to highly suppressed 
$e^-$, $e^+$ and $\gamma$-ray spectra (see the discussion of 
Fig.~\ref{fig:dNdE_nunubar}), which are not detectable in any of the 
observations considered. 
For heavier DM particles, on the other hand, 
final state $\nu\bar{\nu}$ pairs are highly energetic and 
hence the fluxes of $e^-$, $e^+$ and $\gamma$ also become comparable to that 
coming from other SM final states (see Fig.~\ref{fig:dNdE_nunubar}). 
Thus, for such values of $m_\chi$, the constraint on the $\nu\bar{\nu}$ 
channel, become stronger than some of the other SM final states 
(see Figs.~\ref{fig:LAT_IGRB} and~\ref{fig:AMS_positron}) 
and the maximum allowed $\Gamma$ also strengthens. 


The situation changes when the data from the Super-Kamiokande 
observation~\cite{Frankiewicz:2016nyr} are also taken into account. 
\textit{In particular, for $m_\chi \lsim 250\,{\rm GeV}$, Super-Kamiokande 
constraint on the $\nu\bar{\nu}$ final state is considerably stronger 
(see Fig.~\ref{fig:SuperK}) since in this case the final state $\nu\bar{\nu}$ 
pairs themselves are detected instead of the $e^-$, $e^+$ and $\gamma$ 
fluxes produced via electroweak radiation.} 
Therefore, the resulting $\Gamma_{\rm max}$, 
shown by the green solid line (see Fig.~\ref{fig:Gamma_max}; right panel) 
is stronger than what we obtained in the previous case, where we have used the data 
collected by Planck~\cite{Planck:2015fie}, Fermi-LAT~\cite{Fermi-LAT:2014ryh}, 
and AMS-02~\cite{AMS:2014xys}, for $m_\chi  \lsim 250\,{\rm GeV}$. 
\textit{Thus, including $\chi$ decays to $\nu\bar{\nu}$ pairs 
and taking Super-Kamiokande data into account, we obtain 
$\Gamma_{\rm max} \sim 10^{-26}\,{\rm s}^{-1}$ for 
$m_\chi \gsim \mathcal{O}({\rm TeV})$ and 
$\Gamma_{\rm max} \sim 10^{-24}\,{\rm s}^{-1}$ for 
$m_\chi \sim 100\,{\rm GeV}$.}

Comparing the solid lines in left and right panels of Fig.~\ref{fig:Gamma_max}, 
one can clearly see that the inclusion of the $\nu\bar{\nu}$ final state 
changes $\Gamma_{\rm max}$ substantially. Therefore, in the right panel of 
Fig.~\ref{fig:Gamma_max}, we also present $\Gamma_{\rm max}$ for 
Case 3 (dashed dotted line) and Case 4 (dashed line), respectively. In Case 3, 
sum of the branching fractions of the visible SM final states always equals 0.5 and 
the resulting fluxes of $e^+$($e^-$), $\gamma$ and $\nu(\bar{\nu})$ 
are added to half of the corresponding fluxes that would have come if the 
$\nu\bar{\nu}$ final state had $100\%$ branching ratio. Using these fluxes 
we then determine $\Gamma_{\rm max}$ using the data of Planck CMB~\cite{Planck:2015fie}, 
Fermi-LAT IGRB~\cite{Fermi-LAT:2014ryh}, AMS-02 positron~\cite{AMS:2014xys} 
and Super-Kamiokande neutrino~\cite{Frankiewicz:2016nyr} observations. 
In this case, throughout the considered $m_\chi$ range, we obtain 
$\Gamma_{\rm max} \sim 10^{-26}\,{\rm s}^{-1} - 10^{-27}\,{\rm s}^{-1}$ 
which is stronger than the limit obtained in Case 2, especially in the 
range $m_\chi \lesssim \mathcal{O}({\rm TeV})$. This is because, in this case, 
the observed fluxes always receive substantial contributions from the 
visible SM final states.


On the other hand, in Case 4, the fluxes of stable SM particles 
arise from the $\nu\bar{\nu}$ pairs produced in DM decays. Therefore, for any value 
of $m_\chi$ the strongest constraint on the $\nu\bar{\nu}$ final state dictate 
the $\Gamma_{\rm max}$ (dashed line in Fig.~\ref{fig:Gamma_max}; right panel). 
Therefore, for $m_\chi \lsim$ \textit{a few hundreds of GeV}, $\Gamma_{\rm max}$ 
follows the Super-Kamiokande limit on the $\chi \rightarrow \nu\bar{\nu}$ channel 
(see Fig.~\ref{fig:SuperK}) while for higher values of $m_\chi$, $\Gamma_{\rm max}$ 
is determined by the data of Planck~\cite{Planck:2015fie}, 
Fermi-LAT~\cite{Fermi-LAT:2014ryh} and AMS-02~\cite{AMS:2014xys} observations. 
Note that $\Gamma_{\rm max}$ in this case is stronger than that obtained 
in Case 2 for $m_\chi \gsim 2\,{\rm TeV}$. This is because, for such masses 
Fermi-LAT and AMS-02 constraints on the $\nu\bar{\nu}$ final state are stronger 
than the corresponding constraints on some of the visible SM final states 
(see Figs.~\ref{fig:LAT_IGRB} and \ref{fig:AMS_positron}). 

\begin{table}[htb!]
\footnotesize 
\begin{center}
\begingroup
\setlength{\tabcolsep}{8pt}
\renewcommand{\arraystretch}{1.30} 
\begin{tabular}{|c|c|c|cccccccc|}
\hline
& $m_\chi$  & $\Gamma_{\rm max}$ & \multicolumn{8}{c}{Threshold branching ratio $B_f$ ($\%$)} \vline \\
& (GeV) & $({\rm s}^{-1})$ & $gg$ & $b\bar{b}$ & $\gamma\gamma$  & $e^+e^-$ & $\mu^+\mu^-$ & $\tau^+\tau^-$ & $W^+W^-$ & $\nu\bar{\nu}$ \\
\hline
\hline 	
& 10 & $6.1 \times 10^{-27}$ & $77.7$ & $0$ & $0$ & $0$ & $22.3$ & $0$ & - & -  \\ 
& 50 & $1.0 \times 10^{-27}$ & $0$ & $71.6$ & $0$ & $0$ & $28.4$ & $0$ & - & -  \\
& 100 & $7.4 \times 10^{-28}$ & $0$ & $0$ & $0$ & $0$  & $24.3$ & $75.7$  & - & -  \\
& \multirow{ 2}{*}{200} & $\!\!\!\!$\multirow{ 2}{*}{\Bigg\{} $\!\!1.1 \times 10^{-27}$ & $4.2$ & $0$ & $0$ & $2.0$  & $0$ & $93.8$ & $0$ & - \\
& & $1.5 \times 10^{-27}$ & $0$ & $0$ & $0$ & $0$ & $0$ & $93.9$ & $6.1$ & -  \\
Case 1 & 500 & $1.6 \times 10^{-27}$ & $77.9$ & $0$ & $0$ & $0$  & $22.1$ & $0$ & $0$ & -  \\
& \multirow{ 2}{*}{800} & $\!\!\!\!$\multirow{ 2}{*}{\Bigg\{} $\!\!2.1 \times 10^{-27}$ & $79.9$ & $0$ & $0$ & $0$  & $20.1$ & $0$ & $0$ & -  \\
& & $2.0 \times 10^{-27}$ & $0$ & $0$ & $0$ & $0$ & $0$ & $0$ & $100$ & -  \\
& \multirow{ 2}{*}{1000} & $\!\!\!\!$\multirow{ 2}{*}{\Bigg\{} $\!\!2.4 \times 10^{-27}$ & $71.7$ & $0$ & $0$ & $0$  & $28.3$ & $0$ & $0$ & - \\
& & $2.3 \times 10^{-27}$ &  $9.9$ & $0$ & $0$ & $0$ & $0$ & $0$ & $90.1$ & -  \\
&  1500 & $3.2 \times 10^{-27}$ & $0$ & $0$ & $0$ & $0$  & $100$ & $0$ & $0$ & -  \\
&  \multirow{ 2}{*}{5000} & $\!\!\!\!$\multirow{ 2}{*}{\Bigg\{} $\!\!4.6 \times 10^{-27}$ & $0$ & $69.2$ & $30.8$ & $0$  & $0$ & $0$ & $0$ & -  \\
&  & $4.7 \times 10^{-27}$ & $0$ & $0$ & $10.3$ & $0$ & $0$ & $0$ & $89.7$ & -  \\ 
&  10000 & $1.2 \times 10^{-26}$ & $0$ & $0$ & $100$ & $0$  & $0$ & $0$ & $0$ & -  \\
\hline
\hline
& 10 & $2.0 \times 10^{-24}$ & $0$ & $0$ & $0$ & $0$ & $0$ & $0$ & - & $100$ \\
& 50 & $1.1 \times 10^{-24}$ & $0$ & $0$ & $0$ & $0$ & $0$ & $0$ & - & $100$ \\
& 100 & $7.8 \times 10^{-25}$ & $0$ & $0$ & $0$ & $0$ & $0$ & $0$ & - & $100$ \\
&  200 & $5.0 \times 10^{-25}$ & $0$ & $0$ & $0$ & $0$ & $0$ & $0$ & $0$ & $100$ \\
Case 2 & 500 & $1.1 \times 10^{-25}$ & $0$ & $0$ & $0$ & $0$  & $0$ & $0$ & $0$ & $100$ \\
&  800 & $5.8 \times 10^{-26}$ & $0$ & $0$ & $0$ &  $0$ & $0$ & $0$ & $0$ & $100$ \\
&  1000 & $5.6 \times 10^{-26}$ & $0$ & $0$ & $0$ & $0$  & $0$ & $0$ & $0$ & $100$ \\
&  1500 & $3.5 \times 10^{-26}$ & $0$ & $0$ & $0$ & $0$  & $0$ & $0$ & $0$ & $100$ \\
&  \multirow{ 2}{*}{5000} &  $\!\!\!\!$\multirow{ 2}{*}{\Bigg\{} $\!\!4.9 \times 10^{-27}$ & $0$ & $69.2$ & $30.8$ & $0$ & $0$ & $0$ & $0$ & $0$ \\
& & $5.0 \times 10^{-27}$ &  $0$ & $0$ & $10.2$ & $0$ & $0$ & $0$ & $89.8$ & $0$  \\
&  10000 & $1.3 \times 10^{-26}$ & $0$ & $0$ &  $77.7$ & $0$  & $0$ & $0$ & $0$ & $22.3$\\
\hline
\end{tabular}
\endgroup
\end{center}
\caption{Branching fractions of each SM final state corresponding to the threshold branching ratio 
combination $B_f$ that decides $\Gamma_{\rm max}$ (tabulated in the third column), are shown for a  
few benchmark values of $m_\chi$ in the range 10 GeV - 10 TeV. In the upper half we present the  
results for Case 1, while in the lower half the branching ratio combinations obtained for Case 2 
are shown. `-' implies that the corresponding final state is not included while calculating  
$\Gamma_{\rm max}$.}
\label{tab:BPthreshold}
\end{table}

As we have discussed in Sec.~\ref{sec:modelindepexconsmethod}, given any 
$m_\chi$ value, the threshold branching ratio combination is the one for which 
the allowed value of $\Gamma$ is the weakest one, i.e., $\Gamma_{\rm max}$ 
(shown in Fig.~\ref{fig:Gamma_max}). It is evident that this threshold 
combination may be different for different values of $m_\chi$. In fact, 
for any given $m_\chi$, the threshold branching ratio combination need 
not be unique. There may in principle
 be more than one combinations of branching 
fractions which give nearly similar values of $\Gamma_{\rm max}$ for a given 
$m_\chi$. In this sense, the maximum allowed values of 
the total DM decay width (shown in Fig.~\ref{fig:Gamma_max}) are quite robust. 
In Tab.~\ref{tab:BPthreshold}, the threshold branching ratio combinations 
as obtained by us are reported for Case 1 
and Case 2. 

In the upper half of Tab.~\ref{tab:BPthreshold}, we have presented the 
threshold branching ratio combinations for a few benchmark values of 
$m_\chi$ in the range 10 GeV - 10 TeV, considering Case 1. It is clear that, 
for several benchmark values of $m_\chi$, one obtains more than one different 
branching ratio combinations which yield very similar values of 
$\Gamma_{\rm max}$ (see third column of Tab.~\ref{tab:BPthreshold}). 
Such cases are grouped under the 
curly braces in Tab.~\ref{tab:BPthreshold}. 
Note that, in this case, for $m_\chi \sim \mathcal{O}(10{\,\rm TeV})$, 
$\gamma\gamma$ final state dominates the threshold branching ratio combination, 
since this is the least constrained 
channel by the Fermi-LAT data in this DM mass range (see Fig.~\ref{fig:LAT_IGRB}). 
One representative threshold branching ratio combination for each of the $m_\chi$ 
values are also shown in Fig.~\ref{fig:Gamma_max_leftBr} (left panel).       

In the lower half of Tab.~\ref{tab:BPthreshold}, the threshold branching ratio 
combinations for a few benchmark $m_\chi$ values are shown for Case 2. Here, too, 
the threshold branching ratio combination is not necessarily unique for any given 
$m_\chi$. As an example, for $m_\chi = 5\,$TeV, we obtain two different branching 
ratio combinations which give similar values for $\Gamma_{\rm max}$ 
(see Tab.~\ref{tab:BPthreshold}). 
In this case, a generic feature is that, for $m_\chi$ up to $\sim 1.5$ TeV, 
threshold branching ratio combinations are dominated by the $\nu\bar{\nu}$ channel, 
while for larger $m_\chi$, contribution of the $\nu\bar{\nu}$ 
final state to the threshold $B_f$ gradually decreases. This is because, 
$\nu\bar{\nu}$ is the least constrained final state for 
$m_\chi \lesssim \mathcal{O}$(TeV), while, for heavier DM particles Fermi-LAT 
and AMS-02 constraints on the $\nu\bar{\nu}$ channel are stronger than those 
obtained for several other decay modes (see Figs.~\ref{fig:LAT_IGRB} 
and~\ref{fig:AMS_positron}). 
A representative set of threshold branching ratio 
combination for this case is shown in Fig.~\ref{fig:Gamma_max_leftBr}; right panel. 
\section{Radio signals at the SKA}
\label{sec:radiosynchSKA}

In this section, we start with a brief discussion of the generation of 
radio signals from the DM decays occurring inside the dSph 
galaxies~\cite{Colafrancesco:2006he,Natarajan:2013dsa,Natarajan:2015hma,Beck:2015rna,McDaniel:2017ppt}.
It is possible to detect such radio signals in the upcoming 
SKA 
radio telescope. \textit{Our goal is to classify the DM parameter space 
allowed by the existing observational data, into three distinct regions: 
detectable for all possible branching ratio combinations, detectable for 
specific combinations of branching fractions and non-detectable,
assuming 100 hours of observation at the SKA.}
Thus in the next part we go ahead to describe the methodology we have 
adopted for such categorization. 

\subsection{Synchrotron radiation from dwarf spheroidal galaxies: a quick review}
\label{sec:radiosynch}	

The SM particles produced from the decay of the DM particles within a 
galactic structure give rise to $e^+e^-$ pairs via their cascade 
decays. These $e^+(e^-)$ act as the source of radio synchrotron signals 
due to their interaction with the magnetic field of the galaxy. 
The ultrafaint dSph galaxies are the most convenient targets to 
look for such DM induced radio signals. Their low star formation rates 
help in reducing the backgrounds stemming from various unknown 
astrophysical processes while 
high mass-to-light ratios 
(pointing towards a greater abundance of dark matter inside them) and 
closeness to our MW galaxy make them the most widely used targets 
for studying the DM induced radio 
signals~\cite{Colafrancesco:2006he,Natarajan:2013dsa,Natarajan:2015hma,Colafrancesco:2015ola,Regis:2017oet,Kar:2018rlm,Kar:2019cqo,Kar:2019hnj,Ghosh:2020ipv}. 
To present our results, \textit{we consider the dSph Segue 1 (Seg 1) which is only 
$\sim 23\,{\rm kpc}$ away from the Sun and has a significantly high 
mass-to-light ratio (a few thousands times the value estimated for the 
Sun) that makes it one of the ``darkest" dSph galaxy 
found till date}~\cite{Geha:2008zr,Simon_2011}. 

The electrons and the positrons produced from the cascade decays of the 
primary DM decay products propagate through the interstellar medium of the 
parent galaxy facing spatial diffusion and also energy loss by means of 
various electromagnetic processes. The equilibrium distribution of such 
electron (positron) $\left(\frac{dn_e}{dE}(E,\vec{r})\right)$ can be obtained 
by solving the following transport 
equation~\cite{Colafrancesco:2005ji,Colafrancesco:2006he,Beck:2015rna,McDaniel:2017ppt,Storm:2016bfw,Natarajan:2013dsa}:
\begin{equation}
D(E)\nabla^2\left(\frac{dn_e}{dE}(E,\vec{r})\right)+\frac{\partial}{\partial E}\left(b(E)\frac{dn_e}{dE}(E,\vec{r}) \right)+ Q_\chi(E,\vec{r}) = 0
\label{eqn:difflosseqn} ,
\end{equation}	
where we assume that the $e^+$($e^-$) distribution reaches a steady state 
which in general holds for systems such as dwarf galaxies where the 
typical timescales required for the alterations of the DM density and the 
propagation parameters are large enough compared to the propagation 
timescale itself~\cite{Cirelli:2010xx}.

The term $b(E)$ in Eqn.~\ref{eqn:difflosseqn} is the energy loss term for 
the $e^+$($e^-$) and is given 
by~\cite{Colafrancesco:2005ji,McDaniel:2017ppt,Beck:2015rna},
\begin{eqnarray}
b(E)&=&b^0_{\rm IC}\left(\frac{E}{\rm GeV}\right)^2+b^0_{\rm Synch}\left(\frac{E}{\rm GeV}\right)^2\left(\frac{B}{\mu {\rm G}}\right)^2 
+ b^0_{\rm Coul}n_e \left[1+\frac{1}{75}\log\left(\frac{E/m_e}{n_e}\right)\right]\nonumber\\
&& + b^0_{\rm Brem}n_e	\left[0.36+\log\left(\frac{E/m_e}{n_e}\right) \right],
\label{eqn:energyloss}
\end{eqnarray}
with $b^0_{\rm IC}\simeq 0.25\times 10^{-16}\,{\rm GeV\,s}^{-1}$,
$b^0_{\rm Synch}\simeq 0.0254\times 10^{-16}\,{\rm GeV\,s}^{-1}$, 
$b^0_{\rm Coul}\simeq 6.13\times 10^{-16}\,{\rm GeV\,s}^{-1}$ and 
$b^0_{\rm Brem}\simeq 1.51\times 10^{-16}\,{\rm GeV\,s}^{-1}$~\cite{Colafrancesco:2005ji,McDaniel:2017ppt,Beck:2015rna}. 
Here $m_e$ and $n_e$ ($\approx 10^{-6}$~\cite{Colafrancesco:2006he}) 
denote the electron mass and the value of the average thermal 
electron density inside a dSph, respectively.
As can be seen from Eqn.~\ref{eqn:energyloss}, for electron (positron) 
energies $E \gtrsim 1$ $\rm GeV$, the terms $b^0_{\rm IC}$ and 
$b^0_{\rm Synch}$ dominate. Note that since the synchrotron energy loss 
term is proportional to the square of the ambient magnetic field strength 
$B$, the $e^+(e^-)$ loose more energy via synchrotron radiation 
while propagating through the region of high $B$-field. 

Due to the lack of gas and dust particles, the magnetic field strengths 
of the ultrafaint dSphs are hardly known and are also difficult to 
constrain using experiments involving polarization measurements. 
However, there are various astrophysical effects which may give rise to 
significant contributions to the magnetic field strengths inside 
the local dSphs. In fact, a number of theoretical arguments have 
been proposed which predict for the values of $B$ at the $\mu{\rm G}$ level. 
For example, it is sometimes argued that the trend of falling of 
the magnetic field strength inside 
the MW galaxy, as one moves from the center towards the outskirts,  
can be linearly extrapolated to nearby dSph galaxies, leading to 
$B \gsim 1\,\mu{\rm G}$ at the location of Seg 1. 
Such an assumption is based on the observations of giant magnetized 
outflows from the central region of the MW galaxy, which point 
towards a $B$-field value larger than 10 $\mu{\rm G}$ at a distance of 
$\sim$ 7 kpc from the Galactic plane~\cite{Carretti:2013sc}. 
In addition, it is also possible that the dwarf galaxies have their own 
magnetic fields. For further details, 
readers are referred to Ref.~\cite{Regis:2014koa}. 

For most part of our analysis we consider the magnetic field 
$B = 1\,\mu{\rm G}$ for 
Seg 1~\cite{Colafrancesco:2006he,Natarajan:2013dsa,Natarajan:2015hma}, 
but at the same time present our results for a smaller (and hence 
more conservative) value of $B = 0.1\,\mu{\rm G}$. 

The diffusion term $D(E)$ in Eqn.~\ref{eqn:difflosseqn}, 
on the other hand, can be parameterized 
as~\cite{Natarajan:2015hma,McDaniel:2017ppt,Natarajan:2013dsa}: 
\begin{equation}
D(E)=D_0\left(E/{\rm GeV}\right)^\gamma,
\end{equation}
where $D_0$ is the diffusion coefficient for the considered dSph and  
$\gamma$ is the corresponding diffusion index. If the value of $D_0$ 
is smaller, the electrons-positrons loose sufficient energy via synchrotron 
radiation before escaping from the diffusion zone of the dSph galaxy and 
hence the resulting radio signals are larger~\cite{Kar:2019cqo}.

Similar to the case of the magnetic field strengths, 
the diffusion parameters, i.e., $D_0$ and $\gamma$, too, are barely 
constrained for ultrafaint dSphs because of their low luminosity. 
However, analogous to galactic 
clusters~\cite{Rebusco:2005mu,Rebusco:2006fm} one may expect 
inside a dSph $D_0 \propto V \times L$, where $V$ and $L$ represent 
the velocity of the
stochastic gas motions and the associated characteristic length scale, 
respectively. Taking the above-mentioned parameterization, 
Ref.~\cite{Jeltema:2008ax} has shown that the scaling of virial velocity 
dispersions in ultrafaint dSphs with respect to the value corresponding 
to the MW galaxy can be used to infer that, $D_0$ for a dSph 
is either of the same order of magnitude as its value for the Milky Way or 
smaller by an order of magnitude. 

In the context of the dSph galaxy considered in this work, we choose  
$\gamma = 0.7$~\cite{Jeltema:2008ax,Natarajan:2013dsa,Natarajan:2015hma,McDaniel:2017ppt} 
(in analogy with its value used for the MW galaxy~\cite{Buch:2015iya})
and use mostly the value $D_0 = 3\times 10^{28}\,{\rm cm}^2\,{\rm s}^{-1}$ 
which is almost of the same order or one order of magnitude higher than  
the value often predicted for the Milky Way~\cite{Buch:2015iya}. 
In parallel, we have also shown our results for a more conservative 
choice of $D_0$, i.e., $D_0 = 3\times 10^{29}\,{\rm cm}^2\,{\rm s}^{-1}$. 
All the combinations of the diffusion coefficient $D_0$ and the magnetic  
field $B$ assumed for the Seg 1 dSph in this work are still allowed by 
the existing radio observations~\cite{Natarajan:2013dsa,Natarajan:2015hma}. 
The diffusion zone of the Seg 1 dSph has been considered to be spherically 
symmetric with a radius $r_h = 1.6\,{\rm kpc}$~\cite{Natarajan:2015hma}.

The $e^+$($e^-$) source function $Q_\chi(E,\vec{r})$ (in Eqn.~\ref{eqn:difflosseqn}), 
which results from the decay of DM particles inside the target dSph halo, 
is given by~\cite{Regis:2017oet, Ghosh:2020ipv, Dutta:2020lqc}, 
\begin{equation}
Q_\chi(E,\vec{r}) = \frac{\Gamma}{m_\chi}\sum_f\,B_f\,\frac{dN^{e^+}_f}{dE}(E)\rho_d(\vec{r}) .
\label{eqn:sourceterm}
\end{equation}
As mentioned earlier, $dN^{e^+}_f/dE$ is the energy 
spectrum of the $e^+$ (and equivalently $e^-$) produced per DM decay 
for the SM final state (channel) $f$. 
For the dSph Seg 1, $\rho_d(\vec{r})$ is assumed to follow 
the Einasto profile~\cite{Einasto:1965czb,McDaniel:2017ppt}:
\begin{equation}
\rho_d(\vec{r}) = \rho_s\,\exp\,\left\{-\frac{2}{\alpha}\left[\left(\frac{r}{r_s}\right)^\alpha-1\right]\right\},
\label{eqn:Enastoprof}
\end{equation}
where the associated parameters $\rho_s$, $r_s$ and $\alpha$ are set at 
the values used in~\cite{Natarajan:2015hma,McDaniel:2017ppt,MAGIC:2011nta}. 

Now that all terms in Eqn.~\ref{eqn:difflosseqn} are given, 
one can solve this equation using the Green's function method outlined 
in~\cite{Colafrancesco:2005ji,Kar:2019cqo,Colafrancesco:2006he}. 
The solution of Eqn.~\ref{eqn:difflosseqn}, i.e., 
$\frac{dn_e}{dE}(E, \vec{r})$, is then used to obtain the radio synchrotron 
flux $S_\nu$ (resulting from electrons and positrons) as a function of the 
radio frequency ($\nu$):
\begin{equation}
S_\nu(\nu) =  \frac{1}{4\pi}\int_{\Delta\Omega} d\Omega \int_{l.o.s} ds\left(2 \int_{m_e}^{m_\chi/2} dE\frac{dn_e}{dE}(E, r(s,\Omega))P_{\rm Synch}(\nu,E,B)  \right),
\label{eqn:synchflux}
\end{equation}	  
where $P_{\rm Synch}(\nu,E,B)$ is the synchrotron power spectrum 
corresponding to the $e^-$($e^+$) with an energy $E$ in a 
magnetic field $B$ (for the analytic form of the synchrotron power 
spectrum see~\cite{Colafrancesco:2005ji,Colafrancesco:2006he,Beck:2015rna}), 
$s$ is the line-of-sight ($l.o.s$) coordinate and $\Delta\Omega$ denotes 
the size of the emission region of the considered dSph. 
In the forthcoming sections, the observational prospects of 
DM decay induced radio signals from the Seg 1 dSph will be 
studied in the context of the SKA 
radio telescope~\cite{SKA}.
	
\subsection{Projection for the SKA}
\label{sec:SKAmethodology}

Apart from several other fields of cosmology and astrophysics, in 
understanding the properties of dark matter, too, the upcoming 
SKA is expected to play a pivotal role~\cite{Colafrancesco:2015ola,Braun:2015B3,Power:2015G3}. 
SKA operates over a large frequency range, i.e., 50 MHz - 50 GHz, 
which helps in constraining the DM parameter space for a wide range of DM masses.	
The inter-continental baseline lengths of the SKA allow one to efficiently 
resolve the astrophysical foregrounds and thereby enhances its capability. 
In addition, a higher surface brightness sensitivity is possible 
to achieve because of its large effective area compared to any 
other existing radio telescope~\cite{SKA}. 

This work being focused on studying the DM induced diffuse radio 
synchrotron signals from the dSph galaxies, the SKA sensitivity 
corresponding to the surface brightness is calculated 
and compared with the predicted radio signal. In order to 
estimate the approximate values of the sensitivity we have 
utilized the presently accepted baseline design 
given in the documents provided in the SKA website~\cite{SKA}. 
In order to do this estimation we have adopted the methodology 
provided in~\cite{Kar:2019cqo} and the rms noise in observations 
has been evaluated utilizing the formula given in~\cite{Ghara:2015mab}. 
Assuming 100 hours of observation time this estimate suggests that 
the SKA surface brightness sensitivity in the frequency range 
50 MHz - 50 GHz is $10^{-6}$ - $10^{-7}$ Jy with a bandwidth of 
300 MHz~\cite{SKA,Colafrancesco:2015ola,Kar:2018rlm}. Such values 
of the surface brightness sensitivity facilitate the observations 
of very low energy radio signals coming from the ultrafaint dSphs.

The potential of the SKA telescope in detecting the radio signals
originating from DM annihilations/decays is quite 
promising~\cite{Colafrancesco:2015ola,Kar:2018rlm,Kar:2019cqo,Dutta:2020lqc,Chen:2021rea,Ghosh:2020ipv}. 
As mentioned previously, here the detection 
prospects of the radio signals (in the frequency range $\sim$50 MHz to 
$\sim$50 GHz) produced from DM decays occurring inside the Seg 1 dSph 
are investigated assuming 100 hours of observation time at the SKA. 
\textit{Following Ref.~\cite{Kar:2019cqo}, we define the SKA threshold 
sensitivity, which determines the minimum radio flux required for 
detection at the SKA, to be three times above the noise level or the 
sensitivity level~\cite{SKA}, so that the possibility of any spurious 
noise feature being misinterpreted as a potential DM decay signal is 
reduced and the detection of the DM decay induced radio signal, 
if any, is statistically significant.} 

The radio signal predicted for the SKA has been estimated assuming 
that the SKA field of view is larger than the size of the considered 
dSph. As a result, the expected signal at the SKA receives contributions 
from the entire dSph. Such an assumption need not be true for radio 
telescopes like the Murchison Widefield Array (MWA) which is one of 
the precursors of the SKA. In these kind of telescopes the effect of 
the primary beam size is also needed to be taken into account while 
calculating the predicted signal~\cite{Kar:2019hnj}.

\subsubsection{Detectability at the SKA: methodology}


Following our discussions in Sec.~\ref{sec:modelindepexcons}, 
we know that, for any $m_\chi$ value, the upper limit of the allowed 
part of the DM parameter space is set by $\Gamma_{\rm max}$ 
(shown in the left and the right panels of Fig.~\ref{fig:Gamma_max}). 
Hence, for any given $m_\chi$ (in the range 10 GeV - 10 TeV) and 
$\Gamma$ ($\leq \Gamma_{\rm max}$), we scan over all 
possible branching ratio combinations 
of the kinematically allowed two-body SM final states, i.e., 
$\chi \rightarrow$ ${\rm SM}_{1}$ $\overbar{\rm 
SM}_2$ (${\rm SM}_{1}$ $\overbar{\rm SM}_2$ are defined in 
Sec.~\ref{sec:DMdecaytheory}). In this scan, the branching fraction 
for each individual DM decay mode is varied between $0\%$ to $100\%$ 
with an incremental change of $2\%$ in each consecutive step (similar 
to what we have done in Sec.~\ref{sec:modelindepexcons}). For the 
given $m_\chi$ and $\Gamma$ each such combination represents a 
different DM model. 
If the radio flux ($S_\nu$) for any particular branching ratio combination 
is above the SKA sensitivity level (corresponding to a given observation time) 
in at least one frequency bin, then DM decay to that particular final 
state is detectable at the SKA for the considered time of observation. 


For a given ($m_\chi$, $\Gamma$) set lying in the allowed region 
of the DM parameter space, we first obtain {\it the maximum 
radio flux ($S^{\rm max}_\nu$)} and {\it the minimum radio flux 
($S^{\rm min}_\nu$)} in every frequency bin in the range 30 MHz 
to 100 GHz 
by scanning over all possible branching 
ratio combinations of the DM decay modes and hence 
the combinations of branching fractions that give rise to $S^{\rm max}_\nu$ 
(or $S^{\rm min}_\nu$) may be different in each frequency bin. 
Note that in any frequency bin the branching ratio combination that gives 
$S^{\rm max}_\nu$ (or $S^{\rm min}_\nu$) is independent of the chosen value 
of $\Gamma$ and though, $S^{\rm max}_\nu$ (or $S^{\rm min}_\nu$) itself 
changes with $\Gamma$, the corresponding branching ratio combination 
remains the same. 

In each bin, the radio fluxes ($S_\nu$) associated with all branching ratio 
combinations always lie between $S^{\rm max}_{\nu}$ and $S^{\rm min}_{\nu}$. 
Therefore, depending on these maximum and minimum radio fluxes, one can 
determine whether the chosen ($m_\chi$, $\Gamma$) point is \textit{detectable} or 
\textit{non-detectable} at the SKA. 
The DM models which are detectable at the SKA can be further categorized 
into two classes. The first one represents the scenarios which are detectable 
for all possible branching ratio combinations while the second one is composed of the DM models which are detectable only for certain specific branching ratio combinations of the DM decay modes.


\begin{figure*}[ht!]
\centering
\includegraphics[width=7.6cm,height=7.4cm]{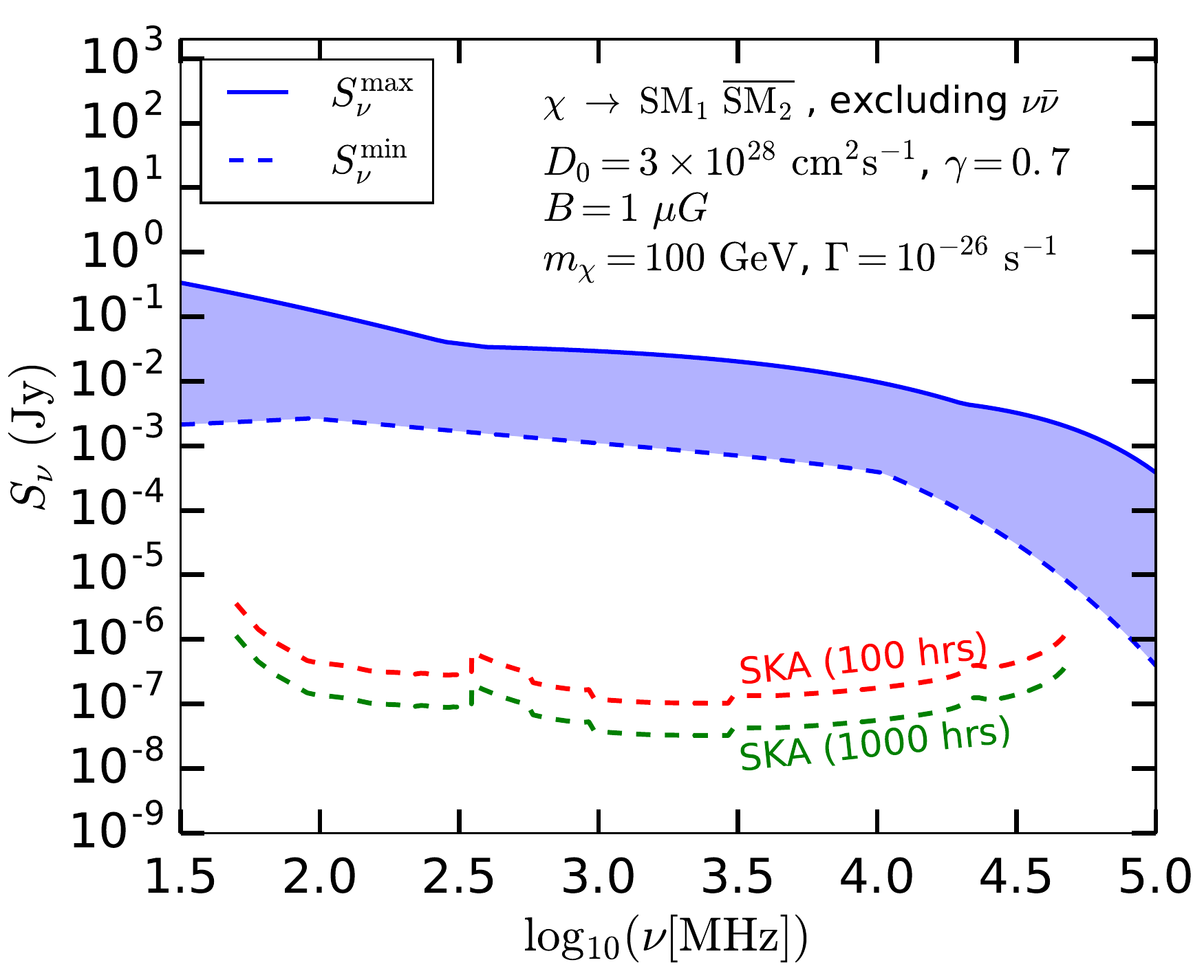}\hspace{0mm}
\includegraphics[width=7.6cm,height=7.4cm]{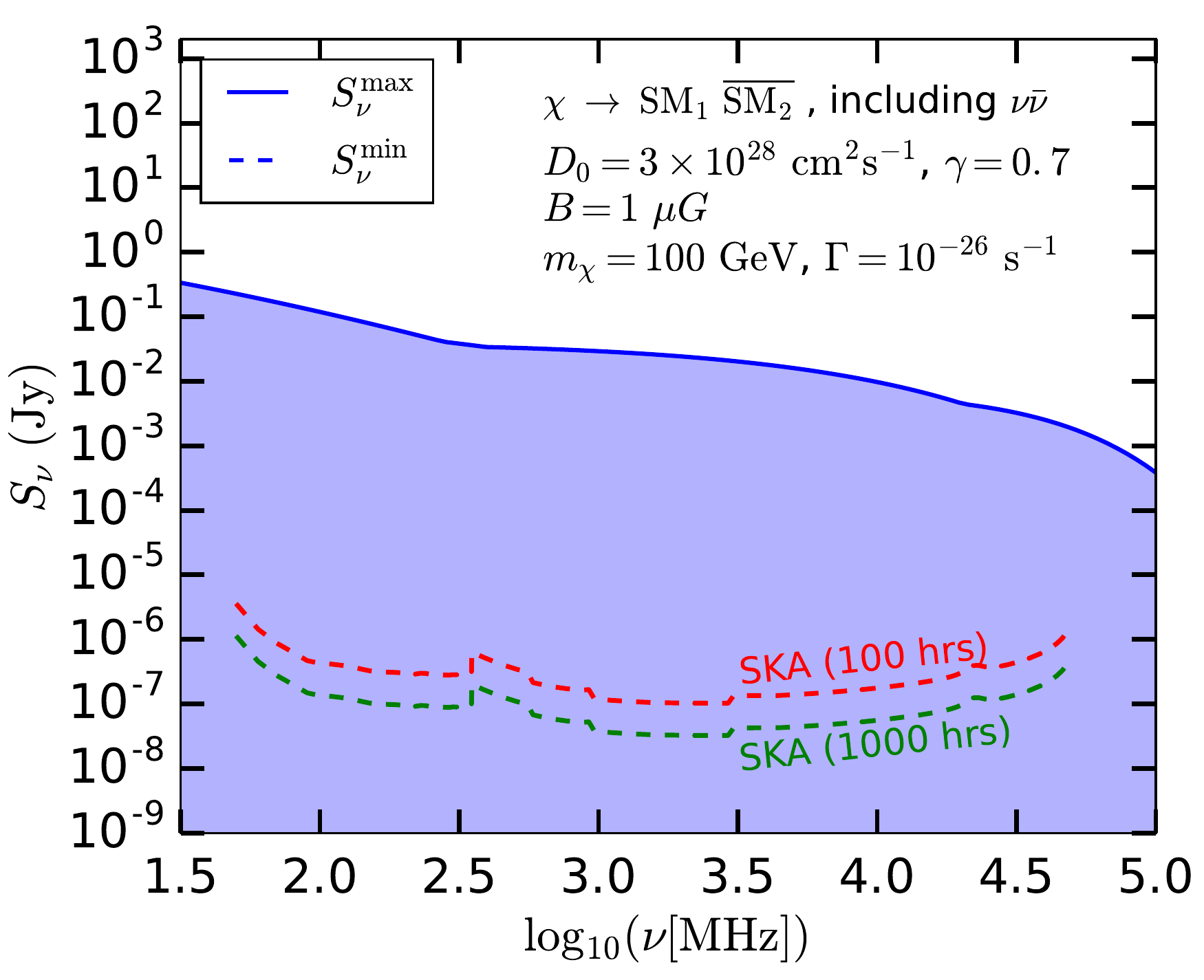}
\includegraphics[width=7.6cm,height=7.4cm]{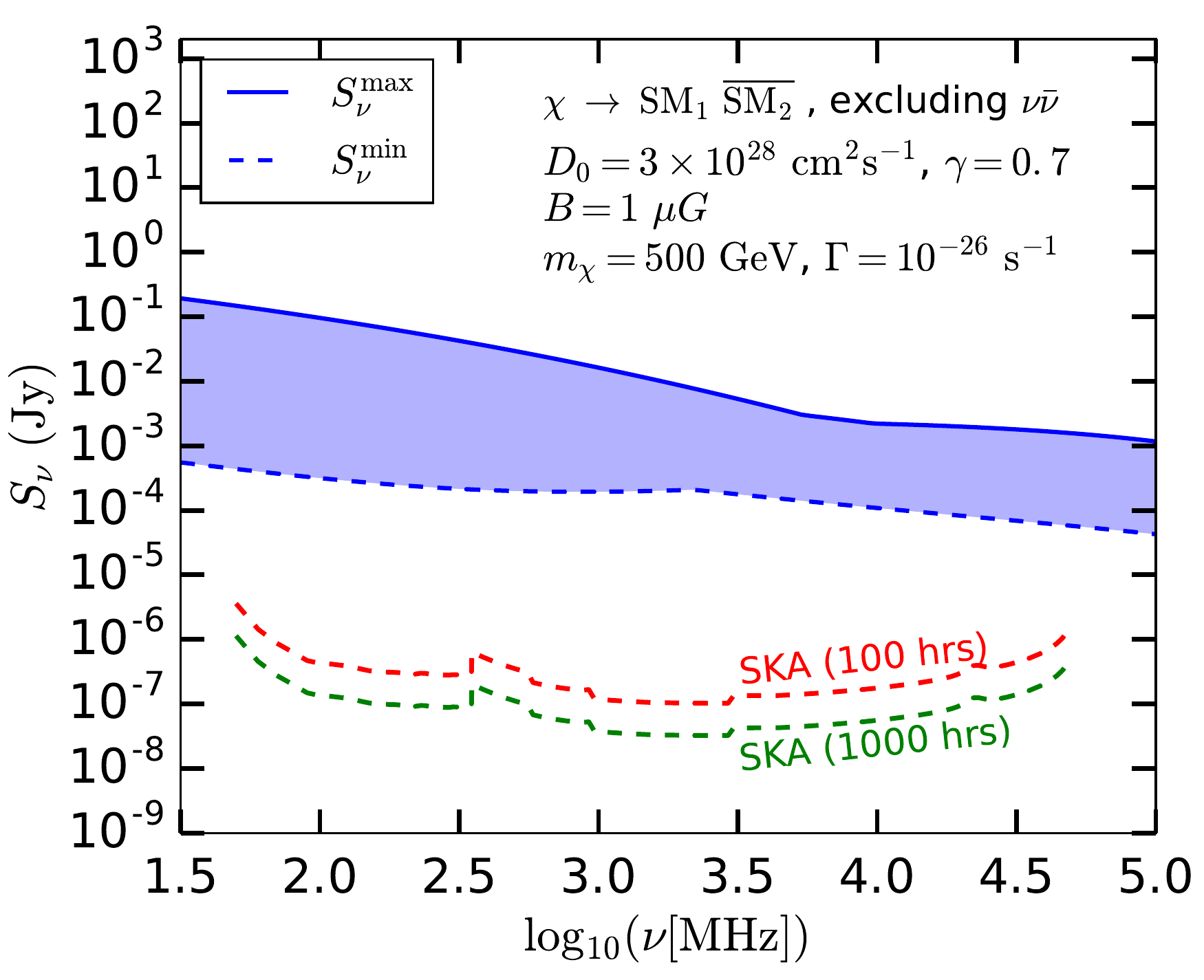}\hspace{0mm}
\includegraphics[width=7.6cm,height=7.4cm]{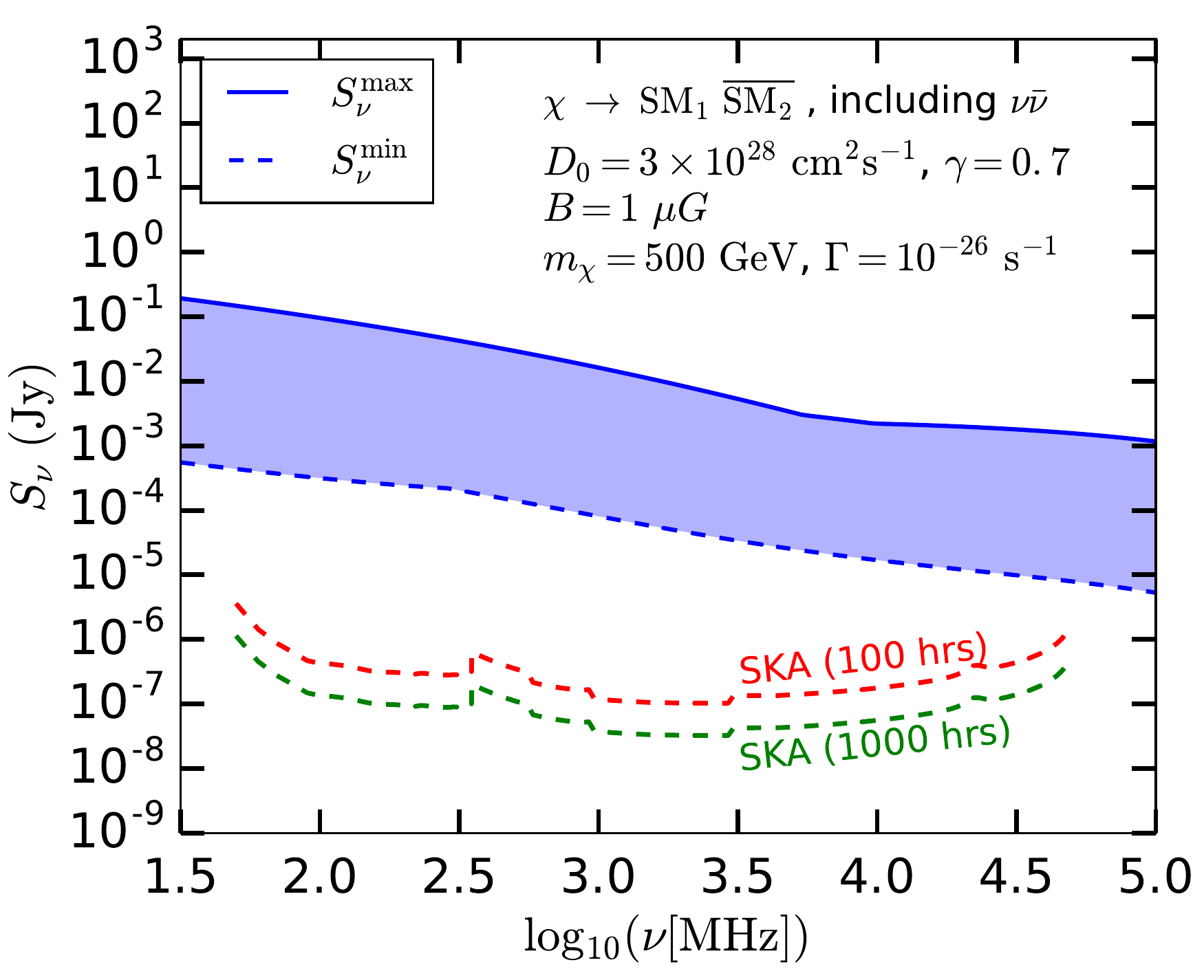}
\caption{Considering the Seg 1 dSph as the target, the maximum ($S^{\rm max}_\nu$) and the minimum ($S^{\rm min}_\nu$) radio synchrotron flux distributions 
(the blue solid and the blue dashed lines, respectively) are shown as functions 
of the radio frequency $\nu$ in the range $30\,$MHz to 
$100\,$GHz, for $m_\chi = 100\,{\rm GeV}$ (top panel) and 
$m_\chi = 500\,{\rm GeV}$ (bottom panel). 
In each panel, the radio fluxes ($S_\nu$) for all branching ratio 
combinations fall within the blue shaded region. 
The results excluding $\chi \rightarrow \nu\bar{\nu}$ are 
shown in the left panel while that including the $\nu\bar{\nu}$ decay 
mode with the corresponding branching ratio varied between 
0$\%$ and 100$\%$ are presented in the right panel. In all cases, an 
illustrative value of the DM decay width $\Gamma = 10^{-26}\,{\rm s}^{-1}$ 
have been used. 
We have taken $D_0 = 3\times 10^{28}\,{\rm cm}^2\,{\rm s}^{-1}$, $\gamma=0.7$ 
and $B=1\mu{\rm G}$ in this figure. The SKA sensitivity curves for 100 
hours (red dashed lines) and 1000 hours (green dashed lines) of observation times are also presented. See the text for details.}
\label{fig:Snu_vs_nu}
\end{figure*}

Considering Seg 1 dSph as the source, in Fig.~\ref{fig:Snu_vs_nu}, 
we show $S^{\rm max}_\nu$ (blue solid line) and 
$S^{\rm min}_\nu$ (blue dashed line) as functions of the 
radio frequency $\nu$ in the range $\sim$30 MHz to $\sim$100 GHz 
for two representative values of the DM mass, i.e., 
$m_\chi = 100\,{\rm GeV}$ (top panel) and $m_\chi = 500\,{\rm GeV}$
(bottom panel). $S_\nu$ for all branching ratio combinations lie 
within the blue shaded region. Here we have assumed 
$D_0 = 3\times 10^{28}\,{\rm cm}^2\,{\rm s}^{-1}$, $\gamma=0.7$ 
and $B=1\mu{\rm G}$.
In the left panel we have shown the fluxes for Case 1 
(mentioned in Sec.~\ref{sec:modelindepexcons}) while in the right panel 
Case 2 (as defined in Sec.~\ref{sec:modelindepexcons}) is considered. 
Both of the chosen $m_\chi$ values, i.e., $100\,{\rm GeV}$ and 
$500\,{\rm GeV}$, are well within the DM mass range we are 
focusing on, i.e., 10 GeV - 10 TeV and have additional 
motivations when one takes $\chi \rightarrow \nu\bar{\nu}$ 
channel into account, which will be evident from the following discussions. 
In each of the cases considered, we have chosen 
$\Gamma = 10^{-26}\,{\rm s}^{-1}$ for illustration. 
In addition, the SKA sensitivities for 100 hours (red dashed line) 
and 1000 hours (green dashed line) of observation times are also presented. 
Note that the SKA noise goes as inverse of the square 
root of the observation time and thus the sensitivity corresponding to 
1000 hours of observation time is enhanced nearly by a factor of three 
compared to that obtained for 100 hours of observation time~\cite{Kar:2019cqo}. 
This is the reason why the green dashed line lies a factor of 
three below the red dashed line in Fig.~\ref{fig:Snu_vs_nu}. 


In the top left panel of Fig.~\ref{fig:Snu_vs_nu}, we have shown the 
radio flux distributions for $m_\chi = 100\,{\rm GeV}$ and 
$\Gamma = 10^{-26}\,{\rm s}^{-1}$ assuming Case 1, i.e., DM decay to 
$\nu\bar{\nu}$ final state is not included. 
For a radio frequency of $\sim 1\,{\rm GHz}$, the maximum flux is 
obtained when $\chi$ dominantly decays into $\mu^+\mu^-$ while DM decays to 
$Z\gamma$, $\gamma\gamma$ or any of their combinations give rise to 
$S^{\rm min}_\nu$. As mentioned earlier, the branching ratio combinations that 
correspond to $S^{\rm max}_\nu$ (or $S^{\rm min}_\nu$) are not the same in 
every frequency bin. As an example, for $\nu \sim 100\,{\rm MHz}$, 
$q\bar{q}$, $b\bar{b}$, $gg$ final states or any of their combinations give 
$S^{\rm max}_\nu$, and for DM decays into $\gamma\gamma$, $e^+e^-$ final states 
or any of their combinations, one obtains the minimum radio flux. 
For our chosen value of $\Gamma$ both $S^{\rm max}_\nu$ and 
$S^{\rm min}_\nu$ are above the SKA sensitivity curves in all frequency 
bins and thus the radio fluxes for all possible branching ratio 
combinations are also above the SKA sensitivity levels everywhere in the 
considered frequency range. Therefore, \textit{this combination of 
$(m_\chi,\Gamma)$ is detectable at the SKA (in both 100 hours and 1000 hours 
of observations) for all possible branching ratio combinations.} 
Although in this case the radio fluxes are above the SKA sensitivity levels 
in all frequency bins, it may also happen that $S_\nu$ go above the SKA 
sensitivity level over a certain frequency range, in which case the corresponding 
radio signals are detectable in that frequency range only. For example, if 
$\Gamma$ is reduced roughly by three orders, radio signals for all branching ratio 
combinations are above the SKA sensitivity levels only in the frequency range 
$60\,{\rm MHz} - 13\,{\rm GHz}$, and hence this $(m_\chi,\Gamma)$ point, too, is 
detectable at the SKA for all possible combinations of the DM decay modes. 


Now if one scales down $\Gamma$ at least by four orders of magnitude 
both $S^{\rm max}_\nu$ and $S^{\rm min}_\nu$ decrease by the 
same amount and consequently, $S^{\rm max}_\nu$ is still above the SKA 
sensitivity levels everywhere but $S^{\rm min}_\nu$ falls below the SKA 
sensitivity curve 
obtained for 100 hours of observation time (red dashed line), in all frequency bins. 
As a result, the red dashed curve lies within the blue shaded region, which implies 
that, $S_\nu$ for certain branching ratio combinations are still above the SKA 
sensitivity level (for 100 hours of observation) while that for other branching ratio 
combinations are below the sensitivity level. \textit{Therefore, such a value of 
$(m_\chi,\Gamma)$ is detectable in the 100 hours of observation at} 
\textit{the SKA depending on the branching ratio combination of the DM decay mode.} 
If $\Gamma$ is further decreased by a factor of 50, both $S^{\rm max}_\nu$ 
and $S^{\rm min}_\nu$ will go below the sensitivity level corresponding 
to 100 hours of the observation time, 
in all frequency bins. As a result, there exist no branching ratio combination 
for which the resulting $S_\nu$ is above the SKA sensitivity level. 
\textit{This $(m_\chi,\Gamma)$ combination is never 
detectable in 100 hours of observation at the SKA.} 

On the other hand, each time, had $\Gamma$ been reduced by an extra 
factor of three, first $S^{\rm min}_\nu$ and then in the next time 
$S^{\rm max}_\nu$ would have gone below the green dashed line. 
Therefore, the resulting $(m_\chi,\Gamma)$ points are 
branching ratio dependently detectable and non-detectable, respectively, 
in the 1000 hours of observation at the SKA. The origin of this factor 
of three can be understood from the relative difference between the 
sensitivity levels obtained for 100 hours and 1000 hours of observation 
times (the red and the green dashed curves in Fig.~\ref{fig:Snu_vs_nu}).

In the bottom left panel of Fig.~\ref{fig:Snu_vs_nu} 
(here, too, $\chi \rightarrow \nu\bar{\nu}$ is not included) results are 
shown for the parameter point $m_\chi = 500\,{\rm GeV}$, 
$\Gamma = 10^{-26}\,{\rm s}^{-1}$. 
In this case, for $\nu \sim 1\,{\rm GHz}$, the branching ratio 
combination that gives $S^{\rm max}_\nu$ is determined by 
$q\bar{q}$, $b\bar{b}$, $gg$, $t\bar{t}$ channels, 
while $S^{\rm min}_\nu$ is obtained when $\chi$ decays 
dominantly into $e^+e^-$ pairs. 
Both $S^{\rm max}_\nu$ and $S^{\rm min}_\nu$ are above the SKA sensitivity 
levels throughout the considered frequency range and thus \textit{this 
$(m_\chi,\Gamma)$ point is detectable at the SKA for all possible branching 
ratio combinations}. 
However, in this case if $\Gamma$ is decreased by nearly 
three orders the resulting ($m_\chi, \Gamma$) point becomes 
detectable at the SKA (100 hours) for certain specific branching ratio 
combinations of the DM decay modes and becomes non-detectable when 
$\Gamma$ is further decreased by a factor of 100. 


As mentioned earlier, the radio fluxes for Case 2, i.e., including 
the $\nu\bar{\nu}$ decay mode (with arbitrary branching fraction in 
the range $0\% - 100\%$) are presented in the right panel of 
Fig.~\ref{fig:Snu_vs_nu}. 
Note that for $m_\chi = 100\,{\rm GeV}$ and 
$\Gamma = 10^{-26}\,{\rm s}^{-1}$ (Fig.~\ref{fig:Snu_vs_nu}; top right panel), 
the maximum radio flux distribution is 
similar to that obtained excluding 
$\nu\bar{\nu}$ final state (see Fig.~\ref{fig:Snu_vs_nu}; top left panel). 
In fact, as in Case 1, here, too, 
for a radio frequency of $\sim 1\,{\rm GHz}$, $S^{\rm max}_\nu$ corresponds to 
DM decays to $\mu^+\mu^-$. 
For this ($m_\chi$, $\Gamma$) point $S^{\rm max}_\nu$ always lies above 
the SKA sensitivity curves. 
However, the minimum radio flux 
is now much below the SKA sensitivity levels. This is because, 
throughout the frequency range, $S^{\rm min}_\nu$ is obtained 
when $\nu\bar{\nu}$ is the dominant decay mode. 

Following our earlier discussions, we know that, for DM mass less 
than \textit{a few hundreds of GeV}, the fluxes of $e^+(e^-)$ produced 
from the $\nu\bar{\nu}$ final state are significantly suppressed 
(see the discussions of Fig.~\ref{fig:dNdE_nunubar}). Thus the resulting radio 
signals 
are also negligibly small for all reasonable values of $\Gamma$. 
As a result, for the considered ($m_\chi,\Gamma$) 
point $S^{\rm min}_\nu$ lies below the SKA sensitivity levels everywhere. 
Clearly, \textit{this ($m_\chi,\Gamma$) point is detectable at the SKA 
only for certain specific branching ratio combinations.} 
Reduction in $\Gamma$ by nearly five orders of magnitude 
causes $S^{\rm max}_\nu$ to go below the red dashed curve 
and hence the resulting ($m_\chi,\Gamma$) point becomes non-detectable 
in the 100 hours of observation at the SKA. Note that if $\Gamma$ is 
reduced by an extra factor of three, the resulting ($m_\chi,\Gamma$) 
point is non-detectable even in the 1000 hours of observation at the SKA. 

On the other hand, for $m_\chi = 500\,{\rm GeV}$ and 
$\Gamma = 10^{-26}\,{\rm s}^{-1}$ (bottom right panel of 
Fig.~\ref{fig:Snu_vs_nu}) both $S^{\rm max}_\nu$ and $S^{\rm min}_\nu$ 
are above the SKA sensitivities in all frequency bins. 
Similar to Case 1 (shown in Fig.~\ref{fig:Snu_vs_nu}; bottom left panel), 
$S^{\rm max}_\nu$ in the $\nu \sim 1\,{\rm GHz}$ bin is obtained for 
DM decays to $q\bar{q}$, $b\bar{b}$, $gg$, $t\bar{t}$ final states or any 
of their combinations. On the other hand, in the same frequency bin, though, 
$S^{\rm min}_\nu$ is obtained when DM decays dominantly into 
$\nu\bar{\nu}$, in this case, 
the DM being heavier, the final state $\nu\bar{\nu}$ pairs are 
comparatively more energetic. Therefore, the resulting fluxes of 
$e^+(e^-)$ coming from the $\nu\bar{\nu}$ final state are comparable to 
that obtained for other SM final states (see Fig.~\ref{fig:dNdE_nunubar}) 
which leads to an enhancement of the associated radio signals. 
Therefore, the chosen ($m_\chi,\Gamma$) point is 
detectable at the SKA for all possible branching ratio combinations. 
Approximately three (five) orders decrease in $\Gamma$ suppresses the 
radio fluxes in a way such that the resulting ($m_\chi,\Gamma$) points 
are detectable for certain branching ratio combinations 
(non-detectable) in the 100 hours of observation at the SKA.


Note that the value $\Gamma = 10^{-26}\,{\rm s}^{-1}$ chosen here, 
is allowed by the existing astrophysical and cosmological data when 
$\chi \rightarrow \nu\bar{\nu}$ is included in the analysis 
(see Fig.~\ref{fig:Gamma_max}; right panel). 
For the analysis performed excluding $\nu\bar{\nu}$, 
$\Gamma = 10^{-26}\,{\rm s}^{-1}$ represents an illustrative value. 
Therefore, in this case, while 
determining the SKA detectability one needs to choose the value of $\Gamma$ 
appropriately so that the resulting $\Gamma$ is allowed by the existing 
observations (see Fig.~\ref{fig:Gamma_max}; left panel).

\subsubsection{SKA detectability criteria: a summary}

Therefore, for any given ($m_\chi,\Gamma$) point, scanning over all 
possible branching ratio combinations of the DM decay modes, one 
obtains $S^{\rm max}_{\nu}$ and $S^{\rm min}_{\nu}$ in every frequency bin. 
Depending on the distributions of $S^{\rm max}_{\nu}$ and 
$S^{\rm min}_{\nu}$, the detectability (for a given observation 
time at the SKA) of any ($m_\chi,\Gamma$) point lying in the allowed 
part of the $m_\chi-\Gamma$ plane is decided by the following 
considerations:     
\begin{itemize}
\item If both $S^{\rm max}_{\nu}$ and $S^{\rm min}_{\nu}$ go above 
the SKA sensitivity level, whose threshold is set at \textit{three times 
the estimated noise level} (corresponding to the chosen value of 
the observation time), \textit{in at least one frequency bin} 
(in the range $\sim$50 MHz to $\sim$50 GHz) then $S_\nu$ 
for all branching ratio combinations are also above the SKA sensitivity 
level in that frequency bin. Therefore, this ($m_\chi, \Gamma$) point 
is detectable at the SKA for all possible branching ratio combinations.

\item If only $S^{\rm max}_{\nu}$ lies above the SKA sensitivity level 
in \textit{at least one frequency bin} but $S^{\rm min}_{\nu}$ falls below 
the sensitivity of the SKA in \textit{all frequency bins} then the 
corresponding ($m_\chi, \Gamma$) point is \textit{detectable only for 
certain specific branching ratio combinations}, i.e., the combinations for which 
$S_\nu$ is above the SKA sensitivity curve in at least one bin. For other 
branching ratio combinations the associated radio fluxes are always 
smaller than the threshold values of the required radio flux. 
		
\item Lastly, if both $S^{\rm max}_{\nu}$ and $S^{\rm min}_{\nu}$ 
lie below the sensitivity level of the SKA in \textit{all frequency bins} 
then there exist no branching ratio combination for which the resulting 
radio signal is observable at the SKA and hence that particular 
($m_\chi, \Gamma$) combination is \textit{never detectable} at the SKA 
for the corresponding length of the observation time. 
\end{itemize}  

Applying the set of criteria discussed above, we now proceed to present 
the classification of 
the allowed region of the DM parameter space based on the detectability 
at the SKA, for 100 hours of observation of the Seg 1 dSph.

\section{Projected sensitivity at the SKA}
\label{sec:SKAresults}


In this section, we determine the detectability of the DM parameter space 
in the 100 hours of observation of the Seg 1 dSph at the SKA, based on the 
criteria outlined in the previous section. We shall present our 
results for the four different cases considered in Sec.~\ref{sec:modelindepexcons}. 
\textit{In each case, the allowed part of the $m_\chi-\Gamma$ plane 
can be classified into three distinct regions: green, yellow and red.} 
\textit{The green region is detectable `DM model independently', 
i.e., detectable for all possible branching ratio combinations. The parameter 
points in the yellow region are detectable only for certain specific combinations 
of branching fractions 
indicating that such points are detectable `depending on DM model'. 
The red region, on the other hand, is `not detectable' 
for any combinations of the DM decay modes.}

\begin{figure*}[htb!]
\centering
\includegraphics[width=7.6cm,height=7.4cm]{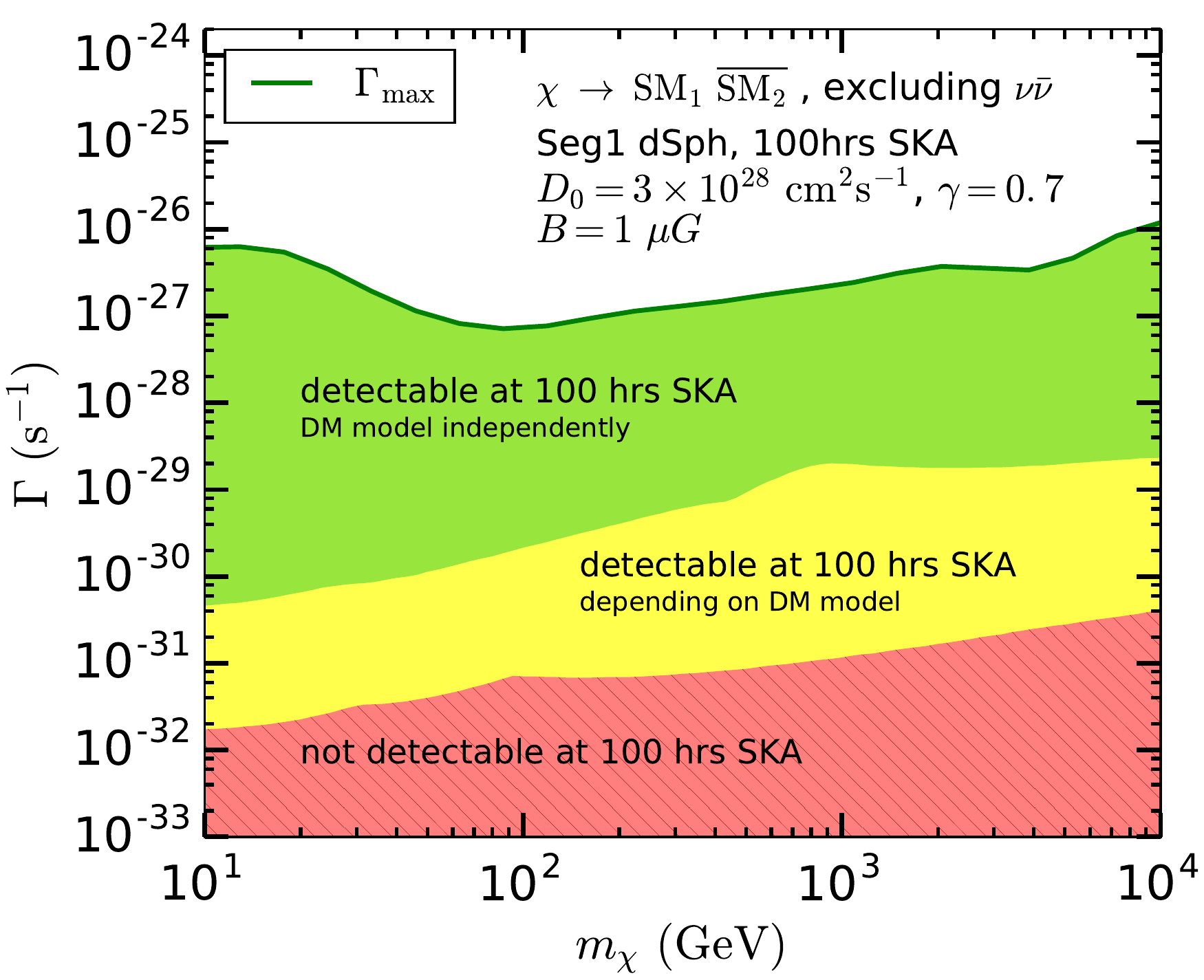} \\
\includegraphics[width=7.6cm,height=7.4cm]{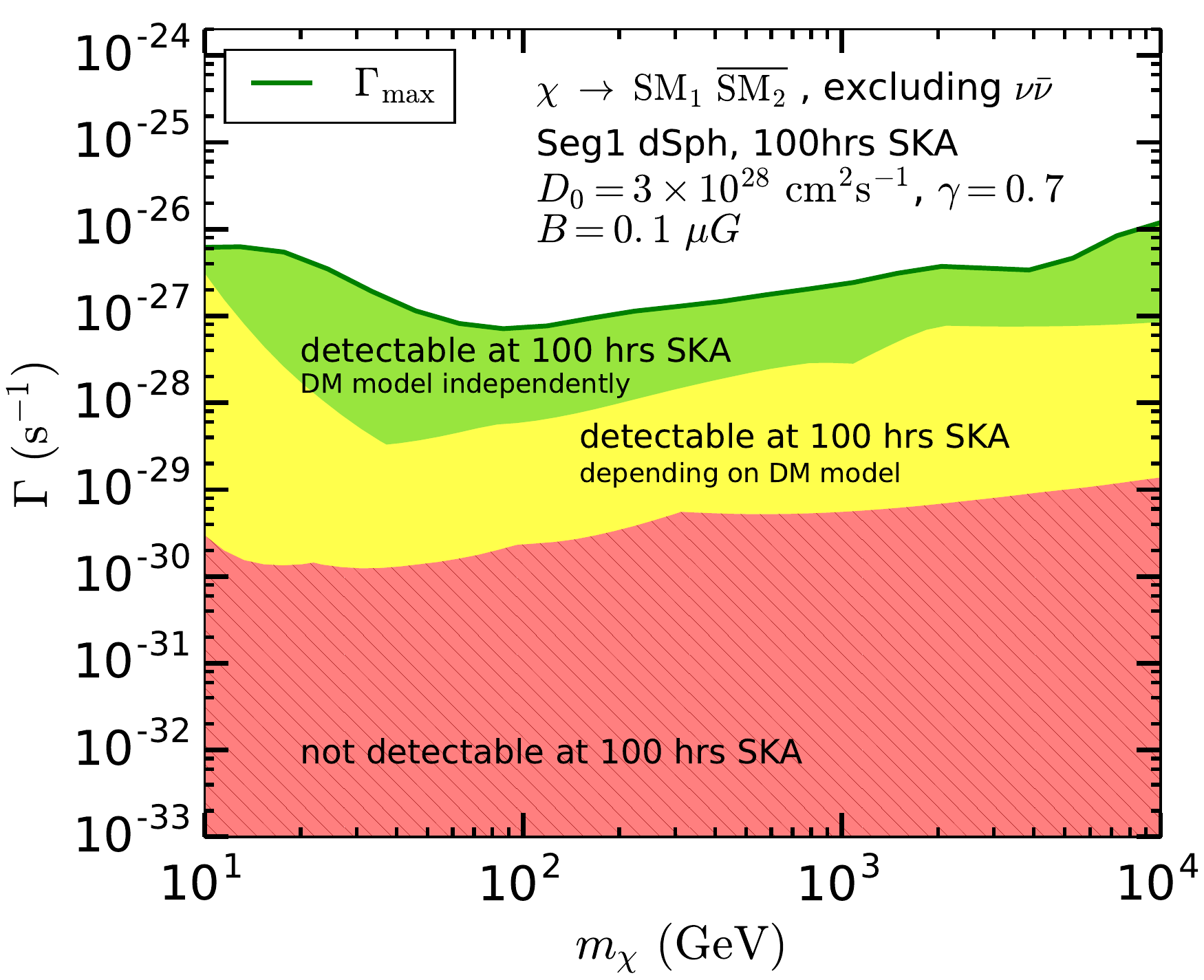} \hspace{0mm}
\includegraphics[width=7.6cm,height=7.4cm]{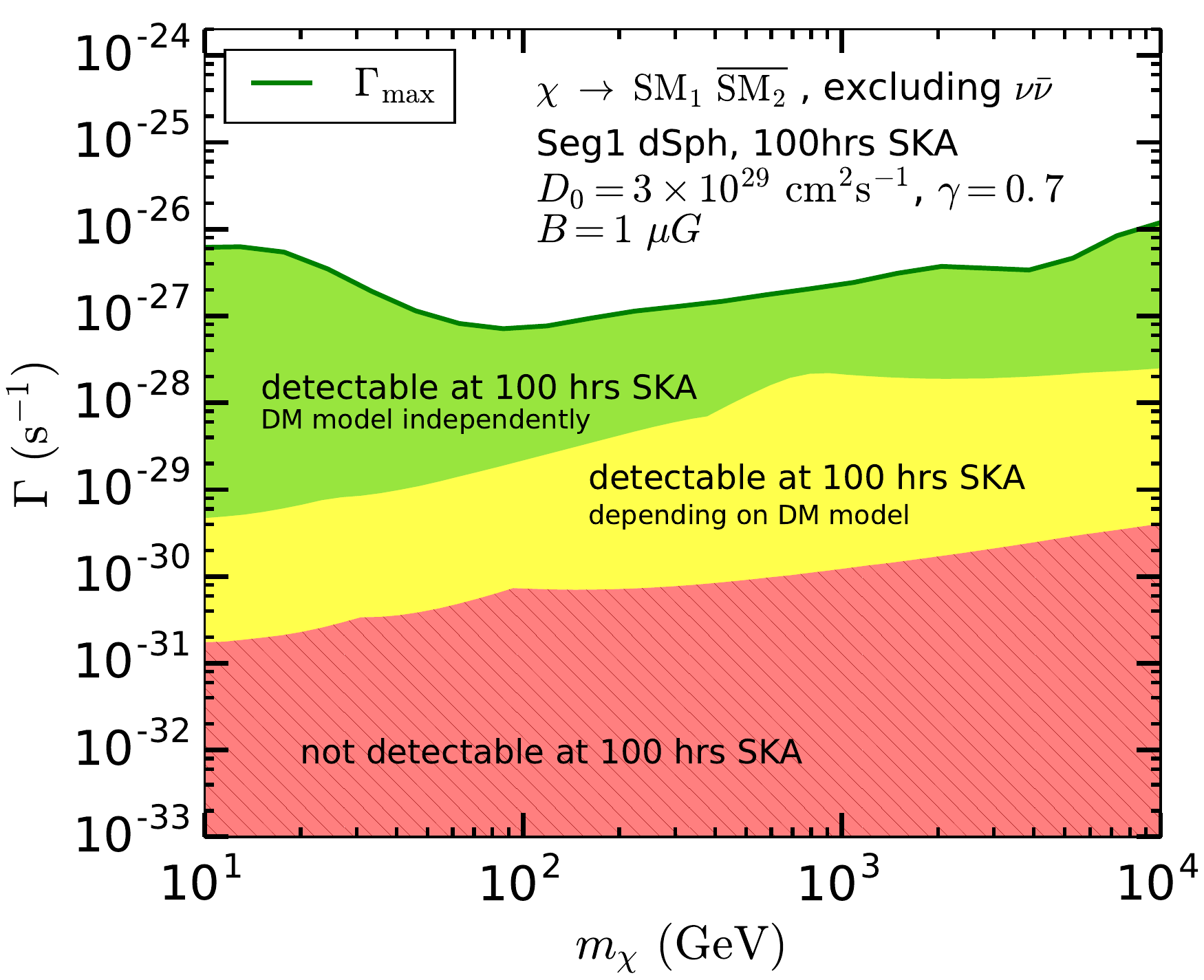}
\caption{
{\it Top:} Regions in the $m_\chi-\Gamma$ plane those are detectable 
in the 100 hours of observation of the Seg 1 dSph at the 
SKA are determined for Case 1, i.e., excluding DM decays to $\nu\bar{\nu}$ 
and varying the branching ratios of the remaining SM final states 
arbitrarily. Here we have considered 
$D_0=3\times10^{28}{\rm cm}^2{\rm s}^{-1}$ and 
$B=1\mu{\rm G}$. $\Gamma_{\rm max}$ (the green solid line) is the 
upper limit on total $\Gamma$, obtained by combining the data of 
Planck CMB, 
Fermi-LAT IGRB 
and AMS-02 
positron flux measurements. 
{\it Bottom:} The corresponding results for two other choices of the 
astrophysical parameters, viz., $D_0=3\times10^{28}{\rm cm}^2{\rm s}^{-1}$, 
$B=0.1\mu{\rm G}$ (bottom left) and $D_0=3\times10^{29}{\rm cm}^2{\rm s}^{-1}$, 
$B=1\mu{\rm G}$ (bottom right), which are even more conservative, are shown. 
The diffusion index $\gamma = 0.7$ is assumed in all the cases. 
See the text for details.
}
\label{fig:SKA_regions_1}
\end{figure*}

In the top panel of Fig.~\ref{fig:SKA_regions_1}, we have presented 
the SKA detectability of Case 1, i.e., DM decays to all possible SM 
final states other 
than $\nu\bar{\nu}$, assuming $D_0=3\times10^{28}{\rm cm}^2{\rm s}^{-1}$, 
$\gamma=0.7$ and $B=1\mu{\rm G}$. The solid green line represents $\Gamma_{\rm max}$ 
obtained by analyzing the data of Planck CMB~\cite{Planck:2015fie}, 
Fermi-LAT IGRB~\cite{Fermi-LAT:2014ryh} and AMS-02 cosmic-ray positron flux~\cite{AMS:2014xys} 
observations (as discussed in Sec.~\ref{sec:modelindepexconscons}). 
We find that, in this case, \textit{DM decays to two-body SM final states are branching 
ratio independently detectable at the SKA (100 hours) for 
$\Gamma \gsim 2 \times 10^{-30}\,{\rm s}^{-1}$ when 
$m_\chi \sim 100\,{\rm GeV}$, and for decay widths greater than 
$2 \times 10^{-29}\,{\rm s}^{-1}$ when $m_\chi \sim 1\,{\rm TeV}$.} 
On the other hand, \textit{depending on the specific branching ratio 
combinations DM decays can be probed at the SKA for 
$\Gamma \gsim 7 \times 10^{-32}\,{\rm s}^{-1}$ when 
$m_\chi \sim 100\,{\rm GeV}$, and for decay rates larger than 
$10^{-31}\,{\rm s}^{-1}$ for $m_\chi \sim 1\,{\rm TeV}$.}

As an example, when $m_\chi \sim 100\,{\rm GeV}$, 
for an allowed value of $\Gamma \sim 10^{-31}\,{\rm s}^{-1}$, 
the radio fluxes coming from $\mu^+\mu^-$, $\tau^+\tau^-$, 
$q\bar{q}$, $b\bar{b}$ and $gg$ final states are above the SKA
 sensitivity level obtained assuming 100 hours of observation time. 
Therefore, for this parameter point DM decay induced radio signals are 
detectable if the DM decays to any one of the above-mentioned 
final states 
or any possible combination of them. On the other hand, for such a 
($m_\chi$, $\Gamma$) point DM decay generated radio 
signals are non-detectable at the SKA when the decay 
occurs into $e^+e^-$, $\gamma\gamma$, $Z\gamma$ final states or 
any of their combinations. 
Therefore, this ($m_\chi$, $\Gamma$) point lies within the yellow 
region in Fig.~\ref{fig:SKA_regions_1} (top panel). 
However, if $\Gamma$ is increased at least by a factor of 20 
the radio fluxes coming from the later mentioned decay modes also become 
detectable at the SKA. As a result, the resulting ($m_\chi$, $\Gamma$) point 
is detectable at the SKA for all possible branching ratio combinations
and thus falls within the green region. On the contrary, if the decay rate is reduced by a factor of two, the resulting ($m_\chi$, $\Gamma$) point become non-detectable 
for all possible branching ratio combinations at the SKA and therefore, lies 
within the red-region
. 


On the other hand, if one considers $m_\chi \sim 1\,{\rm TeV}$, 
a point in the yellow region with 
$\Gamma \sim 2 \times 10^{-31}\,{\rm s}^{-1}$ (allowed by the existing data), 
is detectable at the SKA if the DM decays into $q\bar{q}$, $b\bar{b}$, 
$t\bar{t}$, $gg$, $W^+W^-$, $ZZ$, $Zh$, $hh$ final states or any of 
their combinations. While the fluxes coming from 
$e^+e^-$, $\mu^+\mu^-$, $\tau^+\tau^-$, $\gamma\gamma$, $Z\gamma$ decay 
modes or any of their combinations are non-detectable at the SKA. If we 
increase the DM decay rate by two orders of magnitude, the resulting parameter 
point falls within the green region.
While, if one reduces $\Gamma$ by a factor of two, the resulting 
($m_\chi$, $\Gamma$) point lies within the red region. 
\textit{For $m_\chi \sim 100\,{\rm GeV}$ and $m_\chi \sim 1\,{\rm TeV}$ the 
detectability thresholds of the DM decay rate are 
$7 \times 10^{-32}\,{\rm s}^{-1}$ 
and $10^{-31}\,{\rm s}^{-1}$, respectively, below which the radio fluxes 
are non-detectable for all possible branching ratio combinations at the 
100 hours of the SKA observation. 
Hence, such parameter points fall inside the red region in 
Fig.~\ref{fig:SKA_regions_1} (top panel)}.

As mentioned earlier, apart from $D_0=3\times10^{28}{\rm cm}^2{\rm 
s}^{-1}$ and $B=1\mu{\rm G}$, we have also considered other conservative 
values of these astrophysical parameters to demonstrate how our results 
depend on their variations. For example, in the bottom panel of 
Fig.~\ref{fig:SKA_regions_1} we have shown the regions of the DM parameter 
space, detectable or non-detectable in the 100 hours of observation 
at the SKA for Case 1, considering $D_0=3\times10^{28}{\rm cm}^2{\rm s}^{-1}$, 
$B=0.1\mu{\rm G}$ (Fig.~\ref{fig:SKA_regions_1}; bottom left) and 
$D_0=3\times10^{29}{\rm cm}^2{\rm s}^{-1}$, $B=1\mu{\rm G}$ 
(Fig.~\ref{fig:SKA_regions_1}; bottom right), respectively. 
The diffusion index $\gamma = 0.7$ is assumed in both the cases. 

Note that \textit{if one reduces $B$ the energy emission rate for synchrotron 
radiation decreases and consequently the produced radio signals will also be 
smaller (for details see~\cite{Kar:2019cqo}).}  
This is manifested in the fact that for $m_\chi \sim 100\,{\rm GeV}$, DM decays 
are branching ratio independently detectable at the SKA 
for $\Gamma \gsim 6 \times 10^{-29}\,{\rm s}^{-1}$, while DM decays 
can be probed at the SKA for certain specific combinations of branching 
fractions only if the decay rate is larger than $3 \times 10^{-30}\,{\rm s}^{-1}$ 
(see Fig.~\ref{fig:SKA_regions_1}; bottom left panel). Both of these values 
are more than one order of magnitude larger than the corresponding values 
obtained in the top panel of Fig.~\ref{fig:SKA_regions_1}.
For this DM mass, if the decay width goes below $3 \times 10^{-30}\,{\rm s}^{-1}$, 
the corresponding parameter point falls within the red region. 


On the other hand, as stated earlier, \textit{if one increases 
$D_0$ the DM decay induced electrons (positrons) 
escape from the diffusion zone of the considered dSph galaxy before 
radiating sufficient energies via synchrotron radiation and as a 
result, produced radio signals will also be 
suppressed (for details see~\cite{Kar:2019cqo}).} 
This is evident from the bottom right panel of Fig.~\ref{fig:SKA_regions_1} 
where we have considered $D_0 = 3 \times 10^{29}\,{\rm cm}^2{\rm s}^{-1}$, a value 
one order larger than the $D_0$ value assumed in the top panel of 
Fig.~\ref{fig:SKA_regions_1}. In this case, decay of a 100 GeV DM particle is 
detectable, independent of the branching fractions of each DM decay mode, for 
$\Gamma \gsim 2 \times 10^{-29}\,{\rm s}^{-1}$. 
For this $m_\chi$ value, DM decays are detectable for specific branching ratio 
combinations if the decay width is greater than $7 \times 10^{-31}\,{\rm s}^{-1}$. 
The corresponding values of $\Gamma$ for $D_0 = 3 \times 10^{28}\,{\rm cm}^2{\rm s}^{-1}$ 
were $2 \times 10^{-30}\,{\rm s}^{-1}$ and $7 \times 10^{-32}\,{\rm s}^{-1}$, 
respectively (see Fig.~\ref{fig:SKA_regions_1}; top panel).



In principle, there can be further reduction (enhancement) of the detected flux 
in case one has higher (lower) $D_0$ and lower (higher) $B$. The effects 
of such reduction or enhancement in general can be surmised, for instance,  
from the detectability plots in the $B-D_0$ plane given in the  
Refs.~\cite{Kar:2019cqo,Ghosh:2020ipv}.

\begin{figure*}[ht!]
\centering
\includegraphics[width=7.6cm,height=7.4cm]{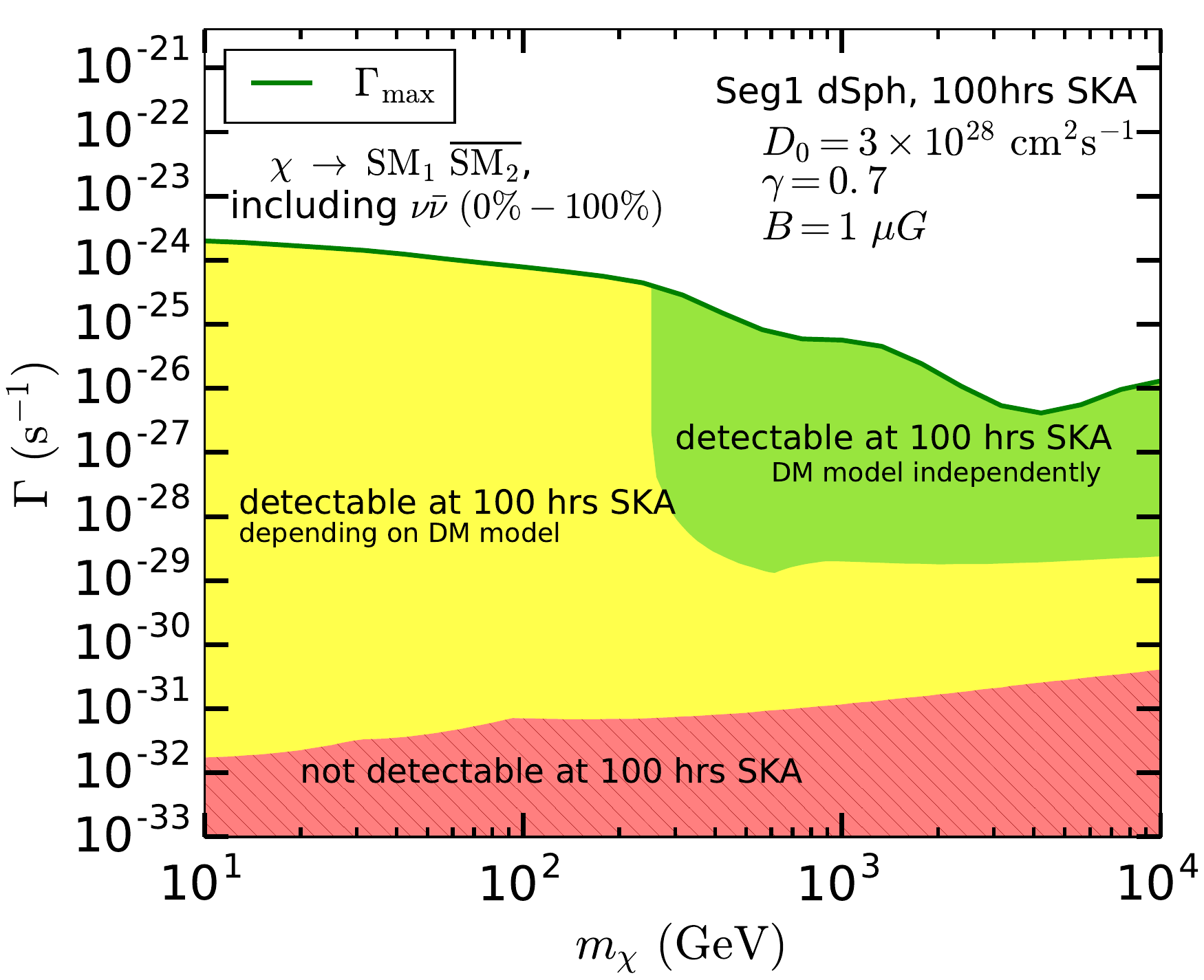}
\caption{
Classification of the DM parameter space (spanned by $m_\chi$ and 
$\Gamma$) based on the detectability at the SKA (for 100 hours of 
observation towards the Seg 1 dSph) is presented for Case 2, i.e., 
including DM decays to $\nu\bar{\nu}$ with arbitrary branching fractions. 
The choice for the astrophysical parameters made here are 
$D_0 = 3\times10^{28}{\rm cm}^2{\rm s}^{-1}$, $\gamma = 0.7$ and $B = 1\,\mu{\rm G}$. 
$\Gamma_{\rm max}$ (the green solid line) is the upper limit obtained 
by using the data of 
Planck, Fermi-LAT, AMS-02 and Super-Kamiokande. 
See the text for details.
}
\label{fig:SKA_regions_2}
\end{figure*}


Fig.~\ref{fig:SKA_regions_2} shows the detectability of the decaying 
DM parameter space for Case 2 assuming 100 hours of observation time 
at the SKA. The results shown here are obtained for 
$D_0=3\times10^{28}{\rm cm}^2{\rm s}^{-1}$, $\gamma=0.7$ 
and $B=1\mu{\rm G}$. In determining $\Gamma_{\rm max}$ 
(green solid line), we have used the data of Planck CMB~\cite{Planck:2015fie}, 
Fermi-LAT IGRB~\cite{Fermi-LAT:2014ryh}, AMS-02 positron flux~\cite{AMS:2014xys} 
and Super-Kamiokande neutrino flux~\cite{Frankiewicz:2016nyr} observations.

For the $\nu\bar{\nu}$ final state, 
when $m_\chi$ is smaller than a few hundreds of GeV, 
final state $\nu\bar{\nu}$ pairs are of low energy and hence 
the resulting radio flux is also suppressed (see the discussions of 
Fig.~\ref{fig:Snu_vs_nu}; top right panel). 
As an example, for $m_\chi \sim 100\,{\rm GeV}$ and 
$\Gamma \sim 10^{-31}\,{\rm s}^{-1}$ (allowed by the existing data), 
$S_\nu$ for DM decays to $\mu^+\mu^-$, $\tau^+\tau^-$, $q\bar{q}$, 
$b\bar{b}$, $gg$ final states or any of their 
combinations are detectable at the SKA. On the other hand, radio fluxes 
for $e^+e^-$, $\gamma\gamma$, $Z\gamma$, $\nu\bar{\nu}$ final states or 
any of their combinations are non-detectable at the SKA. Therefore, this 
($m_\chi$, $\Gamma$) point is detectable at the SKA for certain specific 
combinations of branching fractions. Keeping $m_\chi$ fixed, if we now 
increase $\Gamma$ to $\sim 10^{-25}\,{\rm s}^{-1}$, which is also allowed 
(see right panel of Fig.~\ref{fig:Gamma_max}), DM decay induced radio signals 
are non-detectable at the SKA if DM decays dominantly into 
$\nu\bar{\nu}$ final state. Therefore, this parameter point, too, lies in the 
yellow region. \textit{In fact, for $m_\chi \lesssim 250\,{\rm GeV}$ and 
$\Gamma \gtrsim 10^{-29} - 10^{-30}\,{\rm s}^{-1}$, DM decay generated radio 
flux is non-detectable at the SKA if $\nu\bar{\nu}$ is the dominant decay mode 
of the DM candidate.} This implies that, detectability of any such parameter point 
depends on the branching fractions of each two-body SM final state (including $\nu\bar{\nu}$) 
which is manifested in the fact that the yellow region in Fig.~\ref{fig:SKA_regions_2} 
extends all the way up to $\Gamma_{\rm max}$ for $m_\chi \lsim 250\,{\rm GeV}$.

For higher values of $m_\chi$, $\nu\bar{\nu}$ pairs produced from 
DM decays are highly energetic and hence, the resulting radio flux is also 
detectable at the SKA (see the discussions of Fig.~\ref{fig:Snu_vs_nu}; bottom 
right panel). For example, when $m_\chi \sim 1\,{\rm TeV}$, the radio fluxes 
coming from the $\nu\bar{\nu}$ decay mode are detectable for 
$\Gamma \sim 10^{-29}\,{\rm s}^{-1}$, a value which is allowed by the existing 
observations. However, this point still lies in the yellow region, since the 
radio fluxes coming from the $e^+e^-$ final state are non-detectable at the 
SKA. One needs to increase $\Gamma$ nearly by a factor of two, so that the 
resulting ($m_\chi$, $\Gamma$) point becomes detectable for all possible decay modes. Therefore, when $m_\chi \gsim 250\,{\rm GeV}$, DM decays are detectable at the SKA for all possible branching ratio combinations of the two-body SM final states (including $\nu\bar{\nu}$) for 
$\Gamma \gsim 2 \times 10^{-29}\,{\rm s}^{-1}$ (see Fig.~\ref{fig:SKA_regions_2}). 
On the other hand, similar to top panel of Fig.~\ref{fig:SKA_regions_1}, here, too, 
for $m_\chi \sim 100\,{\rm GeV}$, $\Gamma \lsim 7 \times 10^{-32}\,{\rm s}^{-1}$ 
and for $m_\chi \sim 1\,{\rm TeV}$, any value of the DM decay width smaller than 
$10^{-31}\,{\rm s}^{-1}$ are non-detectable at the SKA.

	
\begin{figure*}[htb!]
\centering
\includegraphics[width=7.6cm,height=7.4cm]{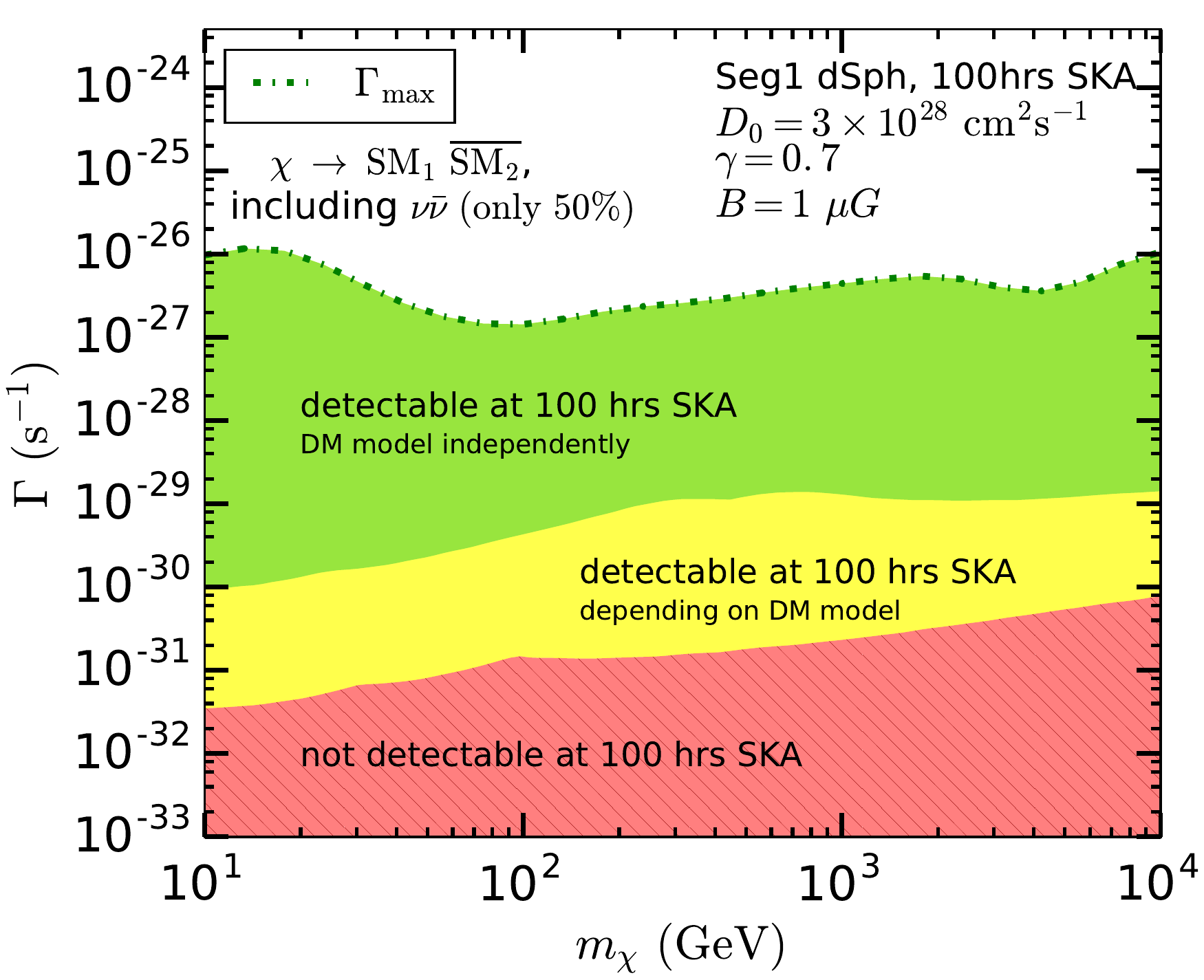}\hspace{0mm}
\includegraphics[width=7.6cm,height=7.4cm]{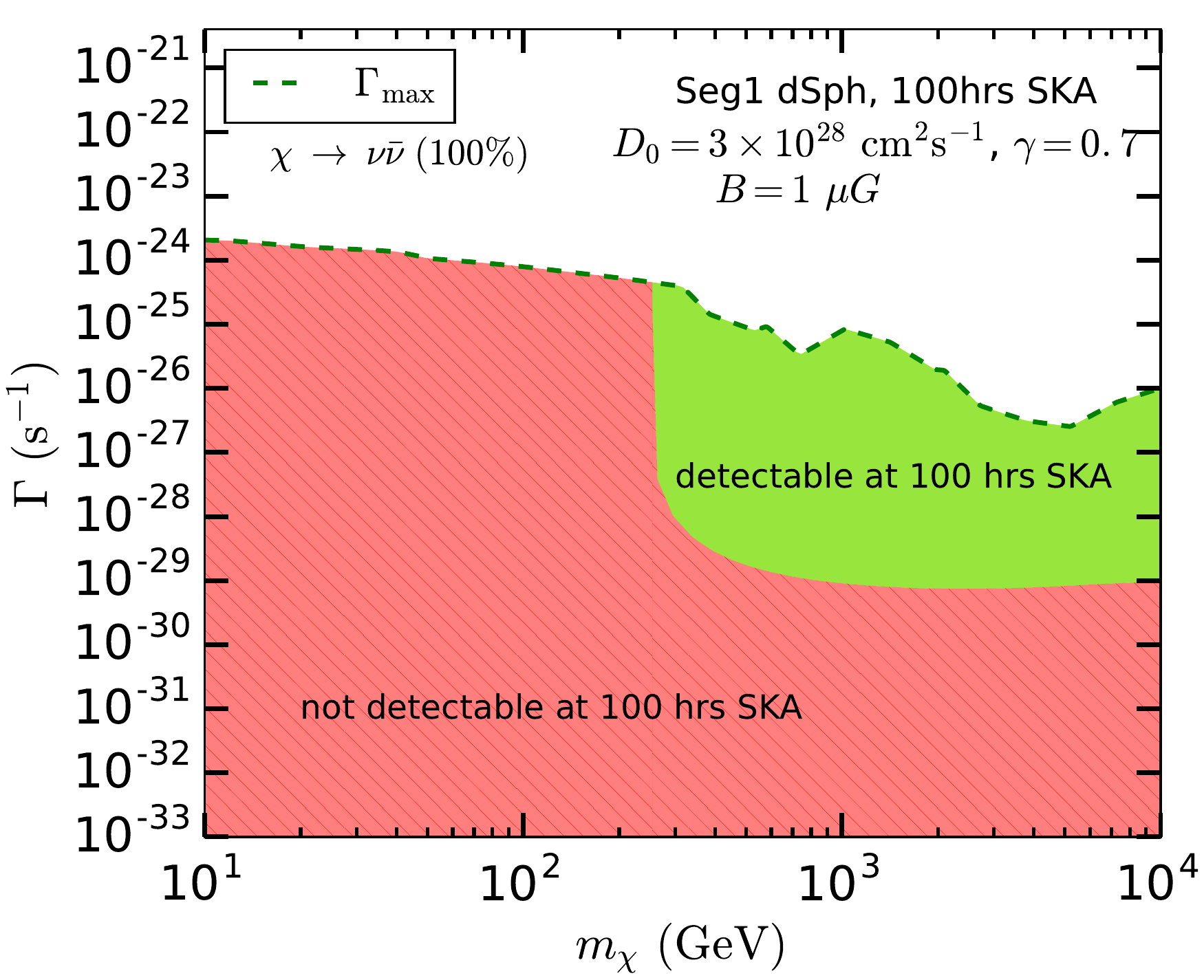}
\caption{
{\it Left:} SKA detectable 
regions (assuming 100 hours observation towards the Seg 1 dSph) 
in the $m_\chi-\Gamma$ plane and the maximum allowed decay width 
$\Gamma_{\rm max}$ (green dashed dotted line) are obtained assuming 
the $\nu \bar{\nu}$ final state to be produced with a branching ratio 
of $50\%$, i.e., Case 3. 
{\it Right:} Same as the left panel but for Case 4, i.e., considering 
exactly 100$\%$ branching ratio for the DM decay to the $\nu\bar{\nu}$ 
final state. The values of $D_0$, $\gamma$ and $B$ are same as the ones 
taken in Fig.~\ref{fig:SKA_regions_1}; top panel and in 
Fig.~\ref{fig:SKA_regions_2}. See the text for details.
}
\label{fig:SKA_regions_3}
\end{figure*}
	
From Fig.~\ref{fig:SKA_regions_2} it is clear that the inclusion of 
the $\nu\bar{\nu}$ decay mode in our analysis affects the detectability 
at the SKA considerably. This fact is further emphasized, in 
Fig.~\ref{fig:SKA_regions_3}, where the results for Case 3 (left panel) and 
Case 4 (right panel) are presented assuming 
$D_0=3\times10^{28}{\rm cm}^2{\rm s}^{-1}$, $\gamma=0.7$ and $B=1\mu{\rm G}$. 
In Case 3 (see Fig.~\ref{fig:SKA_regions_3}; left panel), branching ratio 
for the $\nu\bar{\nu}$ final state being fixed at $50\%$, branching fractions of 
the remaining SM final states are varied in a way such that they add up 
to 0.5. The green dashed dotted line represents the $\Gamma_{\rm max}$ 
(as shown in Fig.~\ref{fig:Gamma_max}; right panel). 
For $m_\chi \lsim 250\,{\rm GeV}$, $S_\nu$ coming from the $\nu\bar{\nu}$ 
final state is negligible and thus the radio fluxes 
are dominated by the contributions of the visible SM final states. 
As a result, throughout the considered mass range, DM decays are branching 
ratio independently detectable at the SKA for 
$\Gamma \gsim 10^{-30} - 10^{-29}\,{\rm s}^{-1}$. 
On the other hand, the parameter point $m_\chi \sim 100\,{\rm GeV}$ and 
$\Gamma = 2 \times 10^{-31}\,{\rm s}^{-1}$ (allowed by 
Fig.~\ref{fig:Gamma_max}; right panel) is detectable for DM 
decays to $\mu^+\mu^-$, $\tau^+\tau^-$, $q\bar{q}$, $b\bar{b}$, $gg$ 
final states or any of their combinations. 
However, such a ($m_\chi$, $\Gamma$) point is non-detectable for DM 
decays to any other channels. Thus, this 
parameter point lies in the yellow region. Note that here, the branching 
ratio of the $\nu\bar{\nu}$ final state being fixed at 50$\%$, the radio 
fluxes coming from the remaining SM final states are halved and the upper 
boundary of the red region has moved upward. As a result, now for 
$m_\chi \sim 100\,{\rm GeV}$, DM decay induced radio signals are 
non-detectable in the 100 hours of observation 
at the SKA for $\Gamma \lsim 10^{-31}\,{\rm s}^{-1}$, while, if 
$m_\chi \sim 1\,{\rm TeV}$, for any value of the decay width smaller than 
$3 \times 10^{-31}\,{\rm s}^{-1}$, DM decay generated radio 
signals are not observable at the SKA. Contrary to Case 2 
(shown in Fig.~\ref{fig:SKA_regions_2}), here the green region is 
extended below $m_\chi \sim 250\,{\rm GeV}$ and the upper boundary of 
the red region has moved upward. Both of these are related to the fact 
that here we have fixed the branching ratio for the $\nu\bar{\nu}$ 
final state to be precisely $50\%$.

In the right panel of Fig.~\ref{fig:SKA_regions_3}, detectability of 
Case 4 is shown for which the green dashed line represents $\Gamma_{\rm max}$.  
Here, DM decay width required for the detection of the $\nu\bar{\nu}$ channel 
dictates the boundary of the red and the green region, since now 
the DM decays entirely into $\nu\bar{\nu}$ 
with precisely $100\%$ branching ratio. Here, it is clear that for 
$m_\chi \lsim 250\,{\rm GeV}$ the radio signals originating from the 
$\nu\bar{\nu}$ final state are negligible enough to be detected at the 
SKA and hence for $m_\chi \lsim 250\,{\rm GeV}$, $\Gamma < \Gamma_{\rm max}$ is 
non-detectable. 
Clearly, the detectability in this case is quite different from 
that obtained in Case 2 (in Fig.~\ref{fig:SKA_regions_2}) and Case 3 
(Fig.~\ref{fig:SKA_regions_3}; left panel).


In the end, we note that every parameter point in the yellow regions 
of Figs.~\ref{fig:SKA_regions_1} - \ref{fig:SKA_regions_3} is detectable 
at the SKA (in 100 hours of observation) {\it for certain combinations of 
branching fractions}. Each such combination can be labelled as a separate 
DM model. 
On the other hand, following the analysis in Sec.~\ref{sec:conseval}, one 
finds constraints from existing astrophysical and cosmological data for each 
such `model'. This may lead to the exclusion of some model(s) corresponding 
to a particular point in the $m_\chi - \Gamma$ plane while other models are 
still allowed for those particular values of $m_\chi$ and $\Gamma$. 
Therefore, it remains meaningful to carry out radio observations in search of 
the latter class of models for any given ($m_\chi$, $\Gamma$) point in the yellow 
regions. For example, in Fig.~\ref{fig:SKA_regions_2}, the point 
$m_\chi \simeq 40\,{\rm GeV}$, $\Gamma \simeq 2 \times 10^{-28}\,{\rm s}^{-1}$ 
is ruled out by the existing data when the DM particle decays dominantly into 
$\gamma\gamma$ (see Fig.~\ref{fig:LAT_IGRB}) or 
$e^+e^-$ (see Fig.~\ref{fig:AMS_positron}) final states, while this parameter 
point is still allowed for DM decay to any other final state, say, 
$b\bar{b}$, $gg$, $\tau^+\tau^-$ etc. However, it is emphasized that radio signals 
corresponding to the allowed visible final states can be detectable in 100 
hours of observation at the SKA for the given ($m_\chi$, $\Gamma$) point. 

\section{Summary and conclusions}
\label{sec:summconc}	

In this work, we have performed a 
study of the indirect detection signals of a decaying scalar DM 
whose mass lies in the range 10 GeV - 10 TeV. We have investigated the 
implications of the existing observational data on the DM parameter 
space which set an upper limit for the allowed part of the DM parameter 
space. Then we explored the projected reach of the upcoming SKA radio telescope 
in discovering the DM decay induced radio signals coming from the Seg 1 
dwarf galaxy. We start by presenting the limits on 
the DM decay width 
obtained from the data of Planck CMB~\cite{Planck:2015fie}, 
Fermi-LAT IGRB~\cite{Fermi-LAT:2014ryh}, AMS-02 positron flux~\cite{AMS:2014xys} 
and Super-Kamiokande neutrino flux~\cite{Frankiewicz:2016nyr} observations 
assuming DM decays to any particular two-body SM final state with 100$\%$ 
branching ratio. 

We have determined the maximum allowed decay width 
by varying the branching ratios of each individual SM final states 
arbitrarily in the range $0\% - 100\%$, such that their sum always equals 
unity. In deriving this limit we have used the data of 
Planck CMB~\cite{Planck:2015fie}, Fermi-LAT IGRB~\cite{Fermi-LAT:2014ryh} 
and AMS-02 positron observation~\cite{AMS:2014xys} while considering 
DM decays to all possible SM final states other than neutrino-antineutrino 
pairs. We found that throughout our chosen range of the DM mass, decay rates smaller 
than $\sim 10^{-27}\,{\rm s}^{-1}$ are allowed. This limit is weaker by nearly 
an order of magnitude compared to the results obtained when 100$\%$ 
branching ratio is attributed to each individual DM decay mode. 
In addition, the Super-Kamiokande neutrino data of the Milky Way 
galaxy~\cite{Frankiewicz:2016nyr} are taken into account 
when DM decay to neutrino-antineutrino pairs is included in our analysis. 
For the $\nu\bar{\nu}$ final state the Super-Kamiokande 
constraints are found to be stronger than Planck, Fermi-LAT and 
AMS-02 limits particularly for DM particles lighter than $500\,{\rm GeV}$. 
On allowing the branching fractions of all SM final states, including 
that of the neutrino pairs, to vary in the range $0\% - 100\%$, 
we found that for $m_\chi \sim \mathcal{O}({\rm TeV})$ or larger, 
the maximum value of the decay width that is allowed is 
$\sim 10^{-26}\,{\rm s}^{-1}$, while, for DM mass less than a hundred GeV, 
$\Gamma \lsim 10^{-24}\,{\rm s}^{-1}$ is consistent with the existing 
data. We have also studied the cases where the branching ratio for the 
neutrino-antineutrino final state is fixed at 50$\%$ and 100$\%$, respectively. 
When a branching fraction of $50\%$ is attributed to the $\nu\bar{\nu}$ 
final state the upper limit on the decay rate is found to be 
$\sim 10^{-27}\,{\rm s}^{-1}$ throughout the DM mass range considered, 
while for 100$\%$ branching to neutrino-antineutrino pairs, 
the maximum allowed decay width roughly follows the constraint obtained 
when an arbitrary branching fraction is allowed for the $\nu\bar{\nu}$ final state. 

Next, we show how the DM parameter space that is allowed by the existing 
observations can be classified into regions which are detectable either 
for all possible branching ratio combinations or for certain specific 
combinations of branching fractions and non-detectable at the SKA, 
assuming 100 hours of observation towards the Seg 1 dSph. 
It is found that for DM decays to 
two-body SM final states (excluding $\nu\bar{\nu}$), 
decay rates greater than $10^{-31}\,{\rm s}^{-1} - 10^{-29}\,{\rm s}^{-1}$ 
are detectable for all possible branching ratio combinations of the DM decay 
modes while $\Gamma \gsim 10^{-32}\,{\rm s}^{-1} - 10^{-31}\,{\rm s}^{-1}$ are 
detectable only for certain specific branching ratio combinations, 
throughout the DM mass range 10 GeV - 10 TeV, assuming 
$D_0=3\times10^{28}{\rm cm}^2{\rm s}^{-1}$ and $B=1\mu{\rm G}$. 
For even more conservative choices of these astrophysical parameters, 
namely, $D_0=3\times10^{28}{\rm cm}^2{\rm s}^{-1}, B=0.1\mu{\rm G}$ and 
$D_0=3\times10^{29}{\rm cm}^2{\rm s}^{-1}, B=1\mu{\rm G}$ we found that, 
the minimum values of the DM decay widths required for branching ratio 
independent detection at the SKA lie in the ranges 
$10^{-28}\,{\rm s}^{-1} - 10^{-27}\,{\rm s}^{-1}$ and 
$10^{-29}\,{\rm s}^{-1} - 10^{-28}\,{\rm s}^{-1}$, respectively.

 
On the other hand, when DM decay to neutrino-antineutrino pairs is included in 
our analysis and the corresponding branching fraction is also varied 
in the range $0\%-100\%$, we found that DM decays are detectable, independent of 
the branching ratio combinations, 
for 
decay widths larger than $\sim 10^{-29}\,{\rm s}^{-1}$, only if 
the DM particle is heavier than $250\,{\rm GeV}$, assuming 
$D_0=3\times10^{28}{\rm cm}^2{\rm s}^{-1}$ and $B=1\mu{\rm G}$. 
For lower values of $m_\chi$ the detectability at the SKA depends on the 
branching fraction of the $\nu\bar{\nu}$ decay mode and if this is 
the dominant decay mode, DM decays are non-detectable at the SKA 
for all allowed values of the DM decay width. 
This fact is further explained by considering exactly 50$\%$ and 
100$\%$ branching ratios for the neutrino-antineutrino final state. 
When the $\nu\bar{\nu}$ decay mode is produced with 
50$\%$ branching ratio, one finds that, throughout the DM mass range 
under consideration, the DM decays are detectable for 
all branching ratio combinations of the non-neutrino SM final states 
only if the decay width is greater than 
$10^{-30} - 10^{-29}\,{\rm s}^{-1}$. 
On the contrary, when 100$\%$ branching ratio is attributed to the 
neutrino final state the detectability at the SKA is entirely 
dictated by the radio fluxes coming from the $\nu\bar{\nu}$ channel 
and hence DM decays are detectable only for DM masses larger than 
$\sim 250\,{\rm GeV}$.  

Note that the results presented here for the Seg 1 dSph would not change 
appreciably had one considered any other dSph such as Draco, Carina, Fornax 
etc.~\cite{Colafrancesco:2006he,Kar:2019hnj,Geringer-Sameth:2014yza,Ghosh:2020ipv,Kar:2018rlm,Kar:2019cqo}, as the target. 
As pointed out in Sec.~\ref{sec:DMdecaytheory}, the constraints on the DM decay 
rate obtained assuming DM decay is the only source of indirect detection signals 
will get even stronger in presence of appreciable DM annihilation rates, which 
would more severely constrain the fluxes coming from the DM decays. Additionally, 
note that the present analysis is applicable to, and consistent with, the 
scenarios discussed in Sec.~\ref{sec:DMdecaytheory} individually, so long as 
the decaying DM particle saturates the relic density. Although our basic 
methodology can be used, with appropriate modifications, to multi-component DM 
scenarios as well, where only some components are liable to decay. However, 
the constraints on the DM decay width tend to get weakened in such cases.



To sum it up, here we have considered a decaying scalar DM of 
mass in the 10 GeV - 10 TeV range. Taking all possible 
kinematically allowed two-body SM decay modes 
into account we have constrained the 
DM parameter space in a purely branching ratio 
independent fashion by utilizing the data of Planck CMB~\cite{Planck:2015fie}, 
Fermi-LAT IGRB~\cite{Fermi-LAT:2014ryh}, AMS-02 positron flux~\cite{AMS:2014xys} 
and Super-Kamiokande neutrino flux~\cite{Frankiewicz:2016nyr} observations. 
The maximum allowed total DM decay width 
for each value of the DM mass has been derived by varying the branching 
ratios of all kinematically allowed DM decay modes arbitrarily so that their 
sum always equals unity. Any value of the DM decay rate larger than this maximum 
allowed value 
is ruled out by at least one existing observation for all possible branching 
ratio combinations. In the next step, the allowed region of the DM parameter space 
obtained this way has been classified into three distinct regions, viz., detectable for all possible branching ratio combinations, 
detectable for specific branching ratio combinations of the two-body SM final 
states and non-detectable, based on the detectability at the SKA assuming 
100 hours of observation time towards the Segue 1 dSph. Our analysis shows 
that, irrespective of whether the DM candidate decays to neutrino-antineutrino 
pairs or not, the SKA can probe much deeper into the parameter space of a generic 
decaying scalar DM, as compared to the existing observations.

\section*{Acknowledgements}
The work of K.D. is partially supported by the Indo-Russian grant 
DST/INT/RUS/RSF/P-21, MTR/2019/000395, and Core Research Grant 
CRG/2020/004347 funded by SERB, DST, Government of India. 
The research of A.K. was supported by the National Research 
Foundation of Korea (NRF) funded by the Ministry of Education 
through the Center for Quantum Space Time (CQUeST) of Sogang University 
with Grant No. 2020R1A6A1A03047877.	
	
\providecommand{\href}[2]{#2}
\addcontentsline{toc}{section}{References}
\bibliographystyle{JHEP}

\bibliography{refs}

\end{document}